\documentclass[11pt]{article}
\pdfoutput=1
\usepackage[whole]{bxcjkjatype}
\usepackage[utf8]{inputenc}
\usepackage[pagebackref]{hyperref}
\usepackage{xcolor}
\usepackage{amsfonts}
\usepackage{amsmath}
\usepackage{amssymb}
\usepackage{bbm}
\usepackage{mathrsfs}  
\usepackage{physics}
\usepackage{multicol}
\usepackage{setspace}
\usepackage{booktabs}
\usepackage[mathscr]{euscript}
\usepackage{stmaryrd}
\usepackage[root radius=.8mm,edge length= 5mm,mark=o]{dynkin-diagrams}

\usepackage{tikz}
\usetikzlibrary{arrows}
\usetikzlibrary{calc}
\usetikzlibrary{cd}
\usetikzlibrary{decorations.pathreplacing} 
\usetikzlibrary{decorations.markings}
\usetikzlibrary{shadows,fadings}

\usepackage[
 a4paper,
 total={160mm,257mm},
 left=25mm,
 top=20mm]{geometry}
\usepackage{amsthm}
\usepackage{authblk}

\hypersetup{colorlinks,breaklinks,
            linkcolor = purple,
            citecolor = teal,
            urlcolor = purple,
            }

\newcommand{\ee}{\mathrm{e}}
\newcommand{\ii}{\mathrm{i}}
\newcommand{\id}{\mathbbm{1}}

\DeclareMathOperator{\str}{\operatorname{str}}
\DeclareMathOperator{\sdet}{\operatorname{sdet}}
\DeclareMathOperator*{\res}{\operatorname{Res}}

\def\Xint#1{\mathchoice
   {\XXint\displaystyle\textstyle{#1}}%
   {\XXint\textstyle\scriptstyle{#1}}%
   {\XXint\scriptstyle\scriptscriptstyle{#1}}%
   {\XXint\scriptscriptstyle\scriptscriptstyle{#1}}%
   \!\int}
\def\XXint#1#2#3{{\setbox0=\hbox{$#1{#2#3}{\int}$}
     \vcenter{\hbox{$#2#3$}}\kern-.5\wd0}}

\def\dashint{\Xint-}

\usepackage[framemethod=TikZ]{mdframed}

\newenvironment{itembox}[1]{\begin{mdframed}[
  roundcorner=5pt,
  skipabove=\topskip,
  frametitleaboveskip=\dimexpr-0.7\baselineskip,
  innertopmargin=\dimexpr-0.25\baselineskip,
  innerbottommargin=\dimexpr0.5\baselineskip,
  frametitle={\tikz{\node[anchor=base,rectangle,fill=white]{\strut #1};}}]
  }
  {%\vspace{.3\baselineskip}
   \end{mdframed}}

\numberwithin{equation}{section}

\title{\textbf{Aspects of Supergroup Gauge Theory}} 
\author{\textsc{Taro Kimura} \\[.5em] \hspace*{-1.75em} 木村~太郎}
\affil{%
Institut de Mathématiques de Bourgogne,
Université de Bourgogne,
France\\
}
\date{}

\begin{document}

\maketitle

\begin{abstract}
%%
%This is a review article on supergroup gauge theory.
%
We provide a survey of recent studies of supergroup gauge theory.
We first discuss supermatrix model as a zero-dimensional toy model of supergroup gauge theory and its geometric and algebraic characterization.
We then focus on four-dimensional Yang--Mills theory with supergroup gauge symmetry and explore its non-perturbative properties, including instanton calculus, Seiberg--Witten geometry, Bethe/gauge correspondence, and its realization with  intersecting defects.

%\keywords{Keyword1; keyword2; keyword3.}
\end{abstract}

%\ccode{PACS numbers:}

\setcounter{tocdepth}{2}
\tableofcontents

%\newpage

\vspace{2em}
\hrule
\vspace{1em}

\section{Introduction}

One of the important features of quantum mechanics is the quantum statistics:
Quantum particles are indistinguishable from one another, and there exist two possible types of particles, \emph{boson} and \emph{fermion},%
\footnote{%
There exists another type of particles, called \emph{anyon}, uniquely in $(2+1)$-dimensional systems. 
} 
obeying the Bose--Einstein statistics~\cite{Bose:1924} and the Fermi--Dirac statistics~\cite{Fermi:1999ncp,Dirac:1926jz}.
\emph{Supersymmetry} is a symmetry between bosons and fermions.
In the context of supersymmetric quantum field theory, it is formulated as a space-time symmetry based on the superfield formalism,%
\footnote{%
The first appearance of supersymmetry was as an internal symmetry introduced to describe both mesons and baryons in a unified way based on the unitary supergroup proposed by Miyazawa~\cite{Miyazawa:1966mfa,Miyazawa:1968zz}.
%Addendum: At the day of submission of this article, I was informed that Prof. Miyazawa passed away in 14 January, 2023.
%Although I couldn't directly talk with him, I saw him just only once in the seminar held at RIKEN when I was a postdoc there. 
%He was indeed one of the star researchers for me in the theoretical physics community in Japan. R.I.P
} 
which provides various theoretical frameworks to understand non-perturbative aspects of quantum field theory.
Although it is typically considered as a global symmetry, one may consider a local version of supersymmetry, which inevitably involves gravitational degrees of freedom with supersymmetry, a.k.a., supergravity. %~\cite{Wess:1992cp,Freedman:2012zz,Tanii:2014gaa,Nath:2016qzm,DallAgata:2021uvl}
In addition to the framework to describe the fundamental interactions in the nature, supersymmetry has so far provided various applications based on its mathematical structure:
In the context of condensed-matter physics, the method of supersymmetry is applied to discuss disorder systems as an alternative approach to the replica trick~\cite{Efetov:1996,Wegner:2016ahw};
The lattice model involving both hopping and spin interactions, called the $t$-$J$ model, realizes supersymmetry by tuning the hopping parameter $t$ and the spin coupling constant $J$~\cite{Sutherland:1975vr,Schlottmann:1987zz};
It has been argued that the critical point of the statistical model with the random field disorder involves emergent supersymmetry (Parisi--Sourlas supersymmetry) together with dimensional reduction~\cite{Parisi:1979ka,Parisi:1982ud};
The critical point of two-dimensional tricritical Ising model is described as a minimal model of super-Virasoro (and also the ordinary Virasoro) algebra of central charge $c = 7/10$~\cite{Friedan:1984rv};
The use of supergroup is also proposed in the context of functional renormalization group to consider a gauge invariant regulator~\cite{Arnone:2000qd}.

The notion of symmetry is well described in terms of group theory, and \emph{supergroup} is an extension of the ordinary groups involving both bosonic and fermionic degrees of freedom.
The main object that we discuss in this article is \emph{supergroup gauge theory}, which is a gauge theory with local supergroup gauge symmetry.
%
%\subsection*{Spin-statistics theorem}
%
Regarding the particle statistics, Fierz and Pauli formulated a connection with the particle spins, which is nowadays known as the \emph{spin-statistics theorem}.
\begin{itembox}{Spin-statistics theorem~\cite{Fierz:1939,Pauli:1940zz}}
Integer-spin particles are bosons, while odd-half-integer–spin particles are fermions. 
\end{itembox}
This theorem is obtained under the following conditions:
\begin{itemize}
    \item Lorentz covariance and relativistic causality
    \item Positive energies and positive norms in the Hilbert space
\end{itemize}
Therefore, for example, this theorem is not applied to the ghost particle appearing in the gauge fixing process, which is a spin-0 fermionic particle.
The fundamental degrees of freedom of supergroup gauge theory are spin-1 boson and fermion, and thus it is not compatible with the spin-statistics theorem.
In fact, the spectrum of supergroup gauge theory is not bounded and there appear negative energy states.
Even in such a situation, one may still apply the method of Lefschetz thimble to evaluate the path integral~\cite{Witten:2010cx,Cristoforetti:2012su} through the analytic continuation.
Furthermore, it has been also known that $\mathrm{U}(N|M)$ gauge theory is not distinguishable with $\mathrm{U}(N-M|0) = \mathrm{U}(N-M)$ theory to all orders in perturbation theory~\cite{Alvarez-Gaume:1991ozb,Yost:1991ht,Berkovits:1999im,Vafa:2001qf,Vafa:2014iua,Dijkgraaf:2016lym}.
However, instability of vacua implies that the perturbative analysis is not reliable and we need a proper non-perturbative treatment of supergroup gauge theory.
In spite of unphysical natures, as discussed in this article, we can discuss various non-perturbative aspects of supergroup gauge theory as a natural extension of the ordinary gauge theory, and it indicates a chance of non-perturbative completion.

\subsection*{Organization of the article}

A purpose of this article is to provide a self-contained overview of supergroup gauge theory.
For this purpose, we start in Sec.~\ref{sec:supermath} with basic notions of supermathematics, including the introduction of Grassmann algebras, supervector space, superalgebra, supermatrix, and Lie supergroup, which could be skipped by experienced readers on this subject.

In Sec.~\ref{sec:super_matrix}, we discuss the supermatrix model, which can be viewed as a zero-dimensional supergroup gauge theory.
In fact, the supermatrix model plays a role of a toy model in the study of supergroup gauge theory, which exhibits various similar properties to higher-dimensional theory discussed in the latter part of this article.
After introducing the eigenvalue integral form of the partition function, in particular we study the asymptotic behavior based on the Coulomb gas method.
We then study the operator formalism that we call the free field realization of the (super)matrix model, and explore the underlying infinite dimensional algebraic structure. 

In Sec.~\ref{sec:supergroup_gauge_theory}, we introduce the supergroup gauge theory and discuss its basic properties.
We discuss realizations of supergroup theory from non-supergroup theory through analytic continuation to unphysical regime, and we show the construction from string/M-theory perspective providing the Seiberg--Witten geometry for supergroup gauge theory.

In Sec.~\ref{sec:instanton}, we explore instantons in supergroup gauge theory, which plays an important role in the study of non-perturbative aspects.
After considering the ADHM construction of instantons, a systematic approach to construct the instanton solutions, for supergroup theory, we study the instanton moduli space and apply the equivariant localization formalism to derive the instanton partition function.
We obtain the instanton partition function in the three-fold way, the equivariant index formula, the contour integral formula, and the combinatorial formula, for supergroup gauge theory.

In Sec.~\ref{sec:non-perturbative_study}, we explore non-perturbative aspects of supergroup gauge theory based on the instanton partition function obtained in advance.
We demonstrate that the instanton partition function is also obtained in the framework of topological string involving both positive and negative branes.
For this purpose, we introduce the negative brane analog of the topological vertex that we call the anti-veretx to compute the supergroup partition function.
We then study the non-perturbative Schwinger--Dyson equation associated with the instanton partition function, which gives rise to doubly quantum Seiberg--Witten geometry, and discuss the free field realization associated with the underlying algebraic structure.
In this case, we show that the instanton partition function obeys a $q$-analog of the Virasoro constraint. 
We also explore a connection with the quantum integrable system, called the Bethe/gauge correspondence, for supergroup gauge theory, and discuss its implications on the quantum integrable system side.
We then discuss realizations of supergroup gauge theory in physical setups.
We in particular study the codimension-two surface defect operators and show that the supergroup structure emerges from intersecting defects.

\subsection*{Notations}

For $N \in \mathbb{N}$, we define the set
\begin{align}
    [N] = \{ 1,\ldots,N \}
    \, .
    \label{eq:set_N}
\end{align}
Throughout the article, the vector bundle $\mathbf{X}$ has the Chern character,
\begin{align}
    \operatorname{ch} \mathbf{X} = \sum_{i \in [\operatorname{rk} \mathbf{X}]} \ee^{x_i}
    \, ,
\end{align}
where $\operatorname{rk} \mathbf{X}$ is the rank of the bundle $\mathbf{X}$.
We denote the one-dimensional bundle by $\mathbf{x}$ with $\operatorname{ch} \mathbf{x} = \ee^{x}$.
We denote the alternating sum of anti-symmetrizations of the bundle $\mathbf{X}$ by
\begin{align}
    \wedge_y \mathbf{X} = \sum_{i =0}^{\operatorname{rk} \mathbf{X}} (-y)^i \wedge^i \mathbf{X}
    \, .
\end{align}
In particular, we apply the notation, $\wedge \mathbf{X} = \wedge_1 \mathbf{X}$.

\paragraph{Special functions}
We define the $q$-shifted factorial ($q$-Pochhammer symbol)
\begin{align}
    (z;q)_n = \prod_{m \in [n]} (1 - z q^{m-1})
    \, ,
    \label{eq:q-factorial}
\end{align}
and the theta function with the elliptic norm $q$,
\begin{align}
    \theta(z;q) = (z;q)_\infty (qz^{-1};q)_\infty
    \, .
\end{align}

\section{Supermathematics}\label{sec:supermath}

In this section, we summarize the basic properties of supermathematics used in this article.
See, e.g.,~\cite{Berezin:1987,DeWitt:1992,Frappat:1996pb,Varadarajan:2004yz,Freund:1986ws,Quella:2013oda,Wegner:2016ahw} for general introductions to supermathematics.

\subsection{Grassmann algebras}

The starting point of supermathematics is the Grassmann algebra, which involves anti-commuting variables (Grassmann variables),
\begin{align}
    \theta_i \theta_j = - \theta_j \theta_i
    \, .
\end{align}
This anti-commutativity immediately leads to nilpotency of the Grassmann variables,
\begin{align}
    \theta_i^2 = 0
    \, .
\end{align}
In general, a product of even number of Grassmann variables is commutative (Grassmann even), while a product of odd number of Grassmann variables is anti-commutative (Grassmann odd).
We remark that the Grassmann even variable is commutative, but still nilpotent: For example, we have $(\theta_1 \theta_2) \theta_3 = \theta_3 (\theta_1 \theta_2)$, while $(\theta_1 \theta_2)^2 = 0$.

In the definition of the complex conjugation for the Grassmann variables, we need a modification compared with the ordinary variables:
Applying the conjugation operator twice, we have an extra sign factor,
\begin{align}
    \bar{\bar{\theta}} = - \theta
    \, .
    \label{eq:Grassmann_var_double_conj}
\end{align}
From this definition, the norm of the Grassmann variable becomes self-conjugate,%
\footnote{%
Another convention is also applied in the literature: 
No sign factor for the double conjugation $\bar{\bar{\theta}} = \theta$, while the conjugation of the product is given by $\overline{\theta_1 \theta_2 \cdots \theta_n} = \bar{\theta}_n \cdots \bar{\theta}_2 \bar{\theta}_1$.
In this convention, we still have the relation $\overline{\bar{\theta} \theta} = \bar{\theta} \theta$.
}
\begin{align}
    \overline{\bar{\theta} \theta} = - \theta \bar{\theta} = \bar{\theta} \theta \, ,
\end{align}
and the Hermitian scalar product behaves as
\begin{align}
    \overline{\bar{\theta}_1 \theta_2} = - \theta_1 \bar{\theta}_2 = \bar{\theta}_2 \theta_1 \, .
\end{align}

\subsubsection*{Derivative and integral}

The derivative and the integral for the Grassmann variable are defined as follows,
\begin{align}
    \dv{\theta_i}{\theta_j} = \delta_{ij}
    \, , \qquad
    \int \dd{\theta}_i \theta_j = \delta_{ij}
    \, , \qquad
    \int \dd{\theta}_i 1 = 0
    \, .
\end{align}
Hence, these operations are essentially equivalent.
Since the Grassmann variables are anti-commutative, we have to be careful of the ordering of the derivative and the integral operation.
Under the linear transformation $\theta_i \to \theta'_i = (A \theta)_i = \sum_j A_{ij} \theta_j$, the integral measure behaves as follows
\begin{align}
    \int \prod_i \dd{\theta}'_i = \int \prod_{i} \dd{\theta}_i (\det A)^{-1}
    \, ,
    \label{eq:Jacobian_Grassmann}
\end{align}
which is opposite to the measure behavior of the commutative variables involving the determinant factor in the numerator.

\subsection{Supervector space}

A (complex) supervector is an element of the (complex) supervector space, which is a $\mathbb{Z}_2$-graded vector space of dimension $(N|M)$, 
\begin{align}
    \Psi = (z_i, \theta_j)_{i \in [N], j \in [M]} \in \mathbb{C}^{N|M} = \mathbb{C}_0^N \oplus \mathbb{C}_1^M \, ,
    \label{eq:supervec_def}
\end{align}
where the index of $\mathbb{C}_\sigma$, $\sigma = 0,1$ denotes the Grassmann parity.
In general, a supervector space consists of even and odd subspaces,
\begin{align}
    V = V_0 \oplus V_1
    \, .
\end{align}
We denote the parity of the element $x$ by $|x| = 0$ if $x \in V_0$ and $|x| = 1$ if $x \in V_1$. 
We define the parity flip operator $\Pi$, such that $(\Pi V)_0 = V_1$ and $(\Pi V)_1 = V_0$.
The superdimension of $V$ is defined as
\begin{align}
    \operatorname{sdim} V = \sum_{\sigma = 0, 1} (-1)^\sigma \operatorname{dim} V_\sigma = \operatorname{dim} V_0 - \operatorname{dim} V_1
    \, ,
    \label{eq:sdim_def}
\end{align}
which may take a negative integer value.

\subsubsection*{Superalgebra}

Superalgebra is a $\mathbb{Z}_2$-graded algebra that consists of the even part and the odd part,
\begin{align}
    \mathfrak{A} = \mathfrak{A}_0 \oplus \mathfrak{A}_1
    \, .
\end{align}
We define the supercommutator for $x,y \in \mathfrak{A}$ as follows,
\begin{align}
    [x,y] = xy - (-1)^{|x||y|} yx
    \, ,
    \label{eq:supercommutator_def}
\end{align}
which is skew-symmetric,
\begin{align}
    [x,y] = - (-1)^{|x||y|} [y,x]
    \, .
\end{align}
Then, we define the Lie superalgebra that obeys the super analog of the Jacobi identity,
\begin{align}
    [x,[y,z]] = [[x,y],z] + (-1)^{|x||y|} [y,[x,z]]
    \, .
    \label{eq:Jacobi_id}
\end{align}

\subsection{Supermatrix}\label{sec:supermatrix}

For supervector spaces $V$ and $W$, we define a linear map, which is described using a supermatrix,
%\begin{subequations}
\begin{align}
    M & = 
    \begin{pmatrix}
    M_{00} & M_{01} \\ M_{10} & M_{11}
    \end{pmatrix}
    \in \operatorname{Hom}(V,W) = \bigoplus_{\sigma,\sigma' = 0,1} \operatorname{Hom}(V_{\sigma'},W_\sigma)
    \, ,
    \label{eq:smat_def}
\end{align}
where each block of the matrix is given by
\begin{align}
    M_{\sigma \sigma'} & \in \operatorname{Hom}(V_{\sigma'},W_\sigma)
    \, .
\end{align}
%\end{subequations}
Since $M_{01}$ and $M_{10}$ change the Grassmann parity, they consists of the Grassmann odd variables.
In particular, if $V = W$, we have $M \in \operatorname{End}(V) = \bigoplus_{\sigma,\sigma' = 0, 1} \operatorname{Hom}(V_\sigma,V_{\sigma'})$, which is invertible if both $M_{00}$ and $M_{11}$ are invertible.
The set of invertible supermatrices defines a general linear supergroup, $\mathrm{GL}(V) = \mathrm{GL}(V_0|V_1)$.

We remark that the definition of supermatrix is not unique to~\eqref{eq:smat_def}.
For example, denoting even and odd variables by $x_{ij}$ and $\xi_{ij}$, we have the following possibilities for $V = W = \mathbb{C}^{2|1}$,
\begin{align}
    M =
    \begin{pmatrix}
        x_{11} & x_{12} & \xi_{13} \\
        x_{21} & x_{22} & \xi_{23} \\
        \xi_{31} & \xi_{32} & x_{33}
    \end{pmatrix}
    \, , \qquad
    \begin{pmatrix}
        x_{11} & \xi_{12} & x_{13} \\
        \xi_{21} & x_{22} & \xi_{23} \\
        x_{31} & \xi_{32} & x_{33}
    \end{pmatrix}    
    \, , \qquad
    \begin{pmatrix}
        x_{11} & \xi_{12} & \xi_{13} \\
        \xi_{21} & x_{22} & x_{23} \\
        \xi_{31} & x_{32} & x_{33}
    \end{pmatrix}    
    \, ,
\end{align}
which correspond to the convention of supervector, $\Psi = (z_1,z_2,\theta)$, $(z_1,\theta,z_2)$, $(\theta,z_1,z_2)$.
This is related to the ambiguity of the root system in Lie superalgebra.
See a related discussion in Sec.~\ref{sec:Hany-Witten_const}.

\subsubsection*{Supertrace}

We define the supertrace operation for the supermatrix,%
\footnote{%
Not to be confused with the symmetrized trace.
We will introduce one-parameter deformation of the supertrace in~\eqref{eq:supertrace_b}.
}%
\begin{align}
    \str M = \sum_{\sigma = 0,1} (-1)^\sigma \tr_\sigma M 
\end{align}
where $\tr_\sigma$ means the trace with respect to the subspace $V_\sigma$: $\tr_\sigma M = \tr M_{\sigma\sigma}$.
Compared with the definition of the superdimension~\eqref{eq:sdim_def}, the supertrace of the identity supermatrix provides the superdimension of the corresponding supervector space, $\str \id_V = \operatorname{sdim} V$.
An important property of the supertrace is the cyclicity,
\begin{align}
    \str M_1 M_2 = \str M_2 M_1
    \, ,
    \label{eq:str_cyclic}
\end{align}
which is analogous to the cyclic property of the ordinary matrix.

\subsubsection*{Supertransposition}

We define the supertransposition operation for the supermatrix,%
\begin{align}
    \begin{pmatrix}
        A & B \\ C & D
    \end{pmatrix}^\text{st}
    =
    \begin{pmatrix}
        A^\text{t} & C^\text{t} \\ -B^\text{t} & D^\text{t}
    \end{pmatrix}
    \, ,
\end{align}
where we denote the ordinary transposition by $A^\text{t}$, etc.
This supertransposition shows an analogous property to the ordinary case, $(M_1 M_2)^\text{st} = M_2^\text{st} M_1^\text{st}$, while it is not an involution, $(M^\text{st})^\text{st} \neq M$ in general.
The Hermitian conjugation is defined as follows,
\begin{align}
    M^\dag = \overline{M}^\text{st}
    \, ,
\end{align}
which is then an involution, $(M^\dag)^\dag = M$.

\subsubsection*{Superdeterminant}

We define the superdeterminant, which is also called the Berezinian,
\begin{align}
        \sdet \begin{pmatrix}
        A & B \\ C & D
    \end{pmatrix}
    = \frac{\det (A - B D^{-1} C)}{\det D}
    = \frac{\det A}{\det (D - C A^{-1} B)}
    \, .
    \end{align}    
    This is analogous to the determinant formula for the block matrix,
    \begin{align}
        \det \begin{pmatrix}
        A & B \\ C & D
    \end{pmatrix}
    = \det (A - B D^{-1} C) \det D
    = \det A \det (D - C A^{-1} B)
    \, .
    \end{align}
    We have the multiplicative property for the superdeterminant, $\sdet (M_1 M_2) = \sdet M_1 \sdet M_2$.
    We remark an identity 
    \begin{align}
        \str \log M = \log \str M
        \, .
    \end{align}  

    Recalling the behavior of the Grassmann variable measure~\eqref{eq:Jacobian_Grassmann}, the integral measure denoted by $\dd{\Psi} = \dd{z}_1 \cdots \dd{z}_N \dd{\theta}_1 \cdots \dd{\theta}_M$ behaves under the linear transform for the supervector~\eqref{eq:supervec_def}, $\Psi' = M \Psi$, as follows,
    \begin{align}
        \dd{\Psi'} = \dd{\Psi} (\sdet M)
        \, .
    \end{align}

\subsection{Supergroup}

\subsubsection*{Unitary supergroup}

For the complex supervector space element \eqref{eq:supervec_def}, $\Psi \in \mathbb{C}^{N|M}$, we consider the squared norm as follows,
\begin{align}
    |\Psi|^2 & = \sum_{i \in [N]} |z_i|^2 + \sum_{j \in [M]} \bar{\theta}_j \theta_j
%    \nonumber \\ &
    = \sum_{i \in [N]} |z_i|^2 - \sum_{j \in [M]} \theta_j \bar{\theta}_j
%    \nonumber \\ &
    = \str (\Psi \Psi^\dag)
    \, .
\end{align}
Then, we define the unitary supergroup $\mathrm{U}(N|M)$ as the isometry group with respect to the supervector space $\mathbb{C}^{N|M}$, such that,
\begin{align}
    |\Psi|^2 = |U \Psi|^2 %= \str (U \Psi \Psi^\dag U^\dag)
    \, , \qquad
    U^\dag = U^{-1}
    \, .
\end{align}
We remark that $|U \Psi|^2 = \str (U \Psi \Psi^\dag U^\dag) = \str (\Psi \Psi^\dag)$.
Hence, we have
\begin{align}
    \mathrm{U}(N|M) = \{ U \in \mathrm{GL}(\mathbb{C}^{N|M}) \mid U^\dag = U^{-1} \} 
    \, .
\end{align}
The even part of the unitary supergroup $\mathrm{U}(N|M)$ is given by $\mathrm{U}(N) \times \mathrm{U}(M)$, and the odd part is given by $N \times \overline{M}$ and $\overline{N} \times M$ representations of the even part with the dimension $2NM$.

\subsubsection*{Orthosymplectic supergroup}

We consider the real supervector space $\mathbb{R}^{N|2M}$ and the squared norm of $\Psi=(x_1,\ldots,x_N,\theta_1,\ldots,\theta_{2M}) \in \mathbb{R}^{N|2M}$ as follows,
\begin{align}
    |\Psi|^2 = \sum_{i \in [N]} x_i^2 + 2 \sum_{j \in [M]} \theta_{2j-1} \theta_{2j} = \Psi^\text{t} \Omega \Psi = \str \left( \Omega \Psi \Psi^\text{t} \right)
    \, ,
\end{align}
where we define the skew-symmetric form for the Grassmann odd sector,
\begin{align}
        \Omega =
    \begin{pmatrix}
        \id_N & 0 \\ 0 & \id_M \otimes \mathbf{j}
    \end{pmatrix}
    \, , \qquad
    \mathbf{j} =
    \begin{pmatrix}
        0 & 1 \\ -1 & 0
    \end{pmatrix}
    \, .
\end{align}
We remark that the standard bilinear form does not work for the Grassmann variables due to its nilpotency, $\sum_{j \in [M]} \theta_j \theta_j = 0$. 
The isometry group for the real supervector space $\mathbb{R}^{N|2M}$ is given by
\begin{align}
    |\Psi|^2 = |U \Psi|^2
    \, , \qquad 
    U^\text{st} 
    \Omega
    U = \Omega
    \, .
\end{align}
We see that $|U \Psi|^2 = \str(\Omega U \Psi \Psi^\text{t} U^\text{st}) = \str(\Omega \Psi \Psi^\text{t})$.
The orthosymplectic supergroup for general field $\mathbb{K}$ is given as follows,
\begin{align}
    \mathrm{OSp}(N|2M,\mathbb{K}) =
    \{ U \in \mathrm{GL}(\mathbb{K}^{N|2M}) \mid U^\text{st}  \Omega U = \Omega \}
    \, .
\end{align}
Precisely speaking, the orthosymplectic group realized as the isometry group of the real supervector space is a compact supergroup, which is in fact a subgroup of the unitary supergroup, denoted by $\mathrm{UOSp}(N|M) = \mathrm{OSp}(N|2M,\mathbb{C}) \cap \mathrm{U}(N|2M)$.
This is analogous to the notation of the compact symplectic group $\mathrm{USp}(2n) = \mathrm{Sp}(2n,\mathbb{C}) \cap \mathrm{U}(2n)$, which is also understood as the quaternionic unitary group $\mathrm{U}(n,\mathbb{H})$.
In this article, we use the notation, $\mathrm{OSp}(N|M) = \mathrm{UOSp}(N|2M)$ and $\mathrm{Sp}(n) = \mathrm{USp}(2n)$ unless it causes confusion.%
\footnote{%
In this notation, we have the isomorphisms at the level of Lie algebra, $\mathfrak{sp}_1 = \mathfrak{su}_2$, $\mathfrak{sp}_2 = \mathfrak{so}_5$.
}
The even part of $\mathrm{OSp}(N|M)$ is thus given by $\mathrm{O}(N) \times \mathrm{Sp}(M)$, and the odd part is given by $N \times 2M$ representation of the even part with the dimension $2NM$.

Another situation that leads to the orthosymplectic supergroup is a subsector of the supervector space $\mathbb{C}^{2N|2M}$,
\begin{align}
    \Psi = 
    \begin{pmatrix}
        z_1 & z_2 \\
        - \bar{z}_2 & \bar{z}_1 \\
        z_3 & z_4 \\
        \vdots & \vdots \\
        - \bar{z}_{2N} & \bar{z}_{2N-1} \\
        \theta_1 & \bar{\theta}_1 \\
        \vdots & \vdots \\
        \theta_M & \bar{\theta}_M
    \end{pmatrix}
    \, , \qquad
    \Psi^\dag =
    \begin{pmatrix}
        \bar{z}_1 & - z_2 & \bar{z}_3 & \cdots & - z_{2N} & \bar{\theta}_1 & \cdots & \bar{\theta}_M \\
        \bar{z}_2 & z_1 & \bar{z}_4 & \cdots & z_{2N-1} & - \theta_1 & \cdots & - \theta_{M}
    \end{pmatrix}
\end{align}
where the two-by-two block in the bosonic part is identified with a quaternion
\begin{align}
    x_i = 
    \begin{pmatrix}
        z_{2i-1} & z_{2i} \\ - \bar{z}_{2i} & \bar{z}_{2i-1}
    \end{pmatrix}
    \in \mathbb{H}
    \, .
\end{align}
Hence, this is an element in $\mathbb{H}^N \oplus \mathbb{R}^{2M} \subset \mathbb{C}^{2N|2M}$.
We denote the norm of $x \in \mathbb{H}$ by $|x|$, such that $\bar{x} x = |x|^2 \id$ where $\id$ is the quaternion identity.
The norm of this supervector is given by
\begin{align}
    |\Psi|^2 = \tr_{\mathbb{H}} \Psi^\dag \Psi = 2 \sum_{i \in [N]} |x_i|^2 + 2 \sum_{j \in [M]} \bar{\theta}_j \theta_j = \str (\Psi \Psi^\dag)
    \, ,
\end{align}
where we denote the trace operation over the quaternion by $\tr_\mathbb{H}$, i.e., for the quaternion units, $\id$, $\mathbf{i}$, $\mathbf{j}$, $\mathbf{k}$, we have $\tr_\mathbb{H} \id = 2$, $\tr_\mathbb{H} \mathbf{i} = \tr_\mathbb{H} \mathbf{j} = \tr_\mathbb{H} \mathbf{k} = 0$.
The isometry group of this supervector is thus given by the orthosymplectic supergroup $\mathrm{U}(2N|2M) \supset \mathrm{OSp}(M|N) \supset \mathrm{Sp}(N) \times \mathrm{O}(M)$.

\subsubsection*{Analytic continuation}

We discuss how to obtain the supergroups considered above through the analytic continuation.
First of all, the dimension of each classical group is given by%
\footnote{%
The orthogonal group is further classified by the parity of the rank, $\dim \mathrm{O}(2N) = N(2N-1)$, $\dim \mathrm{O}(2N+1) = N(2N+1)$.
We remark that $\dim \mathrm{O}(2N+1) = \dim \mathrm{Sp}(N)$.
This dimension formula is also understood from the construction of the adjoint representations for these classical groups~\eqref{eq:O_SP_adj_rep}.
}
\begin{align}
    \dim \mathrm{U}(N) = N^2
    \, , \qquad
    \dim \mathrm{O}(N) = \frac{N(N-1)}{2}
    \, , \qquad
    \dim \mathrm{Sp}(N) = N(2N+1)
    \, ,
\end{align}
from which we observe the following relations,
\begin{align}
    \dim \mathrm{U}(-N) = \dim \mathrm{U}(N)
    \, , \qquad
    \dim \mathrm{O}(-2N) = \dim \mathrm{Sp}(N)
%    \, , \qquad
%    \dim \mathrm{Sp}(-N) = \dim \mathrm{O}(-2N)
    \, .
\end{align}
Such a relation is discussed also at the level of their irreducible representations~\cite{King:1971CJM}. 
From this point of view, these supergroups are obtained from the ordinary classical groups through the analytic continuation,
\begin{subequations}
\begin{gather}
    \mathrm{U}(N+M)
    \ \stackrel{M \leftrightarrow -M}{\longleftrightarrow} \
    \mathrm{U}(N|M)
    \, , \\
    \mathrm{O}(N+2M)
    \ \stackrel{M \leftrightarrow -M}{\longleftrightarrow} \
    \mathrm{OSp}(N|M)
    \, , \qquad
    \mathrm{Sp}(N+M)
    \ \stackrel{M \leftrightarrow -M}{\longleftrightarrow} \
    \mathrm{OSp}(2M|N)
    \, .    
\end{gather}
\end{subequations}
In particular, we have the relations between the superdimension and the dimension of the classical groups,
\begin{subequations}
\begin{align}
    \operatorname{sdim} \mathrm{U}(N|M) & = (N^2 + M^2) - 2 NM = (N-M)^2 = \operatorname{dim} \mathrm{U}(N+M)\Big|_{M \to -M}
    \, , \\
    \operatorname{sdim} \mathrm{OSp}(N|M) & = \left( \frac{N(N-1)}{2} + M(2M+1) \right) - 2 NM = 2 \left( \frac{N}{2} - M \right)^2 - \left( \frac{N}{2} - M \right)
    \nonumber \\
    & = \operatorname{dim} \mathrm{O}(N + 2M)\Big|_{M \to -M} = \operatorname{dim} \mathrm{Sp}\left(\frac{N}{2} + M\right)\Bigg|_{N \to -N}
    \, .
\end{align}
\end{subequations}
This interpretation based on the analytic continuation seems also reasonable from the relation between the ordinary trace and the supertrace.

\section{Supermatrix model}\label{sec:super_matrix}

The matrix model, just given by a matrix integral, is thought of as a zero-dimensional reduction of quantum field theory.
It sounds rather a simple toy model, but it has been playing a role to understand various non-perturbative aspects of quantum field theory.
In this section, we explore the supermatrix model as a toy model that exhibits supergroup symmetry.
See also \cite{Efetov:1996,Guhr:2010nh} for details on this subject.

\subsection{Hermitian supermatrix model: superunitary ensemble}\label{sec:Hermitian_supermatrix_model}

Let $H$ be an $(N|M)$-dimensional Hermitian supermatrix, $H^\dag = H$.
We define the partition function of Hermitian supermatrix model of rank $(N|M)$ as follows~\cite{Alvarez-Gaume:1991ozb,Yost:1991ht},
\begin{align}
    Z_{N|M} = \int \dd{H} \ee^{-\frac{1}{g} \str V(H)}
    \, ,
\end{align}
with a polynomial potential function of degree $(d+1)$,
\begin{align}
    V(x) = \sum_{n = 1}^{d+1} \frac{t_n}{n} \, x^n
    \, .
    \label{eq:pot_fn_supermatrix}
\end{align}
This integral is invariant under superunitary transformation, $H \to U H U^\dag$, $U \in \mathrm{U}(N|M)$:
The potential part is invariant due to the cyclic property of the supertrace~\eqref{eq:str_cyclic}, $\str V(U H U^\dag) = \str V(H)$, and the measure part is due to the unitarity of the superdeterminant $|\sdet U| = 1$.
From this point of view, this model is also called the \emph{superunitary ensemble} as a supermatrix generalization of the unitary ensemble of the ordinary random matrices.
See, e.g.,~\cite{Mehta:2004RMT,Forrester:2010,Eynard:2015aea}.
In fact, this symmetry is interpreted as a remnant of the supergroup gauge symmetry.

Similarly to the ordinary Hermitian matrix, we can diagonalize the Hermitian supermatrix via the superunitary transform,%
\footnote{%
Eigenvalues of Hermitian supermatrices are ordinary commutative numbers (Recall that the diagonal blocks of supermatrices are ordinary matrices).
The matrix model involving Grassmann odd eigenvalues is known as the super-eigenvalue model~\cite{Alvarez-Gaume:1991vno}.
}
\begin{align}
    H = U Z U^\dag
    \, , \qquad
    U \in \mathrm{U}(N|M)
    \, ,
    \label{eq:SU_diaginalization}
\end{align}
where we denote the diagonal supermatrix by
\begin{align}
    Z = 
    \begin{pmatrix}
    X & 0 \\ 0 & Y
    \end{pmatrix}
    \, , \qquad
    X = \operatorname{diag}(x_1,\ldots,x_N)
    \, , \ 
    Y = \operatorname{diag}(y_1,\ldots,y_M)
    \, .
    \label{eq:diagonal_Z}
\end{align}
We remark that the choice of $U$ is not unique in the process of diagonalization: (i) The eigenvalues can be permuted $(x_i,y_j)_{i\in[N], j\in[M]} \to (x_{\sigma(i)},y_{\sigma'(j)})_{i\in[N], j\in[M]}$ where $\sigma \in \mathfrak{S}_N$, $\sigma' \in \mathfrak{S}_M$.
(ii) Since the diagonal superunitary matrix commutes with the diagonal supermatrix, $U_\text{diag} Z U_\text{diag}^\dag = Z$, where $U_\text{diag} \in \mathrm{U}(1)^{N+M}$, the decomposition is invariant under redefinition $U \to U U_\text{diag}$.

Taking the derivative of the relation above, we have the following expression,
\begin{align}
    U^\dag \dd{H} U = \dd{Z} + [U^\dag\dd{U}, Z]
    \, .
    \label{eq:UdHU}
\end{align}
We remark that $U^\dag\dd{U}$ is the Maurer--Cartan one-form with respect to the supergroup $\mathrm{U}(N|M)$, and the Hermitian supermatrix takes a value in the Lie superalgebra $\operatorname{Lie} \mathrm{U}(N|M)$ (up to the imaginary unit).
Then, the right hand side of this equation can be also written using the covariant derivative in the adjoint representation, $D = \dd{} + U^\dag\dd{U}$.
Denoting $I_0 = \{1,\ldots,N\}$ and $I_1 = \{ N+1, \ldots, N+M \}$, and $\mathsf{i}(i) = i$ for $i \in I_0$ and $\mathsf{i}(i) = i - N$ for $i \in I_1$, each component of \eqref{eq:UdHU} is given by
\begin{subequations}\label{eq:UdHU_component}
\begin{align}
    (U^\dag \dd{H} U)_{ii} & =
    \begin{cases}
    \dd{x}_{\mathsf{i}(i)} & (i \in I_0) \\
    \dd{y}_{\mathsf{i}(i)} & (i \in I_1)
    \end{cases}
    \\
    (U^\dag \dd{H} U)_{ij} & = (U^\dag \dd{U})_{ij} \times
    \begin{cases}
    x_{\mathsf{i}(j)} - x_{\mathsf{i}(i)} & (i, j \in I_0) \\
    y_{\mathsf{i}(j)} - y_{\mathsf{i}(i)} & (i, j \in I_1) \\
    y_{\mathsf{i}(j)} - x_{\mathsf{i}(i)} & (i \in I_0, j \in I_1) \\
    x_{\mathsf{i}(j)} - y_{\mathsf{i}(i)} & (i \in I_1, j \in I_0)
    \end{cases}
\end{align}
\end{subequations}
Hence, the supermatrix measure is given in terms of the eigenvalues and the eigenvector components as follows,
\begin{align}
    \dd{H} = |\Delta_{N|M}(X|Y)|^2 \dd{X} \dd{Y} \dd{U}
\end{align}
where we define
\begin{align}
    \dd{X} = \prod_{i \in [N]} \dd{x}_i
    \, , \quad
    \dd{Y} = \prod_{i \in [M]} \dd{y}_i
    \, , \quad
    \dd{U} = \prod_{1 \le i<j \le N+M} \operatorname{Re} (U^\dag \dd{U})_{ij} \operatorname{Im} (U^\dag \dd{U})_{ij}
    \, .
\end{align}
In particular, $\dd{U}$ is the Haar measure on the unitary supergroup $\mathrm{U}(N|M)$.
The Jacobian part is given by the Cauchy determinant
\begin{align}
    \Delta_{N|M}(X|Y) = \prod_{i < j}^N (x_j - x_i) \prod_{i < j}^M (y_j - y_i) \prod_{i \in [N], j \in [M]} (x_i - y_j)^{-1}
    \, ,
\end{align}
which has a determinantal formula for $N \ge M$ as
\begin{align}
    \Delta_{N|M}(X|Y) = \det_{\substack{i \in [N],\, j \in [M] \\ k \in [N-M]}}
    \begin{pmatrix}
        x_i^{k-1} & \displaystyle \frac{1}{x_i - y_j}
    \end{pmatrix}
    \, .
\end{align}
In the limit $M \to 0$, this is reduced to the Vandermonde determinant,
\begin{align}
    \Delta_{N|0}(X) = \Delta_N(X) = \prod_{i<j}^N (x_j - x_i) \, .
\end{align}

Based on these diagonalization process, we obtain the eigenvalue integral form of the partition function,
\begin{align}
    {Z}_{N|M} = \frac{\operatorname{vol}(\mathrm{U}(N|M))}{N! M!} \int \dd{\mu(X)} \dd{\mu(Y)} 
    |\Delta_{N|M}(X|Y)|^2
    \, ,
\end{align}
where the integral measure is given by
\begin{align}
    \dd{\mu(X)} = \prod_{i \in [N]} \frac{\dd{x}_i}{2 \pi} \ee^{-\frac{1}{g}V(x_i)}
    \, , \qquad
    \dd{\mu(Y)} = \prod_{i \in [M]} \frac{\dd{y}_i}{2 \pi} \ee^{+\frac{1}{g}V(y_i)}
    \, .
\end{align}
Constant factors are understood as follows:
The factorial terms are the volumes of $\mathfrak{S}_N$ and $\mathfrak{S}_M$, which are the Weyl group of $\mathrm{U}(N)$ and $\mathrm{U}(M)$, and $(2\pi)^{N+M} = \operatorname{vol}(\mathrm{U}(1)^{N+M})$, the volume of the maximal Cartan torus of $\mathrm{U}(N|M)$.
We have several remarks on this formula: 
\begin{enumerate}
    \item The signatures of the potential term for the $x$-variables and the $y$-variables are opposite.
    Hence, we should consider a complex contour to obtain a converging integral.
    For example, for the Gaussian case $V(x) = \frac{1}{2} x^2$, the $x$-integral is taken along the real axis, $- \infty \to + \infty$, while the $y$-integral should be taken along the imaginary axis, $- \ii \infty \to + \ii \infty$, or vice versa. 
    \item The denominator contribution in the Cauchy determinant is singular in the limit $x_i \to y_j$.
    If there is an intersection of the $x$-contour and the $y$-contour, such a singularity should be regularized  using the principal value prescription.
    \item The volume of the unitary supergroup becomes zero if $NM \neq 0$ due to the Grassmann variable integral (Berezin's theorem. See e.g., \cite{Voronov:2015cya}).
    Hence, we consider the partition function formally normalized by this zero-volume, $\mathcal{Z}_{N|M} := Z_{N|M}/\operatorname{vol}(\mathrm{U}(N|M))$.%
    \footnote{%
    It would be possible that the zero-volume factor cancels the diverging behavior of the eigenvalue integral to give a finite value in the end.
    We do not discuss details of this issue any further in this article.
    }
\end{enumerate}
Taking into account these points, we obtain the eigenvalue integral form of the supermatrix partition function.
\begin{itembox}{Hermitian supermatrix model}
The eigenvalue integral form of the regularized partition function of the Hermitian supermatrix model is given as follows,
    \begin{align}
    \mathcal{Z}_{N|M} =
    \frac{1}{N! M!} \dashint_{\gamma_x^N \times \gamma_y^M} \dd{\mu(X)} \dd{\mu(Y)} 
    |\Delta_{N|M}(X|Y)|^2
    \, ,
\end{align}
where we denote the principal value integral by $\dashint \dd{x} f(x)$, and $\gamma_x$ and $\gamma_y$ are the integration contours on the complex plane that provide a converging integral.
\end{itembox}

\subsection{Real-quaternion supermatrix model: super orthosymplectic ensemble}\label{sec:real-quaternion_supermatrix_model}

We consider an $(N|M)$-dimensional real-quaternion self-conjugate supermatrix,
\begin{align}
    H =
    \begin{pmatrix}
        A & B \\ C & D
    \end{pmatrix}
\end{align}
where $A$ is an $N$-dimensional real symmetric matrix, and $D$ is an $M$-dimensional quaternion self-dual matrix (realized as a $2M$-dimensional Hermitian matrix).
$B$ is a real Grassmann matrix of size $N \times 2M$ and, in order that $H$ is self-conjugate $H^\dag = H$, we have $C = - B^\text{t}$.
In this case, similarly to the Hermitian case~\eqref{eq:SU_diaginalization}, we can diagonalize the supermatrix via the orthosymplectic transformation,
\begin{align}
    H = U Z U^\dag
    \, , \qquad
    U \in \mathrm{OSp}(N|M)
    \, .
    \label{eq:OSp_diaginalization}
\end{align}
The diagonal supermatrix $Z$ is given by
\begin{align}
    Z = 
    \begin{pmatrix}
        X & 0 \\ 0 & Y \otimes \id
    \end{pmatrix}
    %\operatorname{diag}(x_1,\ldots,x_N,y_1 \id,\ldots,y_M\id)
    \, ,
    %\label{eq:diagonal_Z}
\end{align}
where we denote the identity element in quaternion by $\id$.
Namely, if we use the two-by-two matrix realization of quaternion, it is given by the identity matrix of rank two.
Hence, the supertrace is in this case given by
\begin{align}
    \str Z = \sum_{i \in [N]} x_i - 2 \sum_{j \in [M]} y_j
    \, .
\end{align}
We should be careful of the multiplicity in the quaternionic sector, which corresponds to the Kramers doublet.
We will define a deformed supertrace operation respecting this multiplicity (see \eqref{eq:supertrace_b}).

Applying the same argument to the unitary case, we have the relation~\eqref{eq:UdHU} where the corresponding components are given as in~\eqref{eq:UdHU_component}.
In this case, we should be careful of that the $y$-variables are doubly degenerated.
Hence, we should count twice for the mixing terms $y_j - x_i$ and four ($= 2 \times 2$) times for the quaternion part $y_j - y_i$.
Therefore, the real-quaternion supermatrix measure is given as follows,
\begin{align}
    \dd{H} = |\Delta_{N|M}^{(1|4)}(X|Y)| \dd{X} \dd{Y} \dd{U}
\end{align}
where we define
\begin{align}
    \dd{X} = \prod_{i \in [N]} \dd{x}_i
    \, , \qquad
    \dd{Y} = \prod_{i \in [M]} \dd{y}_i
    \, , \qquad
\end{align}
and the corresponding Haar measure $\dd{U}$ of supergroup $\mathrm{OSp}(N|M)$.
We use the following notation for the Jacobian part,
\begin{align}
    \Delta_{N|M}^{(\beta|\beta')}(X|Y) = \frac{\Delta_N(X)^\beta \Delta_M(Y)^{\beta'}}{\prod_{i \in [N], j \in [M]} (x_i - y_j)^2}
\end{align}
In this notation, we have $\Delta_{N|M}(X|Y)^2 = \Delta_{N|M}^{(2|2)}(X|Y)$.
Collecting all the contributions, we obtain the real-quaternion supermatrix. 
\begin{itembox}{Real-quaternion supermatrix model}
The eigenvalue integral form of the regularized partition function of real-quaternion supermatrix model is given as follows,
\begin{align}
    \mathcal{Z}_{N|M} := \frac{Z_{N|M}}{\operatorname{vol}(\mathrm{OSp}(N|M))}  =
    \frac{1}{N! M!} \dashint_{\gamma_x^N \times \gamma_y^M} \dd{\mu(X)} \dd{\mu(Y)} 
    |\Delta_{N|M}^{(1|4)}(X|Y)|
    \, .
\end{align}
\end{itembox}
We study several aspects of the supermatrix models in the following part.

\subsection{Coulomb gas analysis}

In the context of matrix model, we are in particular interested in the asymptotic behavior in the large size limit of the matrix model.
We study such an asymptotic limit of the supermatrix model based on the Coulomb gas analysis.
See also~\cite{Babinet:2022} for details in this part.

We start with the partition function of the $\beta$-deformed supermatrix model,
\begin{align}
    \mathcal{Z}_{N|M} = \frac{1}{N! M!} \int \dd{\mu(X)} \dd{\mu(Y)} 
    |\Delta_{N|M}^{(\beta|\beta')}(X|Y)|
    \, , \qquad
    \beta \beta' = 4
    \, ,
\end{align}
where the integral measure is given by
\begin{align}
    \dd{\mu(X)} = \prod_{i \in [N]} \frac{\dd{x}_i}{2 \pi} \ee^{-\frac{b}{g} V(x_i)}
    \, , \qquad
    \dd{\mu(Y)} = \prod_{i \in [M]} \frac{\dd{y}_i}{2 \pi} \ee^{+\frac{b^{-1}}{g}V(y_i)}
    \, , \qquad
    b = \sqrt{\frac{\beta}{2}}
    \, .
\end{align}
In this notation, the measure factor is given by
\begin{align}
    \Delta_{N|M}^{(\beta|\beta')}(X|Y) = \frac{\Delta_N(X)^{2 b^2} \Delta_M(Y)^{2 b^{-2}}}{\prod_{i \in [N], j \in [M]} (x_i - y_j)^2}
    \, .
\end{align}
For the $\beta$-deformed case, it is convenient to define the $b$-deformed supertrace,
\begin{align}
    \str_b 
    \begin{pmatrix}
        A & B \\ C & D
    \end{pmatrix}
    = b \tr_0 A - b^{-1} \tr_1 D
    \, .
    \label{eq:supertrace_b}
\end{align}
With this operation, the potential factor is concisely written as
\begin{align}
    \str_b V(Z) = b \sum_{i \in [N]} V(x_i) - b^{-1} \sum_{j \in [M]} V(y_j) 
    \, ,
\end{align}
where the diagonal supermatrix $Z$ is given as~\eqref{eq:diagonal_Z}.
Such a deformation is discussed in the context of symmetric polynomial associated with the Lie superalgebra root system. 
See, e.g.,~\cite{Sergeev:2001JNMP,Sergeev:2002TMP,Sergeev:2005AM,Hallnas:2009CA,Desrosiers:2012S}, and also discussions in Secs.~\ref{sec:Bethe/gauge} and \ref{sec:intersecting_defects}.

\subsubsection{Saddle point equation}

To study the asymptotic behavior of the supermatrix model, we rewrite the partition function in the following form,
\begin{align}
    \mathcal{Z}_{N|M} \approx \int \dd{X} \dd{Y} \ee^{-\frac{1}{g^2} S(X|Y)}
\end{align}
where the integral measure is given by
\begin{align}
    \dd{X} = \prod_{i \in [N]} \dd{x}_i
    \, , \qquad 
    \dd{Y} = \prod_{j \in [M]} \dd{y}_j
    \, ,
\end{align}
and we define the effective action
\begin{align}
    S(X|Y) & =
    b g \sum_{i \in [N]} V(x_i) - b^{-1} g \sum_{j \in [M]} V(y_j)
    \nonumber \\
    & \quad 
    - 2 b^2 g^2 \sum_{i < j}^N \log (x_i - x_j)
    - 2 b^{-2} g^2 \sum_{i < j}^M \log (y_i - y_j)
    + 2 g^2 \sum_{i \in [N], j \in [M]} \log (x_i - y_j)
    \, .
    \label{eq:eff_action_supermatrix}
\end{align}
Then, introducing two parameters,
\begin{align}
    \mathsf{t}_0 = bgN 
    \, , \qquad 
    \mathsf{t}_1 = b^{-1}gM
    \, ,
\end{align}
we consider the following asymptotic limit ('t Hooft limit) of the supermatrix model,
\begin{align}
    g \ \longrightarrow \ 0
    \, , \qquad
    N, M \ \longrightarrow \ \infty
    \, , \qquad
    \mathsf{t}_0 , \mathsf{t}_1 = O(1)
    \, .
\end{align}

In the 't Hooft limit, the eigenvalue integral localizes on the configuration that obeys the following saddle point equations,
\begin{subequations}\label{eq:saddle_pt_supermatrix}
    \begin{align}
        0 & = \frac{\partial S}{\partial x_i} = 
        + b g V'(x_i) 
        - 2 b^2 g^2 \sum_{j \in [N]\backslash\{i\}} \frac{1}{x_i - x_j} 
        + 2 g^2 \sum_{j \in [M]} \frac{1}{x_i - y_j}
        \, , \\
        0 & = \frac{\partial S}{\partial y_i} = 
        - b^{-1} g V'(y_i) 
        - 2 b^{-2} g^2 \sum_{j \in [M]\backslash\{i\}} \frac{1}{y_i - y_j} 
        + 2 g^2 \sum_{j \in [N]} \frac{1}{y_i - x_j}
        \, .
    \end{align}
\end{subequations}
We introduce the auxiliary functions,
\begin{subequations}
    \begin{align}
        W_0 (x) & = bg \sum_{i \in [N]} \frac{1}{x - x_i} = bg \tr_0 \left( \frac{1}{x - X} \right)
        \, , \\
        W_1 (x) & = b^{-1}g \sum_{i \in [M]} \frac{1}{x - y_i} = b^{-1} g \tr_1 \left( \frac{1}{x - Y} \right)
        \, , \\
        P_0 (x) & = bg \sum_{i \in [N]} \frac{V'(x) - V'(x_i)}{x - x_i} = bg \tr_0 \left( \frac{V'(x) - V'(X)}{x - X} \right)
        \, , \\
        P_1 (x) & = b^{-1}g \sum_{i \in [M]} \frac{V'(x) - V'(y_i)}{x - y_i} = b^{-1}g \tr_1 \left( \frac{V'(x) - V'(Y)}{x - Y} \right)
        \, .
    \end{align}
\end{subequations}
The auxiliary functions $W_\sigma(x)$ are in particular called the resolvents that involve a pole singularity at $x \in \{x_i\}_{i \in [N]}$ and $x \in \{ y_j \}_{j \in [M]}$, respectively.
Although the other functions $P_\sigma(x)$ look a similar form, they are polynomial functions having no pole singularity.
Recalling that the potential function is given as~\eqref{eq:pot_fn_supermatrix}, the asymptotic behaviors of these auxiliary functions are given by
\begin{align}
    W_\sigma(x) \ \xrightarrow{x \to \infty} \ \frac{\mathsf{t}_\sigma}{x}
    \, , \qquad
    P_\sigma(x) \ \xrightarrow{x \to \infty} \ \mathsf{t}_\sigma t_{d+1} x^{d-1}
    \, .
\end{align}

Using these auxiliary functions, we may rewrite the saddle point equation~\eqref{eq:saddle_pt_supermatrix} as follows,
\begin{subequations}
    \begin{align}
        0 & = - P_0(x) + V'(x) W_0(x) - W_0(x)^2 - bg W_0'(x) + 2 g^2 \sum_{i \in [N], j \in [M]} \frac{1}{(x - x_i)(x_i - y_j)}
        \, , \\
        0 & = + P_1(x) - V'(x) W_1(x) - W_1(x)^2 - b^{-1}g W_1'(x) + 2 g^2 \sum_{i \in [N], j \in [M]} \frac{1}{(x - y_j)(y_j - x_i)}
        \, .
    \end{align}
\end{subequations}
Moreover, we define the supertrace analog of the auxiliary functions,
\begin{subequations}
    \begin{align}
        \mathsf{W}(x) & = W_0(x) - W_1(x) = g \str_b \left( \frac{1}{x - Z} \right)
        \, , \\
        \mathsf{P}(x) & = P_0(x) - P_1(x) = g \str_b \left( \frac{V'(x) - V'(Z)}{x - Z} \right)
        \, ,
    \end{align}
\end{subequations}
which show the following asymptotic behavior,
\begin{align}
    \mathsf{W}(x) \ \xrightarrow{x \to \infty} \ \frac{\mathsf{t}_0 - \mathsf{t}_1}{x}
    \, , \qquad
    \mathsf{P}(x) \ \xrightarrow{x \to \infty} \ (\mathsf{t}_0 - \mathsf{t}_1) t_{d+1} x^{d-1}
    \, .
\end{align}
The total resolvent $\mathsf{W}(x)$, that we call the superresolvent, has poles with the residue $+bg$ for $x \in \{x_i\}$ and $-b^{-1}g$ for $x \in \{y_i\}$, while $\mathsf{P}(x)$ is again a polynomial function.
Then, combining the two equations, we obtain 
\begin{align}
    0 & = \mathsf{P}(x) - V'(x) \mathsf{W}(x) + \mathsf{W}(x)^2 + g (b W_0'(x) + b^{-1} W_1'(x))
    \, .
    \label{eq:saddle_pt_supermatrix_finite}
\end{align}
We study this equation in detail in Sec.~\ref{sec:sp_curve}.

We remark that the two saddle point equations \eqref{eq:saddle_pt_supermatrix} are written as a single equation using the superresolvent, 
\begin{align}
    0 = V'(x) - 2 \mathsf{W}_\text{reg}(x) 
    \, , \qquad
    x \in \{ x_i, y_j \}
    \, ,
    \label{eq:saddle_pt_supermatrix_single_form}
\end{align}
where we define the regularized one by
\begin{align}
    \mathsf{W}_\text{reg}(x) = 
    \begin{cases}
        \displaystyle
        bg \sum_{j \in [N]\backslash\{i\}} \frac{1}{x - x_j} - W_1(x) & (x = x_i) \\
        \displaystyle
        W_0(x) - b^{-1} g \sum_{j \in [M]\backslash\{i\}} \frac{1}{x - y_j} - W_1(x) & (x = y_i) 
    \end{cases}
\end{align}
We remark that the expression of the saddle point equation~\eqref{eq:saddle_pt_supermatrix_single_form} is identical to the standard matrix model by replacing the superresolvent with the original one.
However, its analytic property should be different since the superresolvent may contain both positive and negative residues, while the original resolvent only involves positive one.

\subsubsection{Functional method}

Let us discuss an alternative approach based on the functional method.
We define the density functions $\rho_\sigma(x)$ for $\sigma = 0,1$.
Then, we rewrite the effective action~\eqref{eq:eff_action_supermatrix} using these density functions, 
\begin{align}
    S[\rho_{0,1}] & = \sum_{\sigma = 0, 1} \left[ (-1)^\sigma \mathsf{t}_\sigma \int \dd{x} \rho_\sigma (x) V(x) - 2 \mathsf{t}_\sigma^2 \int_{x < y} \dd{x} \dd{y} \rho_\sigma(x) \rho_\sigma(y) \log |x - y| \right]
    \nonumber \\
    & \qquad +
    2 \mathsf{t}_0 \mathsf{t}_1 \dashint \dd{x} \dd{y} \rho_0(x) \rho_1(y) \log |x - y|
    + \sum_{\sigma = 0,1} \sum_{i = 1}^{m_\sigma} \mathsf{t}_\sigma \ell_{\sigma,i} \left( \epsilon_{\sigma,i} - \int_{\mathcal{C}_{\sigma,i}} \dd{x} \rho_\sigma(x) \right)
    \, ,
\end{align}
where we consider the $m_\sigma$-cut solution for each sector, $\sigma = 0, 1$: 
We added the Lagrange multiplier that imposes the condition
\begin{align}
    \int_{\mathcal{C}_{\sigma,i}} \dd{x} \rho_\sigma(x) = \epsilon_{\sigma, i}
    \, ,
\end{align}
where we denote the cut by 
\begin{align}
    \mathcal{C}_\sigma = \bigsqcup_{i \in [m_\sigma]} \mathcal{C}_{\sigma,i}
    \, , \qquad
    \sigma = 0, 1
    \, ,
\end{align}
and the corresponding filling fraction $(\epsilon_{\sigma,i})_{\sigma = 0, 1, i \in [m_\sigma]}$ with $\displaystyle \sum_{i \in [m_\sigma]} \epsilon_{\sigma,i} = 1$.
Taking the functional derivative of the effective action, we obtain
\begin{align}
    x \in \mathcal{C}_{\sigma,i} \ : \quad 
    \mathsf{t}_\sigma^{-1} \frac{\delta S[\rho_{0,1}]}{\delta \rho_\sigma(x)} & = 
        (-1)^\sigma \left( V(x) - 2 \dashint \dd{y} \bar{\rho}(y) \log |x-y| \right) - \ell_{\sigma,i}
    \nonumber \\
    & =: 
    (-1)^\sigma V_\text{eff}(x) - \ell_{\sigma,i}
    \, ,
\end{align}
where we define the effective potential $V_\text{eff}(x)$ and the effective density functions,
\begin{align}
    \bar{\rho}(x) = \mathsf{t}_0 \rho_0(x) - \mathsf{t}_1 \rho_1(x) 
    \, .
\end{align}
The functional version of the saddle point equation~\eqref{eq:saddle_pt_supermatrix_single_form} is obtained by the derivative of the effective potential,
\begin{align}
    \dv{V_\text{eff}(x)}{x} = V'(x) - 2 \dashint \dd{y}  \frac{\bar{\rho}(y)}{x - y}
    \, .
\end{align}
Writing the integral form of the superresolvent,
\begin{align}
    \mathsf{W}(x) = \int \dd{y} \frac{\bar{\rho}(y)}{x - y}
    \, ,
\end{align}
the regularized one is given by the principal value integral
\begin{align}
    \mathsf{W}_\text{reg}(x) & = \dashint \dd{y} \frac{\bar{\rho}(y)}{x - y} 
    \nonumber \\
    & = \operatorname{Re} \mathsf{W}(x \pm \ii 0) := \lim_{\epsilon \to 0^+} \frac{\mathsf{W}(x + \ii \epsilon) + \mathsf{W}(x - \ii \epsilon)}{2}
    \, .
\end{align}
Hence, we obtain the functional version of \eqref{eq:saddle_pt_supermatrix_single_form} as follows,
\begin{align}
    \dv{V_\text{eff}(x)}{x} = 0
    \ \implies \ 
    V'(x) - 2 \operatorname{Re} \mathsf{W}(x \pm \ii 0) = 0
    \, , \qquad
    x \in \mathcal{C}_\sigma
    \, .
\end{align}

\subsubsection{Spectral curve and quantization}\label{sec:sp_curve}

We have seen that the saddle point equation of the supermatrix model gives rise to the relation among the resolvents as shown in \eqref{eq:saddle_pt_supermatrix_finite}.
Further taking the limit $g \to 0$, the relation~\eqref{eq:saddle_pt_supermatrix_finite} is written in a closed form of the superresolvent,
\begin{align}
    0 = \mathsf{W}(x)^2 - V'(x) \mathsf{W}(x) + \mathsf{P}(x)
    \, ,
\end{align}
which defines the spectral curve of the supermatrix model:
\begin{itembox}{Spectral curve of supermatrix model}
Given the potential function $V'(x)$ and the polynomial function $\mathsf{P}(x)$, the spectral curve of the supermatrix is given as follows,
\begin{align}
    \Sigma = \{ (x,y) \in \mathbb{C} \times \mathbb{C} \mid \mathcal{H}(x,y) = 0 \}
    \, , \qquad
    \mathcal{H}(x,y) = y^2 - V'(x) y + \mathsf{P}(x)
    \, .
    \label{eq:sp_curve_supermatrix}
\end{align}
\end{itembox}
This is formally identical to the spectral curve of the standard matrix model (see, e.g.~\cite{Eynard:2015aea}), but as we mentioned before, we should be careful of its analytic property. 

While the spectral curve is based on the closed equation for the superresolvent, the saddle point equation~\eqref{eq:saddle_pt_supermatrix_finite} itself is not written as a closed form.
In order to discuss an alternative form, we rewrite the resolvents as follows,
\begin{align}
    W_0(x) = bg \dv{}{x} \log \psi_0(x)
    \, , \qquad
    W_1(x) = b^{-1}g \dv{}{x} \log \psi_1(x)
\end{align}
where we define the wave functions (characteristic polynomials),
\begin{align}
    \psi_0(x) = \prod_{i \in [N]} (x - x_i)
    \, , \qquad
    \psi_1(x) = \prod_{i \in [M]} (x - y_i)
    \, .
\end{align}
Then, the superresolvent is given by the logarithmic derivative
\begin{align}
    \mathsf{W}(x) = g \dv{}{x} \log \frac{\psi_0(x)^b}{\psi_1(x)^{b^{-1}}} %\frac{\prod_{i \in [N]} (x - x_i)^b}{\prod_{j \in [M]} (x - y_j)^{b^{-1}}}
    \, .
\end{align}
Together with these wave functions, we can recast the saddle point equation~\eqref{eq:saddle_pt_supermatrix_finite} in the following form,
\begin{align}
    \left[ D_x^{(b)2} - V'(x) D_x^{(b)} + \mathsf{P}(x) \right] \psi_0 \cdot \psi_1 = 0
    \, ,
    \label{eq:quantum_curve_supermatrix}
\end{align}
where we define the $b$-deformed Hirota derivative,
\begin{align}
    D_x^{(b)} \psi \cdot \phi 
    = g \left( b \pdv{}{x} - b^{-1} \pdv{}{x'} \right) \psi(x) \phi(x')\Big|_{x = x'}
    = g b \psi'(x) \phi(x) - g b^{-1} \psi(x) \phi'(x)
    \, .
\end{align}
The standard Hirota derivative corresponds to the case $b = 1$ ($\beta = 2$): $D_x = D_x^{(1)}$.
In fact, this bilinear equation is interpreted as a quantization of the spectral curve~\eqref{eq:sp_curve_supermatrix}: 
\begin{itembox}{Quantum curve for $\beta$-supermatrix model}
Based on the two-variable function $\mathcal{H}(x,y)$ that defines the spectral curve~\eqref{eq:sp_curve_supermatrix}, we have the quantum curve for $\beta$-supermatrix model involving the Hirota derivative,
\begin{align}
    \mathcal{H}(\hat{x},\hat{y}) \psi_0 \cdot \psi_1 = 0
    \, , \qquad
    \hat{x} = x
    \, , \ 
    \hat{y} = g D_x^{(b)}
    \, .
\end{align}
\end{itembox}
Recalling the definition of the Hirota derivative, the canonical commutation relation is given by
\begin{align}
    [\hat{y}, \hat{x}] =
    \begin{cases}
        + g b & (\text{for} \ \psi_0) \\
        - g b^{-1} & (\text{for} \ \psi_1) \\
    \end{cases}
\end{align}
From this point of view, we call the bilinear equation~\eqref{eq:quantum_curve_supermatrix} the quantum curve for the supermatrix model.
We now have the both positive and negative quantum parameters (Planck constants) for quantization of the supermatrix spectral curve, corresponding to that the superresolvent has both positive and negative residues.

Moreover, we now relate the supermatrix parameters to the so-called $\Omega$-background parameters (see Sec.~\ref{sec:instanton}),
\begin{align}
    (\epsilon_1, \epsilon_2) = \left(g, - \frac{g}{b^2} \right)
    \, .
    \label{eq:Omega_background_matrix_parameter}
\end{align}
In this notation, we have $b^2 = - \epsilon_1 / \epsilon_2$, and thus the condition $b = 1$ is equivalent to $\epsilon_1 + \epsilon_2 = 0$.
Then, the $b$-Hirota derivative is rewritten using $\epsilon_{1,2}$ as follows,
\begin{align}
    D_x^{(b)} \psi \cdot \phi & = b \epsilon_1 \psi'(x) \phi(x) + b \epsilon_2 \psi(x) \phi'(x) =: b D_x^{(\epsilon_1, \epsilon_2)} \psi \cdot \phi
    \, ,
\end{align}
where $D_x^{(\epsilon_1, \epsilon_2)}$ is the $(\epsilon_1,\epsilon_2)$-deformed Hirota derivative defined in~\cite{Nakajima:2003pg}.
This operator is reduced to the ordinary derivative in the limit $\epsilon_1 \to 0$ or $\epsilon_2 \to 0$, which is called the Nekrasov--Shatashvili (NS) limit,
\begin{align}
    D_x^{(\epsilon_1, \epsilon_2)} \psi \cdot \phi
    \ \longrightarrow \
    \begin{cases}
        \epsilon_1 \psi'(x) \phi(x) & (\epsilon_2 \to 0) \\
        \epsilon_2 \psi(x) \phi'(x) & (\epsilon_1 \to 0)
    \end{cases}
\end{align}
See also Sec.~\ref{sec:Bethe/gauge} for a related discussion.

\subsubsection{Gaussian model}

We consider the simplest example with the quadratic potential $V(x) = \frac{1}{2} x^2$, which is called the Gaussian matrix model.
In this case, the quantum curve is given by
\begin{align}
    \left[ D_x^{(b)2} - x D_x^{(b)} + (\mathsf{t}_0 - \mathsf{t}_1) \right] \psi_0 \cdot \psi_1 = 0
    \, .
\end{align}
Hence, in particular for the unitary case $b = 1$ ($\beta = 2$), we have
\begin{align}
    \left[ D_\xi^2 - \xi D_\xi + (N - M) \right] \psi_0 \cdot \psi_1 = 0
    \, , \qquad
    x = \sqrt{g} \xi
    \, .
\end{align}
This bilinear equation is known to be (a part of) the bilinear equations for the $\tau$-functions in the symmetric form of the Painlevé IV equation~\cite{Noumi:1999NMJ}.
In this case, the polynomial solution is given by the generalized Hermite polynomial, which is given through specialization of Schur functions.
In the NS limit, as mentioned above, the Hirota derivative is reduced to the ordinary derivative, where this bilinear equation is accordingly reduced to the differential equation for the Hermite polynomial.

\subsection{Free field realization}\label{sec:free_field_realization}

We turn to discuss algebraic aspects of supermatrix model.
In particular, we show that the supermatrix partition function has a realization in terms of the chiral boson fields (free field realization).

\subsubsection{Operator formalism}\label{sec:op_formalism}

In order to discuss the free field formalism, we consider the ordinary $\beta$-deformed matrix model of rank $N$,
\begin{align}
    \mathcal{Z}_N = \int \dd{X} \ee^{-\frac{b}{g} \tr V(X)} \Delta_N(X)^{2 b^2}
    \, .
\end{align}
In this case, the matrix moment (the power-sum average of the eigenvalues) is given as follows,
\begin{align}
    \left< \tr X^n \right> = \frac{1}{\mathcal{Z}_N} \int \dd{X} \left( \tr X^n \right) \ee^{-\frac{b}{g} \tr V(X)} \Delta_N(X)^{2 b^2}
    = - \frac{b}{g} n \pdv{}{t_n} \log \mathcal{Z}_N
    \, .
\end{align}
Hence, the derivative with the coupling constant $\{t_n\}$ plays a similar role to the multiplication of the matrix power $\{ \tr X^n \}$.
This is because the potential factor $\ee^{-\frac{b}{g} \tr V(X)}$ plays a role of the plane wave factor $\ee^{\ii p x}$ in the Fourier transform (FT).
In order to have this correspondence for all $n \in \mathbb{N}$, we consider the potential with infinitely many coupling constants, $V(x) = \sum_{n=1}^\infty \frac{t_n}{n} x^n$.
For the supermatrix model defined by
\begin{align}
    \mathcal{Z}_{N|M} = \frac{1}{N! M!} \int \dd{X} \dd{Y} \ee^{-\frac{1}{g} \str_b V(Z)} \frac{\Delta_N(X)^{2b^2} \Delta_M(Y)^{2b^{-2}}}{\prod_{i \in [N], j \in [M]}(x_i - y_j)^2}
    \, ,
\end{align}
the same argument is applied for $\langle \str_b X^n \rangle \leftrightarrow - \frac{1}{g} n \pdv{}{t_n}$.

Then, we define oscillator operators,
\begin{align}
    a_n = \sqrt{2} g n \pdv{}{t_n} \left( \stackrel{\text{FT}}{\longleftrightarrow} \ - \frac{1}{\sqrt{2}b} \tr X^n \right)
    \, , \qquad
    a_{-n} = \frac{1}{\sqrt{2}g} t_n
    \, , \qquad
    (n > 0)
\end{align}
which obeys the commutation relation of the Heisenberg algebra,
\begin{align}
    [a_n, a_m] = n \delta_{n+m,0}
    \, .
\end{align}
In addition, we also add the zero modes $(a_0,\bar{a}_0)$ with the commutation relation,
\begin{align}
    [a_n, \bar{a}_0] = \delta_{n,0}
    \, ,
\end{align}
where we interpret $a_0 \stackrel{\text{FT}}{\longleftrightarrow} -(\tr X^0)/\sqrt{2}b = - N /\sqrt{2}b$.
In this formalism, there exist infinitely many operators $\{ a_n \}_{n \in \mathbb{Z}}$.
They are independent operators if the matrix size is taken to be infinite $N \to \infty$.
For example, if $N = 1$, we have relations among these operators, $\tr X^n = (\tr X)^n$.

We define the vacuum state, which is annihilated by the positive modes,
\begin{align}
    a_{n} | 0 \rangle = 0 
    \, , \qquad
    n \ge 0
    \, .
\end{align}
In this sense, the positive modes are the annihilation operators, and the negative modes are the creation operators.
We also define the charged vacuum using the zero mode,
\begin{align}
    | \alpha \rangle = \ee^{\alpha \bar{a}_0} | 0 \rangle
    \, , \qquad
    a_0 | \alpha \rangle = \alpha | \alpha \rangle
    \, .
\end{align}

Based on the operators defined above, we define the operators, called the \emph{chiral boson} and the \emph{U(1) current}, as follows,
\begin{subequations}
    \begin{align}
    \phi(x) & = - \sum_{n \in \mathbb{Z}_{\neq 0}} \frac{a_n}{n} x^{-n} + a_0 \log x + \bar{a}_0
    & \implies \ - \frac{1}{\sqrt{2}b} \left[ \tr \log (x - X) - \frac{b}{g} V(x) \right]
    \, , \\
    J(x) & = \partial \phi(x) = \sum_{n \in \mathbb{Z}} \frac{a_n}{x^{n+1}} 
    & \implies \ - \frac{1}{\sqrt{2}b} \left[  \tr \left( \frac{1}{x - X} \right) - \frac{b}{g} V'(x) \right]
    \, .
\end{align}
\end{subequations}
Furthermore, we also define the energy-momentum tensor 
\begin{align}
    T(x) = \frac{1}{2} {:JJ:}(x) + \rho \partial J(x) =: \sum_{n \in \mathbb{Z}} \frac{L_n}{x^{n+2}}
    \, ,
    \qquad
    \rho = \sqrt{2}(b - b^{-1})
    \, ,
\end{align}
where we denote the normal ordering symbol by ${: - :}$, the annihilation operators are placed to right, the creation operators are placed to left in this symbol.
Recalling the operator product expansion (OPE) of the current operators,
\begin{align}
    J(x) J(x') = \frac{1}{(x - x')^2} + \text{regular}
    \, ,
\end{align}
the normal ordering is given as follows,
\begin{align}
    {:JJ:}(x) = \lim_{x' \to x} \left[ J(x) J(x') - \frac{1}{(x - x')^2} \right]
    \, .
\end{align}
It turns out that the generators $\{L_n\}_{n \in \mathbb{Z}}$ written in terms of the oscillators,
\begin{align}
    L_n = \frac{1}{2} \sum_{m \in \mathbb{Z}} {: a_m a_{n-m} :} - \rho (n+1) a_n
    \, ,
    \label{eq:Virasoro_gen_free_field}
\end{align}
obey the algebraic relation of the Virasoro algebra,
\begin{align}
    [L_n, L_m] = (n-m) L_{n+m} + \frac{c}{12} n (n^2 - 1) \delta_{n+m,0}
    \, , 
\end{align}
with the central charge
\begin{align}
    c = 1 - 6 (b - b^{-1})^2
    =
    \begin{cases}
        1 & (\beta = 2) \\
        - 2 & (\beta = 1, 4)
    \end{cases}
    \, .
\end{align}
Hence, the energy-momentum tensor plays a role of the generating current of the Virasoro algebra, and the construction of the energy-momentum tensor as a bilinear form of the current operator is called the Sugawara construction.

\subsubsection{Vertex operators}

We define the vertex operator from the chiral boson,
\begin{align}
    \mathsf{V}_\alpha (x) = {: \ee^{\alpha \phi(x)} :}    
    \, .
\end{align}
Then, the OPE with the energy-momentum tensor is given by
\begin{align}
    T(x) \mathsf{V}_\alpha (x') = \frac{\Delta_\alpha}{(x - x')^2} \mathsf{V}_\alpha (x') + \frac{1}{x - x'} \pdv{}{x'} \mathsf{V}_\alpha (x') + \text{regular}
\end{align}
where the coefficient called the conformal weight $\Delta_\alpha$ is given by
\begin{align}
    \Delta_\alpha = \frac{1}{2} \alpha (\alpha + \rho)
    \, .
\end{align}
From the Virasoro algebra point of view, the conformal weight is given by the eigenvalue of the operator $L_0$.
Moreover, the operator annihilated by $L_{n>0}$ is called the primary operator, and the vertex operator $\mathsf{V}_\alpha (x)$ is actually primary.
We remark that there are two possibilities to provide $\Delta_\alpha = 1$,  
\begin{align}
    \alpha_0 = - \sqrt{2} b
    \, , \qquad
    \alpha_1 = \sqrt{2} b^{-1}
    \quad \implies \quad
    \Delta_{\alpha_{0,1}} = 1
    \, .
\end{align}    
Hence, defining the screening current having the conformal weight one,
\begin{align}
    S_{\sigma}(x) = {: \ee^{\alpha_{\sigma} \phi(x)} :}
    \, , \qquad
    \sigma = 0, 1
    \, ,
\end{align}
the singular part of the OPE with the energy-momentum tensor is written as a total derivative,
\begin{align}
    T(x) S_\sigma (x') 
    & = \frac{1}{(x - x')^2} S_\sigma (x') + \frac{1}{x - x'} \pdv{}{x'} S_\sigma (x') + \text{regular}
    \nonumber \\
    & = \pdv{}{x'} \left[ \frac{1}{x - x'} S_\sigma (x') \right] + \text{regular}
    \, .
\end{align}
This implies that the screening charge defined by
\begin{align}
    Q_\sigma = \oint \dd{x} S_\sigma(x)
    \, ,
\end{align}
does not provide a singular contribution in the OPE, and thus it commutes with the energy-momentum tensor
\begin{align}
    [T(x), Q_\sigma] = 0
    \, .
\end{align}
This is a crucial property that characterizes the Virasoro algebra:
In fact, the Virasoro algebra is defined as a commuting sub-algebra of the Heisenberg algebra, and such a characterization can be applied to more generalized situations (W-algebras).
See, e.g.,~\cite{Bouwknegt:1992wg}.

\subsubsection{Construction of matrix model}

Let us discuss how to construct the matrix model based on the operator formalism discussed above.
Recalling that the vertex operator product is given by
\begin{align}
    \frac{\mathsf{V}_\alpha (x) \mathsf{V}_{\alpha'} (x')}{: \mathsf{V}_\alpha (x) \mathsf{V}_{\alpha'} (x') :} = (x - x')^{\alpha \alpha'}
    \, ,
\end{align}
the screening charge product is given by
\begin{align}
    Q_0^N & = \oint_{|x_1| < \cdots < |x_N|} \dd{X} \Delta_N(X)^{2b^2} {: \prod_{i\in[N]} S_0(x_i) :} 
    \nonumber \\
    & = \frac{1}{N!} \oint \dd{X} \Delta_N(X)^{2b^2} {: \prod_{i \in [N]} S_0(x_i) :} 
    \, ,
\end{align}
where the integration contour is initially taken in the radial ordering, and then analytically continuated to obtain the second expression.
Hence, defining the $\mathcal{Z}$-state
\begin{align}
    | \mathcal{Z}_N \rangle = Q_0^N | 0 \rangle
    \, ,
\end{align}
and the modified dual charged vacuum,%
\footnote{%
This modified vacuum is realized as a coherent state with respect to the Heisenberg algebra.
}
\begin{align}
    \langle \alpha; d+1 | a_{-n} = 
    \begin{cases}
        \displaystyle
        \langle \alpha; d+1 | \alpha & (n = 0) \\ \displaystyle
        \langle \alpha; d+1 | \frac{\mathfrak{t}_n}{\sqrt{2}g} & (n \in [d+1])
        \\
        0 & (\text{otherwise})
    \end{cases}
\end{align}
we obtain the matrix model partition function as a correlation function of the vertex operators,
\begin{align}
    \langle N \alpha_0; d+1 | \mathcal{Z}_N \rangle = \langle N \alpha_0; d+1 | Q_0^N | 0 \rangle = \frac{1}{N!} \oint \dd{X} \ee^{-\frac{b}{g} \tr V(X)} \Delta_N(X)^{2b^2}
    \, ,
\end{align}
where the potential function is now given by a finite polynomial function, $V(x) = \sum_{n=1}^{d+1} \frac{\mathfrak{t}_n}{n} x^n$.
Considering both screening charges $Q_{0,1}$, we instead obtain
\begin{align}
    | \mathcal{Z}_{N|M} \rangle & := Q_0^N Q_1^M | 0 \rangle
    \nonumber \\
    & = 
    \frac{1}{N! M!} \oint \dd{X} \dd{Y} \frac{\Delta_N(X)^{2 b^2} \Delta_M(Y)^{2 b^{-2}}}{\prod_{i \in [N], j \in [M]} (x_i - y_j)^2}
    {: \prod_{i \in [N]} S_0(x_i) \prod_{j \in [M]} S_1(y_j) :}
    \, ,
\end{align}
which gives rise to the supermatrix model partition function.
\begin{itembox}{Free field realization of supermatrix partition function}
The eigenvalue integral form of supermatrix model partition function is realized using two types of screening charges,
\begin{align}
    &
    \langle N\alpha_0 + M\alpha_1; d+1 | \mathcal{Z}_{N|M} \rangle 
    \nonumber \\
    & = 
    \frac{1}{N! M!} \oint \dd{X} \dd{Y} \ee^{-\frac{b}{g} \tr V(X)} \ee^{+\frac{b^{-1}}{g}\tr V(Y)} \frac{\Delta_N(X)^{2 b^2} \Delta_M(Y)^{2 b^{-2}}}{\prod_{i \in [N], j \in [M]} (x_i - y_j)^2}
    \, .
\end{align}
\end{itembox}
We remark that such an integral form of the correlation function is known as the \emph{Dotsenko--Fateev integral formula} in the context of two-dimensional CFT.

\subsubsection{Virasoro constraint}

Let us consider the action of the energy-momentum tensor on the vacuum state,
\begin{align}
    T(x) | 0 \rangle = \sum_{n \in \mathbb{Z}} \frac{L_n}{x^{n+2}} | 0 \rangle
    \, .
\end{align}
Requiring the regularity at $x = 0$, we should have
\begin{align}
    L_n | 0 \rangle = 0
    \, , \qquad
    n \ge -1
    \, .
    \label{eq:Virasoro_const_vac}
\end{align}
This condition is rephrased as follows:
The vacuum is interpreted as the primary field with the conformal weight zero, hence $L_{n \ge 0}$ annihilates the vacuum.
Furthermore, $L_{-1}$ is a generator of the translation realized by the derivative $\pdv{}{x}$.
Hence, $L_{-1}$ annihilates the vacuum since it is translation invariant.

Recalling that the $\mathcal{Z}$-state is constructed from the vacuum with the screening charges, that commute with the energy-momentum tensor (and thus the Virasoro generators).
This implies that the $\mathcal{Z}$-state shows the same behavior as in the case of the vacuum \eqref{eq:Virasoro_const_vac}.
\begin{itembox}{Virasoro constraint}
The $\mathcal{Z}$-state constructed by the screening charges obeys the following relation,
\begin{align}
    L_n | \mathcal{Z}_{N|M} \rangle = 0
    \, , \qquad
    n \ge -1
    \, .
    \label{eq:Virasoro_const}
\end{align}
This is called the Virasoro constraint for the (super)matrix model.
\end{itembox}
We see how this Virasoro is obtained in the context of the matrix model.
We start with the following identity,
\begin{align}
    0 & = \int \dd{X} \sum_{\ell \in [N]} \pdv{}{x_\ell} \left[ x_\ell^k \ee^{-\frac{b}{g} \tr V(X)} \Delta_N(X)^{2b^2} \right]
    \nonumber \\
    & = \int \dd{X} \sum_{\ell \in [N]} \left[ k x_\ell^{k-1} - \frac{b}{g} V'(x_\ell) x_\ell^k + 2b^2 \sum_{j (\neq \ell)} \frac{x_\ell^k}{x_\ell - x_j} \right] \ee^{-\frac{b}{g} \tr V(X)} \Delta_N(X)^{2b^2}
    \, .
\end{align}
Recalling the identity
\begin{align}
    \sum_{i \in [N]} \sum_{j (\neq i)} \frac{2 x_i^k}{x_i - x_j} & = \sum_{i \neq j} \frac{x_i^k - x_j^k}{x_i - x_j}
    \nonumber \\
    & = \sum_{m=0}^{k-1} \tr X^m \tr X^{k-m-1} - k \tr X^{k-1}
    \, ,
\end{align}
we obtain the following relation among the expectation values of the matrix moments, called the \emph{loop equation},
\begin{align}
    \sum_{m=0}^{k-1} \langle \tr X^m \tr X^{k-m-1} \rangle + (b^{-2} - 1) k \langle \tr X^{k-1} \rangle - \frac{b^{-1}}{g} \langle \tr X^k V'(X) \rangle = 0
    \, .
\end{align}
In the operator formalism discussed in Sec.~\ref{sec:op_formalism}, we may write this relation as
\begin{align}
    L_{k-1} \mathcal{Z}_N = 0
    \, , \qquad 
    k \ge 0
    \, ,
\end{align}
where we apply the free field realization of the Virasoro generator \eqref{eq:Virasoro_gen_free_field}.
See~\cite{Desrosiers:2009pz} for the loop equation of the supermatrix model in the case $b = 1$.

\section{Supergroup gauge theory}\label{sec:supergroup_gauge_theory}

In this section, we introduce supergroup gauge theory, gauge theory having supergroup gauge symmetry, and discuss fundamental perspectives.

\subsubsection*{Differential forms}

Let $G$ be a Lie supergroup, and the corresponding Lie superalgebra $\mathfrak{g} = \operatorname{Lie} G$.
Let $M$ be a $d$-dimensional space-time manifold.
The fundamental degrees of freedom of supergroup gauge theory is the one-form connection that takes a value in $\mathfrak{g}$, $A \in \Omega^1(M,\mathfrak{g})$.
Then, we define the covariant derivative $D = d + A$ and the curvature two-form given by $D^2 = dA + A^2 \in \Omega^2(M,\mathfrak{g})$.
The $G$-gauge transformation is given by
\begin{align}
    G \ : \
    D \longmapsto g D g^{-1}
    \, , \qquad
    F \longmapsto g F g^{-1}
    \, , \qquad
    A \longmapsto g A g^{-1} + g d g^{-1}
    \, , \qquad
    g \in G
    \, .
    \label{eq:gauge_transf}
\end{align}
Formally, these expressions are parallel with the ordinary (non-supergroup) gauge theory.
In the $\mathbb{Z}_2$-graded situation, the connection is then called the \emph{superconnection}~\cite{Quillen:1985vya,Mathai:1986tc}.
Let us discuss the supermatrix realization of these differential forms.
Let $G = \mathrm{U}(n_0|n_1)$ for the moment.
In this case, the connection $A$ is given by an anti-Hermitian supermatrix,
\begin{align}
    A = 
    \begin{pmatrix}
        A^{(0)} & \psi \\ \psi^\dag & A^{(1)}
    \end{pmatrix}
    \quad \text{with} \quad
    A^{(\sigma)\dag} = - A^{(\sigma)}
    \quad \implies \quad
    A^\dag = - A
    \, .
\end{align}
We recall that $\bar{\bar{\theta}} = -\theta$ for the Grassmann variable (see~\eqref{eq:Grassmann_var_double_conj}).
We remark that $A^{(\sigma)}$ is in the adjoint representation of $\mathrm{U}(n_\sigma)$, and $\psi$ is in the bifundamental representation, $\mathrm{U}(n_0) \times \overline{\mathrm{U}(n_1)}$.
Moreover, each component is given by
\begin{align}
    A^{(\sigma)} = A_\mu^{(\sigma)} dx^\mu 
    \, , \qquad
    \psi = \psi_\mu dx^\mu
    \, .
\end{align}
Hence, $\psi_\mu$ is a spin-1 fermionic degree of freedom, that is not compatible with the spin-statistics theorem.
The curvature two-form is given by
\begin{align}
    F = dA + A \wedge A =
    \begin{pmatrix}
        dA^{(0)} + A^{(0)} \wedge A^{(0)} + \psi \wedge \psi^\dag & d \psi + A \wedge \psi + \psi \wedge B \\
        d \psi^\dag + \psi^\dag \wedge A + B \wedge \psi^\dag & dA^{(1)} + A^{(1)} \wedge A^{(1)} + \psi^\dag \wedge \psi
    \end{pmatrix}
    \, .
\end{align}
We may write the wedge product using the anti-symmetrization symbol,
\begin{align}
    a \wedge b = \frac{1}{2} \left(a_\mu b_\nu - a_\nu b_\mu \right) dx^\mu dx^\nu = \frac{1}{2} a_{[\mu} b_{\nu]} dx^\mu dx^\nu
    \, .
\end{align}
Therefore, the component of the curvature two-form is given by
\begin{align}
    F = \frac{1}{2} F_{\mu \nu} dx^\mu dx^\nu
\end{align}
where we have
\begin{align}
    F_{\mu \nu} & = 
%    \begin{pmatrix}
%        \partial_\mu A^{(0)}_\nu - \partial_\nu A^{(0)}_\mu & \partial_\mu \psi_\nu - \partial_\nu \psi_\mu \\
%        \partial_\mu \psi_\nu^\dag - \partial_\nu \psi_\mu^\dag & \partial_\mu A^{(1)}_\nu - \partial_\nu A^{(1)}_\mu
%    \end{pmatrix}
%    +
%    \begin{pmatrix}
%        [A^{(0)}_\mu, A^{(0)}_\nu] + \psi_{[\mu} \psi^\dag_{\nu]} & A^{(0)}_{[\mu} \psi_{\nu]} + \psi_{[\mu} A^{(1)}_{\nu]} \\
%        \psi_{[\mu}^\dag A^{(0)}_{\nu]} + A^{(1)}_{[\mu} \psi^\dag_{\nu]} & [A^{(1)}_\mu, A^{(1)}_\nu] + \psi_{[\mu}^\dag \psi_{\nu]}
%    \end{pmatrix}
%    \nonumber \\
%    & = 
    \begin{pmatrix}
        F_{\mu\nu}^{(0)} + \psi_{[\mu} \psi^\dag_{\nu]} & \partial_{[\mu} \psi_{\nu]} + A^{(0)}_{[\mu} \psi_{\nu]} + \psi_{[\mu} A^{(1)}_{\nu]} \\
        \partial_{[\mu} \psi_{\nu]}^\dag + \psi_{[\mu}^\dag A^{(0)}_{\nu]} + A^{(1)}_{[\mu} \psi^\dag_{\nu]} & F_{\mu\nu}^{(1)} + \psi_{[\mu}^\dag \psi_{\nu]}
    \end{pmatrix}
    \, .
    \label{eq:super_curvature}
\end{align}
We denote the curvature associated with the bosonic subgroup of $G$ by $F^{(\sigma)} = d A^{(\sigma)} + A^{(\sigma)} \wedge A^{(\sigma)} = \frac{1}{2} F^{(\sigma)}_{\mu\nu} dx^\mu dx^\nu$, e.g., for $G = \mathrm{U}(n_0|n_1)$, $F^{(\sigma)} \in \Omega^2(M,\mathfrak{u}_{n_\sigma})$.

In general, we define the commutator for the differential forms as follows:
Let $X = X^a \otimes t^a \in \Omega^p(M,\mathfrak{g})$ and $Y = Y^a \otimes t_b \in \Omega^q(M,\mathfrak{g})$ with the generators of the Lie superalgebra denoted by $(t^a)_{a \in [\operatorname{dim} \mathfrak{g}]}$.
Recalling $X^a \wedge Y^b = (-1)^{pq} (-1)^{|a||b|} Y^b \wedge X^a$, denoted by $|a| = |X_a|$, etc, the commutator for the differential forms is defined by
\begin{align}
    [X, Y] & = t^a t^b \otimes X^a \wedge Y^b - (-1)^{pq} t^b t^a \otimes Y^b \wedge X^a
    \nonumber \\
    & = (t^a t^b - (-1)^{|a||b|} t^b t^a) \otimes X^a \wedge Y^b
    \nonumber \\
    & = [t^a, t^b] \otimes X^a \wedge Y^b
    \, .
\end{align}
The commutator in the last line is the supercommutator with respect to the superalgebra $\mathfrak{g}$ defined in~\eqref{eq:supercommutator_def}.
For the one-form connection, $A \in \Omega^1(M,\mathfrak{g})$, we have
\begin{align}
    A \wedge A = \frac{1}{2} [A,A] = \frac{1}{2} [t^a, t^b] \otimes A^a \wedge A^b
    \, .
\end{align}
Therefore, the component of the curvature two-form is given in the same way as the ordinary case,
\begin{align}
    F_{\mu\nu} = \partial_\mu A_\nu - \partial_\nu A_\mu + [A_\mu, A_\nu]
    \, ,
\end{align}
which takes a value in the Lie superalgebra $\mathfrak{g}$.

\subsection{Supergroup Yang--Mills theory}

We consider the Yang--Mills (YM) action for the supergroup theory,
\begin{align}
    S_\text{YM} = - \frac{1}{g^2} \int_M \str (F \wedge \star F)
    \, ,
\end{align}
where we define the Hodge star operator, $\star$ : $\Omega^p(M) \to \Omega^{d-p}(M)$.
Hence, we have the volume form, $F \wedge \star F = \frac{1}{4} F_{\mu\nu} F^{\mu\nu} d\!\operatorname{vol}(M)$. 
In order to impose the invariance under the $G$-gauge transform~\eqref{eq:gauge_transf}, we replace the ordinary trace with the supertrace.

From the expression of the two-form curvature~\eqref{eq:super_curvature}, we can explicitly write down the YM action in terms of the gauge fields, $(A^{(\sigma)}, \psi)$, which would be complicated.
Hence, for the moment, we just write the leading contributions, 
\begin{align}
    S_\text{YM} = - \frac{1}{g^2} \int_M \tr_0 (F^{(0)} \wedge \star F^{(0)}) + \frac{1}{g^2} \int_M \tr_1 (F^{(1)} \wedge \star F^{(1)}) + \cdots
    \, ,
    \label{eq:S_YM_leading}
\end{align}
which are the standard YM actions of the bosonic subgroups $\mathrm{U}(n_\sigma)$ in $G = \mathrm{U}(n_0|n_1)$.
Now we observe that the kinetic term signatures are opposite due to the supertrace property, that means that the spectrum of supergroup YM theory is not bounded.
Such a property was also found in the supermatrix model discussed in Sec.~\ref{sec:super_matrix}, and thus it seems to be a universal behavior in supergroup theory.
Due to this unboundness, the notion of vacua is not well-defined in this case, and thus we would need a non-perturbative completion for supergroup theory:%
\footnote{%
Such a situation resembles unstable vacua. 
In this case, we should find true (stable) vacua non-perturbatively.
}
In fact, it has been known that supergroup theory is pertubatively equivalent to the ordinary gauge theory through the analytic continuation, but there would be an essential difference in the non-perturbative regime.

Even though there are no bounded vacua, one can still consider the equation of motion with respect to the YM action, which is given by
\begin{align}
    D(\star F) = 0
    \, .
\end{align}
This is a second order non-linear PDE on $M$, which is in general difficult to solve.
In the four-dimensional case, we have a special class of solutions, called the (anti-)instanton, given by the solution of the (anti-)self-dual ((A)SD) YM equation,
\begin{align}
    \star F = \mp F
    \, .
    \label{eq:ASDYMeq}
\end{align}
One can check that the instanton provides a solution of the equation of motion using the Bianchi identity, 
\begin{align}
    D(\star F) \stackrel{\text{(A)SD}}{=} \mp D F \stackrel{\text{Bianchi}}{=} 0
    \, .
\end{align}
We remark that the Bianchi identity still holds for the supergroup case due to the Jacobi identity~\eqref{eq:Jacobi_id}.
In fact, the instanton plays an essential role in non-perturbative aspects of supergroup gauge theory.
We will discuss the details of instantons in Sec.~\ref{sec:instanton}.

\subsection{Quiver gauge theory realization}\label{sec:quiver_realization}

As observed in~\eqref{eq:S_YM_leading}, the YM action of supergroup gauge theory consists of two ordinary YM actions.
In fact, supergroup gauge theory has a realization as a quiver gauge theory through analytic continuation.

Quiver gauge theory is a class of gauge theories involving multiple gauge degrees of freedom: The gauge group is given by a product form, $G = \prod_i G_i$.
In addition to the gauge field, there is another degree of freedom, called the \emph{bifundamental matter}, that connects different gauge fields, transforming in the bifundamental representation of the connecting gauge groups, $G_i \times G_j$, $G_i \times \overline{G}_j$, etc.
In the case of $\mathrm{U}(n_0|n_1)$ supergroup theory, in addition to the gauge fields of $\mathrm{U}(n_\sigma)$ subgroup, there are also fermionic degrees of freedom that transform in $\mathrm{U}(n_0) \times \overline{\mathrm{U}(n_1)}$ and the conjugate.
Hence, these fermionic fields are interpreted as bifundamental matters from this point of view.
Since we have two such fields $(\psi,\psi^\dag)$, this theory is identified with $\widehat{A}_1$ quiver gauge theory.%
\footnote{%
This classification is based on the identification of quiver diagram and (affine) Dynkin diagram.
$\widehat{A}_1$ quiver consists of two nodes and two connecting edges.
}

An important feature of supergroup gauge theory is the signature of the kinetic term. 
In order to realize this situation based on $\widehat{A}_1$ quiver theory, we have to assign the coupling constants as follows,
\begin{align}
    \left( \frac{1}{g_0^2} \, , \, \frac{1}{g_1^2} \right)
    \ \longrightarrow \ 
    \left( + \frac{1}{g^2} \, , \, - \frac{1}{g^2} \right)
    \, .
    \label{eq:coupling_tuning_A1}
\end{align}
Otherwise, we impose the condition $\frac{1}{g_0^2} + \frac{1}{g_1^2} = 0$.
This assignment is in fact unphysical since we have a negative coupling. 
Therefore, the supergroup gauge theory is realized in unphysical parameter regime, which could be interpreted as the analytic continuation of physical $\widehat{A}_1$ quiver gauge theory~\cite{Dijkgraaf:2016lym}.

\subsubsection*{Orthosymplectic theory}

From this point of view, the orthosymplectic supergroup theory is realized similarly by $\widehat{A}_1$ quiver with O and Sp gauge nodes.
In fact, the combinations of O $\times$ O and Sp $\times$ Sp are not compatible with their flavor symmetry, and O $\times$ Sp is a unique choice among O and Sp theories.

\subsubsection*{Chern--Simons theory}

A similar construction is also available for Chern--Simons theory.
In this case, the Chern--Simons level plays a role of the coupling constant.
In fact, the supersymmetric Chern--Simons--matter theory, a.k.a., Aharony--Bergman--Jafferis--Maldacena (ABJM) theory~\cite{Aharony:2008ug,Aharony:2008gk}, completely fits these conditions:
It is a quiver gauge theory involving two nodes and two bifundamental matters, and two couplings (levels) with opposite signs.
Actually, the partition function of ABJM theory obtained through the localization formalism supports its connection with supergroup theory: 
The partition function of $\mathrm{U}(N)_k \times \mathrm{U}(M)_{-k}$ theory takes a form~\cite{Drukker:2009hy,Marino:2009jd},
\begin{align}
    \mathcal{Z}_{N|M} = \frac{1}{N!M!} \int \dd{X} \dd{Y} \ee^{-\frac{1}{2g} \tr X^2 + \frac{1}{2g} \tr Y^2} \frac{\prod_{i < j}^{N} \sinh (x_i - x_j)^2 \prod_{i < j}^{M} \sinh (y_i - y_j)^2}{\prod_{j \in [M]}^{i \in [N]} \cosh(x_i - y_j)^2}
    \, ,
\end{align}
where we have the pure imaginary coupling $g = 2 \pi \ii / k$.
This is interpreted as a trigonometric analog of the supermatrix model discussed in Sec.~\ref{sec:super_matrix}.
We remark that since the Chern--Simons theory is a topological theory, a negative coupling (negative level) does not imply unphysical behavior.
See also \cite{Horne:1989ue,Bourdeau:1991hn,Rozansky:1992zt} \cite{Mikhaylov:2014aoa,Mikhaylov:2015nsa} \cite{Okazaki:2017sbc} \cite{Aghaei:2018cbn} \cite{Cassia:2021uly} for related works on supergroup Chern--Simons theory.

\subsubsection*{Supergroup quiver gauge theory}

We can generalize this argument for quiver gauge theory of supergroups.
In this case, since each gauge node is of supergroup, we need copies of $\widehat{A}_1$ quiver theory.
For $\Gamma$-quiver supergroup gauge theory, the total structure is given by $(\widehat{A}_1, \Gamma)$ quiver.
Such a theory characterized by a pair of quivers is called the double quiver theory, which has a natural geometric origin in eight dimensions.
See \cite{Kimura:2022zsm}.

\subsection{String/M-theory perspective}\label{sec:string_theory_perspective}

String/M-theory provides various insights on non-perturbative aspects of gauge theory.
In particular, considering a stack of D-branes, non-Abelian gauge theory is realized as a low-energy effective theory.
Open strings ending on the brane provide matrix degrees of freedom as homomorphism of the Chan--Paton vector space associated with the boundary condition of open string:
$n$ stack of branes gives rise to $n$-dimensional vector space.
From this point of view, we require two different types of branes to realize a $\mathbb{Z}_2$-graded vector space to construct supergroup gauge theory.
A natural candidate is an anti-brane.
Even though a brane--anti-brane system exhibits similar properties, it has been known that this configuration does not yield supergroup gauge theory:
Open strings connecting brane--anti-brane give rise to the \emph{tachyon}, that is a bosonic degree of freedom transforming in the bifundamental representation of two gauge groups associated with branes and anti-branes.
See, e.g.,~\cite{Witten:1998cd,Kraus:2000nj,Takayanagi:2000rz,Alishahiha:2000du} for details.
Therefore, we need a different type of branes required for realizing a fermionic degree of freedom.
Such an object is known as the \emph{negative brane} (also called the \emph{ghost brane})~\cite{Vafa:2001qf,Okuda:2006fb,Vafa:2014iua,Dijkgraaf:2016lym}, and it has been shown that a stack of brane--negative-brane actually yields a supergroup gauge theory.
More recently, it has been pointed out that such a negative brane plays an essential role in the resurgence~\cite{Marino:2022rpz,Schiappa:2023ned}.

Comparing the anti-brane and the negative brane, while the anti-brane has a positive tension (positive energy density; source of gravity), the negative brane has a negative tension (negative energy density; source of anti-gravity) although both have negative RR charges.
From this point of view, the negative brane is associated with an unphysical open string boundary condition.
On the other hand, an advantage of negative branes is that a bound state with branes does not violate further supersymmetry, which is still BPS, while that for anti-branes is not BPS.
This property also plays an important role to discuss the instanton solution in supergroup gauge theory.
See Sec.~\ref{sec:instanton}.

\subsubsection{Hanany--Witten construction}\label{sec:Hany-Witten_const}

A stack of brane--negative-brane yields sixteen supercharges supergroup gauge theory similarly to the configuration of ordinary branes.
In order to reduce the supersymmetries, one can consider the Hanany--Witten setup~\cite{Hanany:1996ie} that involves NS5 branes and D4 branes suspended between them in the type IIA setup.
Adding the negative D4 branes, we can realize four-dimensional $\mathcal{N} = 2$ supergroup gauge theory that preserves eight supercharges.
Each brane is extended in the following directions:%
\footnote{%
We denote the ordinary $p$-brane (positive brane) by D$p$ or D$p^+$, the negative brane by D$p^-$, the anti-brane by $\overline{\text{D}p}$ or $\overline{\text{D}p}^+$, the anti-negative brane by $\overline{\text{D}p}^-$.
}
\begin{align}
    \begin{tabular*}{.8\textwidth}{@{\extracolsep{\fill}}ccccccccccc}\toprule
        &0&1&2&3&4&5&6&7&8&9 \\\midrule
        NS5 &--&--&--&--&--&--&&& \\
        D4$^\pm$ &--&--&--&--&&&--&& \\\toprule
    \end{tabular*}
\end{align}
In this configuration, the positions of D4$^\pm$ branes in 45-direction are identified with the Coulomb moduli of $\mathcal{N}=2$ gauge theory (two coordinates are combined into a single complex coordinate) and the distance between NS5 branes in 6-direction $L$ is interpreted as the gauge coupling constant:
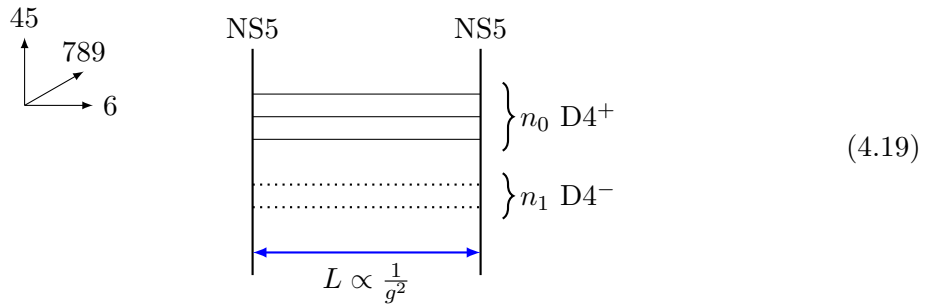
\begin{align}
    \begin{tikzpicture}[scale=1.5,baseline=(current  bounding  box.center)]
    % axis
    \begin{scope}[shift={(-3,.5)},scale=.6]
     \draw[-latex] (0,0) -- ++(1,0) node [right] {6};
     \draw[-latex] (0,0) -- ++(0,1) node [above] {45};
     \draw[-latex] (0,0) -- ++(30:1.) node [above] {789};
    \end{scope}
    % NS5
   \draw[thick] (-1,-1) -- ++(0,2) node [above] {NS5};
   \draw[thick] (1,-1) -- ++(0,2) node [above] {NS5};
    \draw[latex-latex,thick,blue] (-1.,-.8) -- ++(1,0) node [below,black] {$L \propto \frac{1}{g^2}$} -- ++(1,0);
    % D4+
   \draw (-1,.2) -- ++(2,0);
   \draw (-1,.4) -- ++(2,0);
   \draw (-1,.6) -- ++(2,0);
    % D4-
   \draw [dotted,thick] (-1,-.2) -- ++(2,0);
   \draw [dotted,thick] (-1,-.4) -- ++(2,0);
   \draw [thick,decorate,decoration={brace,amplitude=4pt,mirror,raise=4pt},yshift=0pt] (1.1,.1) -- ++(0,.6) node [black,midway,xshift=2.6em] {$n_0$ D4$^+$};
   \draw [thick,decorate,decoration={brace,amplitude=4pt,mirror,raise=4pt},yshift=0pt] (1.1,-.5) -- ++(0,.4) node [black,midway,xshift=2.6em] {$n_1$ D4$^-$};   
   \end{tikzpicture}
\end{align}
Let us remark an ambiguity of the diagram presented above.
Since we have two different types of branes, we have possibly different configurations, for example, for U($2|1$) theory, as follows,
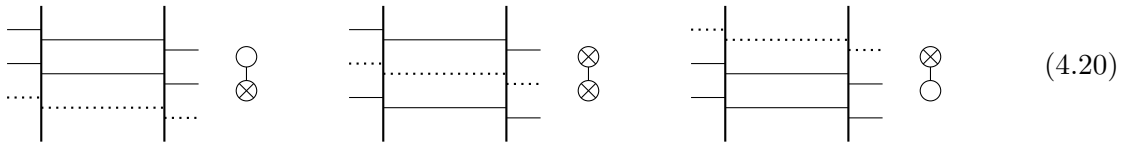
\begin{align}
 \begin{tikzpicture}[baseline=(current bounding box.center),scale=.9]
%  \node at (-.2,2.5) {(a)};
  \draw[thick] (0,0) -- ++(0,2);
  \draw[thick] (1.8,0) -- ++(0,2);
  \draw (0,1.5) -- ++(1.8,0);
  \draw (0,1) -- ++(1.8,0);
  \draw[dotted,thick] (0,.5) -- ++(1.8,0);
  \draw (3,1.25) -- ++(0,-.5);
  \draw (-.5,1.65) -- ++(.5,0); %matter
  \draw (1.8,1.35) -- ++(.5,0); %matter
  \draw (-.5,1.15) -- ++(.5,0); %matter
  \draw (1.8,.85) -- ++(.5,0); %matter
  \draw[dotted,thick] (-.5,.65) -- ++(.5,0);  %matter
  \draw[dotted,thick] (1.8,.35) -- ++(.5,0);  %matter
  \filldraw[fill=white,draw=black] (3,1.25) circle (.15);
  \filldraw[fill=white,draw=black] (3,.75) circle (.15);
   \draw (3,.75)++(135:.15) -- ++(-45:.3);
   \draw (3,.75)++(45:.15) -- ++(-135:.3);     
  \begin{scope}[shift={(5,0)}]
%   \node at (-.2,2.5) {(b)};
  \draw[thick] (0,0) -- ++(0,2);
  \draw[thick] (1.8,0) -- ++(0,2);
  \draw (0,1.5) -- ++(1.8,0);
  \draw[dotted,thick] (0,1) -- ++(1.8,0);
  \draw (0,.5) -- ++(1.8,0);
  \draw (3,1.25) -- ++(0,-.5);
  \draw (-.5,1.65) -- ++(.5,0); %matter
  \draw (1.8,1.35) -- ++(.5,0); %matter
  \draw[dotted,thick] (-.5,1.15) -- ++(.5,0); %matter
  \draw[dotted,thick] (1.8,.85) -- ++(.5,0); %matter
  \draw (-.5,.65) -- ++(.5,0);  %matter
  \draw (1.8,.35) -- ++(.5,0);  %matter
   \filldraw[fill=white,draw=black] (3,1.25) circle (.15);
   \filldraw[fill=white,draw=black] (3,.75) circle (.15);
   \draw (3,1.25)++(135:.15) -- ++(-45:.3);
   \draw (3,1.25)++(45:.15) -- ++(-135:.3);
   \draw (3,.75)++(135:.15) -- ++(-45:.3);
   \draw (3,.75)++(45:.15) -- ++(-135:.3);     
  \end{scope}
  \begin{scope}[shift={(10,0)}]
%   \node at (-.2,2.5) {(c)};
  \draw[thick] (0,0) -- ++(0,2);
  \draw[thick] (1.8,0) -- ++(0,2);
  \draw[dotted,thick] (0,1.5) -- ++(1.8,0);
  \draw (0,1) -- ++(1.8,0);
  \draw (0,.5) -- ++(1.8,0);
  \draw (3,1.25) -- ++(0,-.5);
  \draw[dotted,thick] (-.5,1.65) -- ++(.5,0); %matter
  \draw[dotted,thick] (1.8,1.35) -- ++(.5,0); %matter
  \draw (-.5,1.15) -- ++(.5,0); %matter
  \draw (1.8,.85) -- ++(.5,0); %matter
  \draw (-.5,.65) -- ++(.5,0);  %matter
  \draw (1.8,.35) -- ++(.5,0);  %matter
  \filldraw[fill=white,draw=black] (3,1.25) circle (.15);
  \filldraw[fill=white,draw=black] (3,.75) circle (.15);
   \draw (3,1.25)++(135:.15) -- ++(-45:.3);
   \draw (3,1.25)++(45:.15) -- ++(-135:.3);
  \end{scope}
 \end{tikzpicture}
 \label{eq:HW_U(2|1)}
\end{align}
This ambiguity corresponds to that for the simple roots of Lie superalgebra.
We also present the corresponding Dynkin diagrams aside the brane diagrams.
See also Sec.~\ref{sec:supermatrix} for a related discussion.

\subsubsection{Gauging trick}\label{sec:gauging_trick}

Starting with the Hanany--Witten configuration for supergroup gauge theory, we in addition include infinitely extended D4$^+$ branes.
Since they have infinite length, their couplings are zero (non-dynamical).
Then, we consider gauging this configuration: 
The middle part of D4$^+$ branes can be moved in 45-direction, while the external parts are still frozen, that plays a role of the flavor brane.
Further tuning the position of D4$^+$ branes, one can remove D4$^-$ branes via pair annihilation between D4$^+$ and D4$^-$ branes, and the resulting configuration does not involve D4$^-$ branes any longer:
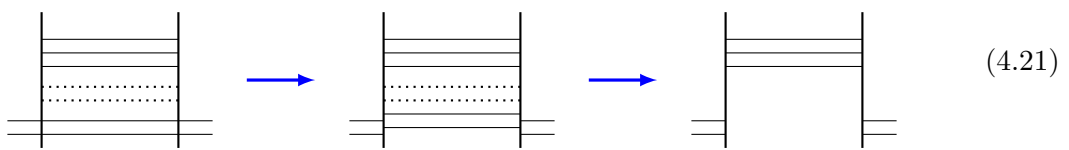
\begin{align}
  \begin{tikzpicture}[scale=.9,baseline=(current  bounding  box.center)]
   \draw[thick] (-1,-1) -- ++(0,2) node [above] {};% {NS5};
   \draw[thick] (1,-1) -- ++(0,2) node [above] {};% {NS5};
   \draw (-1,.2) -- ++(2,0);
   \draw (-1,.4) -- ++(2,0);
   \draw (-1,.6) -- ++(2,0);
   \draw [dotted,thick] (-1,-.1) -- ++(2,0);
   \draw [dotted,thick] (-1,-.3) -- ++(2,0);
   \draw (-1.5,-.6) -- ++(3,0);
   \draw (-1.5,-.8) -- ++(3,0);   
   \draw[-latex,blue,very thick] (2,0) -- ++(.5,0) node [above] {} -- ++(.5,0);
   \begin{scope}[shift={(5,0)}]
   \draw[thick] (-1,-1) -- ++(0,2) node [above] {};% {NS5};
   \draw[thick] (1,-1) -- ++(0,2) node [above] {};% {NS5};
   \draw (-1,.2) -- ++(2,0);
   \draw (-1,.4) -- ++(2,0);
   \draw (-1,.6) -- ++(2,0);
   \draw [dotted,thick] (-1,-.1) -- ++(2,0);
   \draw [dotted,thick] (-1,-.3) -- ++(2,0);
    \draw (-1.5,-.6) -- ++(.5,0);
    \draw (-1.5,-.8) -- ++(.5,0);
    \draw (1,-.6) -- ++(.5,0);
    \draw (1,-.8) -- ++(.5,0);       
    \draw (-1,-.5) -- ++(2,0);
    \draw (-1,-.7) -- ++(2,0);    
    \draw[-latex,blue,very thick] (2,0) -- ++(.5,0) node [above] {} -- ++(.5,0);   \end{scope}
   \begin{scope}[shift={(10,0)}]
   \draw[thick] (-1,-1) -- ++(0,2) node [above] {};% {NS5};
   \draw[thick] (1,-1) -- ++(0,2) node [above] {};% {NS5};
   \draw (-1,.2) -- ++(2,0);
   \draw (-1,.4) -- ++(2,0);
   \draw (-1,.6) -- ++(2,0);
   %
%   \draw [dotted] (-1,-.1) -- ++(2,0);
%   \draw [dotted] (-1,-.3) -- ++(2,0);
   %    
    \draw (-1.5,-.6) -- ++(.5,0);
    \draw (-1.5,-.8) -- ++(.5,0);
    \draw (1,-.6) -- ++(.5,0);
    \draw (1,-.8) -- ++(.5,0);           
   \end{scope}      
  \end{tikzpicture}  
\end{align}
This configuration is identical to U$(n_0)$ gauge theory with $n_F = 2n_1$ flavor degrees of freedom~\cite{Dijkgraaf:2016lym}.
This process implies that the (Coulomb branch of) moduli space of vacua of supergroup gauge theory has an intersection with that for the ordinary $\mathcal{N}=2$ gauge theory with flavors (SQCD).

\subsubsection{$\widehat{A}_1$ quiver realization}

As discussed in Sec.~\ref{sec:quiver_realization}, supergroup gauge theory has a realization as $\widehat{A}_1$ quiver gauge theory.
In fact, $\widehat{A}_1$ quiver theory is similarly realized in this framework.
In this case, we compactify 6-direction and put D4 branes in this direction.
Then, we have two domains suspended by NS5 branes, that realize two distinct gauge nodes.
This configuration can be converted to the previous configurations through analytic continuation as follows.

We recall that supergroup gauge theory can be obtained by tuning the couplings as in~\eqref{eq:coupling_tuning_A1} in $\widehat{A}_1$ quiver theory, and in this setup, each coupling $1/g_\sigma^2$ is interpreted as length of D4 branes between NS5 branes, $L_\sigma \propto 1/g_\sigma^2$.
From this point of view, a negative brane would be interpreted as a brane with negative length, that would be thought of as analytic continuation: 
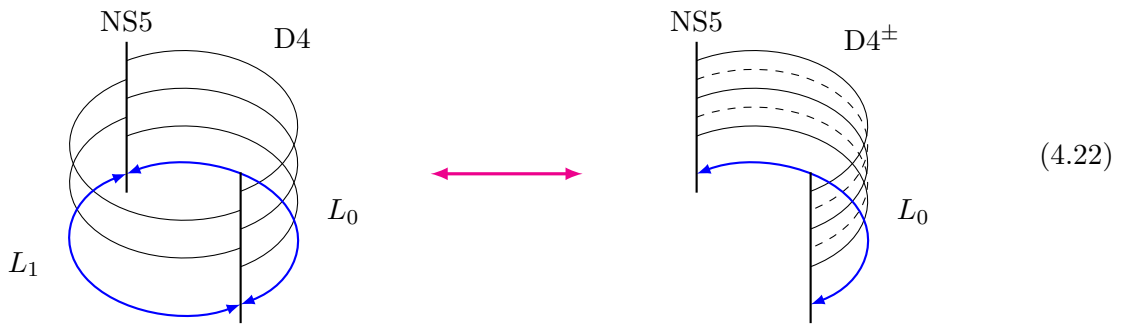
\begin{align}
  \begin{tikzpicture}[baseline=(current  bounding  box.center)]
  % D4
  \draw (0,0) arc [x radius = 1.5, y radius = 1, start angle = 120, end angle = -60];
  \draw (0,-.5) arc [x radius = 1.5, y radius = 1, start angle = 120, end angle = -60];
    \draw (0,-1) arc [x radius = 1.5, y radius = 1, start angle = 120, end angle = -60];
    \draw (0,-.25) arc [x radius = 1.5, y radius = 1, start angle = 120, end angle = 300];    
    \draw (0,-.75) arc [x radius = 1.5, y radius = 1, start angle = 120, end angle = 300];        
    \draw[blue,latex-latex,thick] (0,-1.5) arc [x radius = 1.5, y radius = 1, start angle = 120, end angle = -60];
    \draw[blue,latex-latex,thick] (0,-1.5) arc [x radius = 1.5, y radius = 1, start angle = 120, end angle = 300];
    % NS5
    \draw[thick] (0,-1.75) -- ++(0,2) node [above] {NS5};
    \draw[thick] (0,-1.75)++(1.5,-1.73) -- ++(0,2);
    \node at (1.8,.3) [right] {D4};    
    \node at (2.5,-2.) [right] {$L_0$};
    \node at (-1,-2.7) [left] {$L_1$};
    \draw[very thick,magenta,latex-latex] (4,-1.5) -- ++(2,0);
    \begin{scope}[shift = {(7.5,0)}]
      % D4
  \draw (0,0) arc [x radius = 1.5, y radius = 1, start angle = 120, end angle = -60];
  \draw (0,-.5) arc [x radius = 1.5, y radius = 1, start angle = 120, end angle = -60];
    \draw (0,-1) arc [x radius = 1.5, y radius = 1, start angle = 120, end angle = -60];
    \draw[dashed] (0,-.25) arc [x radius = 1.5, y radius = 1, start angle = 120, end angle = -60];    
    \draw[dashed] (0,-.75) arc [x radius = 1.5, y radius = 1, start angle = 120, end angle = -60];        
    \draw[blue,latex-latex,thick] (0,-1.5) arc [x radius = 1.5, y radius = 1, start angle = 120, end angle = -60];
%    \draw[blue,latex-latex] (0,-1.5) arc [x radius = 1.5, y radius = 1, start angle = 120, end angle = 300];    
    % NS5
    \draw[thick] (0,-1.75) -- ++(0,2) node [above] {NS5};
    \draw[thick] (0,-1.75)++(1.5,-1.73) -- ++(0,2);
    \node at (1.8,.3) [right] {D4$^\pm$};    
    \node at (2.5,-2.) [right] {$L_0$}; 
    \end{scope}
  \end{tikzpicture}
\end{align}

Let us comment a possible connection to the gauging trick discussed above.
Recalling that the supergroup condition~\eqref{eq:coupling_tuning_A1}, namely $L_0 + L_1 = 0$, it would correspond to the shrinking limit in 6-direction.
In fact, this is the strong coupling limit, that also implies necessity of non-perturbative treatment in supergroup gauge theory.
Then, applying the T-dual in 6-direction, this direction is infinitely extended, and the gauge coupling of either first or second gauge node has to be zero.
This means that the resulting configuration is a single node gauge theory with a flavor node, which is consistent with the gauging trick.
This argument is extended to supergroup quiver gauge theory in general.
See~\cite{Kimura:2019msw,Kimura:2020jxl} for details.

\subsubsection{Seiberg--Witten theory}\label{sec:SW_theory}

Seiberg--Witten theory provides a geometric description of the moduli space of supersymmetric vacua of four-dimensional $\mathcal{N}=2$ gauge theory.
In fact, one can concisely extract the Seiberg--Witten geometry from the Hanany--Witten brane configuration~\cite{Witten:1997sc}.

In this setup, we specify the positions of D4 branes and NS5 branes using two complex variables $(x,y) \in \mathbb{C} \times \mathbb{C}^\times$ as follows,
\begin{align}
    y + \frac{\mathfrak{q}}{y} = \det(x - \Phi)
    \, ,
\end{align}
where we denote the exponentiated complexified gauge coupling by $\mathfrak{q} = \exp (2 \pi \ii \tau)$, $\tau = \frac{\theta}{2\pi} + \frac{4 \pi \ii}{g^2}$, with the $\theta$-angle, and $\Phi$ is the adjoint complex scalar field in the vector multiplet that parametrizes the Coulomb branch of the moduli space.
In order to obtain the Seiberg--Witten geometry of supergroup gauge theory, we replace the characteristic polynomial with the supercharacteristic function,
\begin{align}
    y + \frac{\mathfrak{q}}{y} = \sdet(x - \Phi)
    \, .
\end{align}
Hence, we have the Seiberg--Witten curve for $\mathrm{U}(n|m)$ pure SYM theory as follows.
\begin{itembox}{Seiberg--Witten curve for $\mathrm{U}(n|m)$ theory}
The Seiberg--Witten curve for $\mathrm{U}(n|m)$ pure SYM theory is given by the supercharacteristic function of the complex adjoint scalar $\Phi \in \operatorname{Lie} \mathrm{U}(n|m)$ in the vector multiplet,
\begin{align}
    \Sigma = \{ (x,y) \in \mathbb{C} \times \mathbb{C}^\times \mid y + \frac{\mathfrak{q}}{y} = \sdet(x - \Phi) \}
    \, .
\end{align}
\end{itembox}
In terms of the eigenvalues of $\Phi$, $\{ a_\alpha^\sigma \}_{\sigma=0,1,\alpha \in [n_\sigma]}$ for $G = \mathrm{U}(n_0|n_1)$, we may write it as follows,
\begin{align}
    y + \frac{\mathfrak{q}}{y} = \frac{T_0(x)}{T_1(x)}
    \, , \qquad
    T_\sigma(x) = \prod_{\alpha \in [n_\sigma]} (x - a_\alpha^\sigma)
    \, ,
\end{align}
which can be further rewritten as
\begin{align}
    T_1(x) y + \mathfrak{q} \frac{T_1(x)}{y} = T_0(x)
    \, .
\end{align}
This algebraic equation characterizes the Seiberg--Witten curve of U$(n_0)$ gauge theory with $2 n_1$ flavors, which is consistent with the gauging trick.
This algebraic equation shows the Riemann surface of genus $g = n_0 - 1$ with $2 n_1$ punctures.
Imposing the special unitary condition, $\sum_{\alpha \in [n_1]} a_\alpha^1 = 0$, the last two cycles are not independent, hence we have $(n_0 - 1)$ $A$ and $B$ cycles and $2(n_1 - 1)$ cycles associated with the punctures.
The Euler characteristics of this Riemann surface is given by $\chi = 2 - 2 (n_0 - 1) + 2 (n_1 - 1) = 2 - 2 (n_0 - n_1)$, where we have the superdimension $\operatorname{sdim} \mathbb{C}^{n_0|n_1} = n_0 - n_1$.
This relation is also obtained from $\widehat{A}_1$ quiver point of view~\cite{Dijkgraaf:2016lym}.
We can also incorporate the superflavor factor by imposing the supercharacterisitic function with respect to $G_F = \mathrm{U}(n_0^F|n_1^F)$.
We will derive the Seiberg--Witten geometry for supergroup gauge theory from the microscopic instanton counting in Sec.~\ref{sec:instanton}.

\section{Supergroup instanton counting}\label{sec:instanton}

As in the case of the ordinary non-supergroup gauge theory, the instanton plays an important role in the study of non-perturbative aspects of supergroup gauge theory.
In this section, we explore several aspects of instantons and their applications to study dynamics of supergroup gauge theory.

\subsection{Instanton moduli space}

The instanton is a solution of the ASDYM equation~\eqref{eq:ASDYMeq}.
Writing the SD/ASD part of the curvature two-form as
\begin{align}
    F_\pm = \frac{F \pm \star F}{2}
    \, ,
\end{align}
such that $F = F_+ + F_-$, the ASDYM equation is rewritten as 
\begin{align}
    F_+ = 0
    \, ,
\end{align}
which would be thought of as a ``half'' flat connection, and hence the corresponding moduli space is expected to have a rich mathematical structure.

Let $M = \mathbb{C}^2$ and $G = \mathrm{U}(n)$.
In order that the YM action becomes finite, we require that the curvature behaves as $F \to 0$ as $x \to \infty$.
This implies that the connection approaches to the pure gauge at the boundary, $A \xrightarrow{x \to \infty} g d g^{-1}$, where we obtain a map, $g$: $\partial \mathbb{C}^2 = \mathbb{S}^3 \to G$.
Knowing that $\pi_3(G) = \mathbb{Z}$ for any compact simple Lie groups, we can apply a classification on the gauge field with respect to the topological charge 
\begin{align}
    k = \frac{1}{8 \pi^2} \int_M \tr F \wedge F = c_2[M] \in \mathbb{Z}
    \, ,
\end{align}
which is called the instanton number, given by the integral of the second Chern class over $M$.%
\footnote{%
Precisely speaking, the four-manifold $M$ should be compact to have $c_2[M] \in \mathbb{Z}$.
In this case, $\mathbb{C}^2 \simeq \mathbb{R}^4$ can be mapped to a four-sphere $\mathbb{S}^4$ through the stereographic projection.
This map does not violate the ASD property since it is a conformal map.
Under this map, the boundary of $\mathbb{C}^2$ is mapped to the north pole (marked point) on $\mathbb{S}^4$, where we fix the gauge as $A = gdg^{-1}$ (called the framing).
}
Here, we focus on the ASD instanton solution having a positive charge $k \ge 0$.
The SD anti-instanton solution is similarly obtained by changing the orientation of the manifold.
Hence, the moduli space of instantons has a decomposition
\begin{align}
    \mathfrak{M}_G = \bigsqcup_{k=0}^\infty \mathfrak{M}_{G,k}
    \, ,
\end{align}
where each topological sector of the moduli space is given by
\begin{align}
    \mathfrak{M}_{G,k} = \{ A \in \Omega^1(M,\mathfrak{g}) \mid F_+ = 0, c_2[M] = k \} / \mathcal{G}
    \, .
    \label{eq:mod_sp_def_naive}
\end{align}
We remark that the $\mathcal{G}$-quotient means the quotient with respect to the equivalent class under $G$-gauge transformation.

\subsection{ADHM construction of instanton}\label{sec:ADHM_const}

Although the definition of the moduli space~\eqref{eq:mod_sp_def_naive} is conceptually reasonable, it is still difficult to analyze in practice.
In order to discuss such an instanton moduli space, we can apply the ADHM construction, which is a systematic approach to obtain instanton solutions~\cite{Atiyah:1978ri}.
See, e.g.,~\cite{Dorey:2002ik} for an extended review on this subject.
For the $k$-instanton solution in $\mathrm{U}(n)$-YM theory, we define two vector spaces,
\begin{align}
    N = \mathbb{C}^n
    \, , \qquad
    K = \mathbb{C}^k
    \, .
\end{align}
The fundamental degrees of freedom in this construction are the linear maps associated with these vector spaces,
\begin{align}
    X & = \operatorname{Hom}(K,K) \oplus \operatorname{Hom}(K,K) \oplus \operatorname{Hom}(N,K) \oplus \operatorname{Hom}(K,N)
    \nonumber \\
    & \ni (B_1, B_2, I, J)
    \, ,
\end{align}
where we impose the so-called ADHM equations,
\begin{subequations}\label{eq:ADHMeqs}
    \begin{align}
        0 & = \mu_\mathbb{R} := [B_1, B_1^\dag] + [B_2, B_2^\dag] + II^\dag - J^\dag J
        \, , \\
        0 & = \mu_\mathbb{C} := [B_1, B_2] + IJ
        \, .
    \end{align}
\end{subequations}
We remark that these equations are invariant under the $\mathrm{U}(k)$ action,
\begin{align}
    g \cdot (B_1, B_2, I, J) = (g B_1 g^{-1}, g B_2 g^{-1}, g I, J g^{-1})
    \, , \qquad
    g \in \mathrm{U}(k)
    \, .
\end{align}
We call $\mu_{\mathbb{R},\mathbb{C}}$ the \emph{moment maps} that take a value in the dual of Lie algebra, $\mu_{\mathbb{R},\mathbb{C}} :$ $X \to \mathfrak{u}_k^\vee \otimes \mathbb{R}^3$.
Based on these degrees of freedom, we have an alternative description of the instanton moduli space,
\begin{align}
    \mathfrak{M}_{G,k} = \{ (B_1,B_2,I,J) \in X \mid \mu_\mathbb{R} = 0, \mu_\mathbb{C} = 0 \} /\!\!/\!\!/ \mathrm{U}(k)
    \, .
\end{align}
The triple slash means the hyper-Kähler quotient with respect to three moment maps, $(\mu_\mathbb{R}, \operatorname{Re} \mu_\mathbb{C}, \operatorname{Im} \mu_\mathbb{C})$.

We show that the ASD connection can be constructed from these ADHM variables.
We define the zero-dimensional Dirac operator,
\begin{align}
    D^\dag = 
    \begin{pmatrix}
        B_1 - z_1 & B_2 - z_2 & I \\
        - B_2^\dag + \bar{z}_2 & B_1^\dag - \bar{z}_1 & - J^\dag
    \end{pmatrix}
    \ : \ K \otimes \mathbb{C}^2 \oplus N \to K \otimes \mathbb{C}^2
    \, .
\end{align}
We remark that the space-time dependence in the Dirac operator is associated with the quaternion structure
\begin{align}
    \begin{pmatrix}
        z_1 & z_2 \\ - \bar{z}_2 & \bar{z}_1
    \end{pmatrix}
    = 
    \begin{pmatrix}
        x_4 + \ii x_3 & x_2 + \ii x_1 \\ - x_2 + \ii x_1 & x_4 - \ii x_3
    \end{pmatrix}
    = x \cdot \sigma
    \, , \quad
\end{align}
where we define the quarternion basis,
\begin{align}
    \sigma = (\ii \vec{\sigma}, \id_{\mathbb{C}^2})
    \, , \qquad
    \bar{\sigma} = (-\ii \vec{\sigma}, \id_{\mathbb{C}^2})    
    \, .
\end{align}
Their product is given by
\begin{align}
    \sigma_\mu \bar{\sigma}_\nu = \delta_{\mu\nu} + \ii \eta_{\mu\nu}^{(+)}
    \, , \qquad
    \bar\sigma_\mu {\sigma}_\nu = \delta_{\mu\nu} + \ii \eta_{\mu\nu}^{(-)}
    \, ,
\end{align}
where $\eta_{\mu\nu}^{(\pm)}$ is an (A)SD tensor, called the 't Hooft symbol, $\star \eta_{\mu\nu}^{(\pm)} = \pm \eta_{\mu\nu}^{(\pm)}$.

The ADHM equations~\eqref{eq:ADHMeqs} are equivalent to the condition such that the Dirac operator squared is diagonal with respect to $\mathbb{C}^2$,
\begin{align}
    D^\dag D = \Delta \otimes \id_{\mathbb{C}^2}
    \, ,
\end{align}
where $\Delta$ : $K \to K$ is explicitly given by
\begin{align}
    \Delta & = 
    (B_1 - z_1)(B_1^\dag - \bar{z}_1) + (B_2 - z_2)(B_2^\dag - \bar{z}_2) + II^\dag 
    \nonumber \\
    & = 
    (B_1^\dag - \bar{z}_1)(B_1 - z_1) + (B_2^\dag - \bar{z}_2)(B_2 - z_2) + J^\dag J
    \, .
\end{align}
Therefore, the asymptotic behavior is given by $\Delta \xrightarrow{z \to \infty} |z|^2 \id_K$.
Since the Dirac operator is a rectangular matrix of a rank $2k$, we consider the complementary space, which is given by the set of normalized zero modes (the Dirac operator kernel),
\begin{align}
    \operatorname{Ker} D^\dag = \{ \Psi : N \to K \otimes \mathbb{C}^2 \oplus N \mid D^\dag \Psi = 0, \Psi^\dag \Psi = \id_N \}
    \, .
\end{align}
We define the projector from $K \otimes \mathbb{C}^2 \oplus N$ to $N$ using the zero modes,
\begin{align}
    P := \Psi \Psi^\dag = \id_{K \otimes \mathbb{C}^2 \oplus N} - D (\Delta^{-1} \otimes \id_{\mathbb{C}^2}) D^\dag
    \, ,
\end{align}
where we remark $PD = \Psi \Psi^\dag D = 0$.
Then, we obtain a connection from the zero modes,
\begin{align}
    A = \Psi^\dag d \Psi
    \, .
\end{align}
Due to the normalization condition, it turns out to be anti-Hermitian, $A^\dag = - A$.
The curvature two-form constructed from this connection is given as follows,
\begin{align}
    F = dA + A \wedge A = d \Psi^\dag (\id_{K \otimes \mathbb{C}^2 \oplus N} - \Psi \Psi^\dag) d \Psi = \Psi^\dag (d D) (\Delta^{-1} \otimes \id_{\mathbb{C}^2}) (d D^\dag) \Psi 
    \, .
\end{align}
Recalling
\begin{align}
    d D =
    \begin{pmatrix}
        - \id_K \otimes \bar{\sigma} \\ 0
    \end{pmatrix}
    \, , \qquad
    d D^\dag = 
    \begin{pmatrix}
        - \id_K \otimes \sigma  & 0
    \end{pmatrix}
    \, ,
\end{align}
it turns out that the curvature is ASD,
\begin{align}
    F = \Psi^\dag 
    \begin{pmatrix}
        \Delta^{-1} \otimes 2 \ii \eta^{(-)} & 0 \\ 0 & 0
    \end{pmatrix}
    \Psi
    \, ,
\end{align}
where we define the (A)SD two-form $\eta^{(\pm)} = \frac{1}{2} \eta^{(\pm)}_{\mu\nu} dx^\mu dx^\nu$, such that $\star \eta^{(\pm)} = \pm \eta^{(\pm)}$.
In addition, applying Osborn's formula~\cite{Osborn:1978rn}, we see that the instanton number is given as follows,
\begin{align}
    \frac{1}{8\pi^2} \int_M \tr F \wedge F = \frac{1}{16 \pi^2} \int_M \dd[4]{x} \partial^2 \partial^2 \tr_K \log \Delta^{-1} = \tr_K \id_K = k
    \, .
    \label{eq:Osborn_formula}
\end{align}
In this calculation, we use the asymptotic behavior of $\Delta \xrightarrow{x \to \infty} |x|^2 \id_K$ and evaluate the boundary contribution.

\subsubsection{String theory perspective}\label{sec:ADHM_string_perspective}

As discussed in Sec.~\ref{sec:string_theory_perspective}, one can obtain gauge theory from a stack of D-branes.
In particular, $k$-instanton configuration in $\mathrm{U}(n)$-YM theory is realized as $n$ D$p$ \& $k$ D$(p-4)$ brane system.
Let us put $p = 4$.
In this context, the two vector spaces $(N,K)$ are identified with the Chan--Paton spaces of $n$ D4 and $k$ D0 branes, and thus the ADHM variables $(B_1,B_2,I,J)$ are interpreted as degrees of freedom of open strings connecting D0-D0, D4-D0, and D0-D4 branes, respectively.
The ADHM equations are obtained as the BPS equations of this configuration.
We remark that the anti-instanton is realized by $\overline{\text{D}0}$ brane that violates the BPS condition.

\subsubsection{Regularization of moduli space}

We would study the moduli space defined here, however it has been known that it is non-compact and singular that should be reguralized for the latter purpose.

For the sake of compactification of the instanton moduli space, we add the point-like instantons, $\bigsqcup_{\ell=0}^k \mathfrak{M}_{n,k-\ell} \times \operatorname{Sym}^\ell \mathbb{C}^2$, which is called the \emph{Uhlenbeck compactification} (see, e.g.,~\cite{Donaldson:1997}).
This is analogous to the compactification of $\mathbb{R}^4$ to $\mathbb{S}^4$ by adding the infinity.
Another regularization that we need is resolution of singularity.
This can be done by modifying the ADHM equations, $(\mu_\mathbb{R}, \mu_\mathbb{C}) = (\zeta_{\neq 0} \id_K, 0)$, and the corresponding moduli space is given by~\cite{Nakajima:1993jg}
\begin{align}
    \mathfrak{M}_{n,k}^\zeta = \{ (B_1, B_2, I, J) \in X \mid \mu_\mathbb{R} = \zeta_{\neq 0} \id_K, \mu_\mathbb{C} = 0 \}/\!\!/\!\!/\mathrm{U}(k)
    \, .
\end{align}
This modification has been also discussed in the context of the instanton moduli space on non-commutative $\mathbb{C}^2$~\cite{Nekrasov:1998ss}.
Another moduli space is for the rank-$n$ framed torsion free sheaves on $\mathbb{CP}^2$ given by the geometric invariant theory (GIT) quotient (see, e.g.,~\cite{Nakajima:1999}),
\begin{align}
    \widetilde{\mathfrak{M}}_{n,k} = \{ (B_1, B_2, I, J) \in X \mid \mu_\mathbb{C} = 0, \text{(co-)stability} \}/\!\!/\mathrm{GL}(K)
    \, ,
\end{align}
which is known to be isomorphic to the resolved moduli space $\mathfrak{M}_{n,k}^\zeta$.
The (co-)stability condition is given as follows,
\begin{align}
    K = 
    \begin{cases}
        \mathbb{C}[B_1, B_2] I(N) & (\zeta > 0;\ \text{stability}) \\
        \mathbb{C}[B_1, B_2] J(N)^\vee  & (\zeta < 0;\ \text{co-stability}) \\
    \end{cases}
\end{align}
One can obtain the (co-)stability condition from the real part of the ADHM equation, $\mu_\mathbb{R} = \zeta \id_K$, depending on the sign of the deformation parameter $\zeta$.
See, e.g.,~\cite{Kimura:2020jxl} for details.

\subsection{ADHM construction of super instanton}

We turn to the construction of instantons in supergroup gauge theory~\cite{Taniguchi:2009zz,Kimura:2019msw}.
Let $G = \mathrm{U}(n_0|n_1)$.
Recalling that the ADHM moduli space is given by the $\mathrm{U}(k)$-quotient for the $k$-instanton sector, it seems natural to consider the supergroup quotient, i.e., $\mathrm{U}(k_0|k_1)$-quotient in the supergroup setup.
This implies that the topological sector is characterized by $k = (k_0|k_1)$ in this case.
From this point of view, we replace the vector spaces $(N,K)$
 with the supervector spaces,
\begin{align}
    N = N_0 \oplus N_1 = \mathbb{C}^{n_0|n_1} = \mathbb{C}^{n_0} \oplus \mathbb{C}^{n_1}
    \, , \qquad
    K = K_0 \oplus K_1 = \mathbb{C}^{k_0|k_1} = \mathbb{C}^{k_0} \oplus \mathbb{C}^{k_1}
    \, .
    \label{eq:N_K_svec_sp}
\end{align}
Then, the ADHM variables are obtained as supermatrices representing the linear maps for these supervector spaces.

The remaining steps are formally parallel with the standard case discussed in Sec.~\ref{sec:ADHM_const}.
In this case, Osborn's formula~\eqref{eq:Osborn_formula} yields
\begin{align}
    \frac{1}{8\pi^2} \int_M \str F \wedge F = \str_K \id_K = k_0 - k_1
    \, ,
\end{align}
which implies that $k_0$ and $k_1$ count the positive and negative charge instantons. 
This is reasonable from string theory perspective discussed in Sec.~\ref{sec:ADHM_string_perspective}:
The current $(k_0|k_1)$-instanton configuration in $\mathrm{U}(n_0|n_1)$-YM theory would be realized as $k_0$ D0$^+$, $k_1$ D0$^-$, $n_0$ D4$^+$, and $n_1$ D4$^-$ brane system, which relate the supervector spaces~\eqref{eq:N_K_svec_sp} to the Chan--Paton spaces.
In contrast to the brane--anti-brane system, the current situation does not violate the ASD property, which is compatible with the BPS equation.
Meanwhile, the resolved moduli space is given for $(n,k) = (n_0|n_1,k_0|k_1)$ by
\begin{align}
    \mathfrak{M}_{n,k}^\zeta = \{ (B_1, B_2, I, J) \in X \mid \mu_\mathbb{R} = (+\zeta \id_{K_0}) \oplus (-\zeta \id_{K_1}), \mu_\mathbb{C} = 0 \}/\!\!/\!\!/\mathrm{U}(k_0|k_1)
    \, .
\end{align}
We remark that the deformation parameter is assigned with opposite signs for the positive and negative sectors.
Physically speaking, this is because the positive and negative instantons (positive and negative D0 branes) are oppositely charged under the flux yielding the non-commutativity of the space-time manifold via the Seiberg--Witten map~\cite{Seiberg:1999vs}.
Hence, fixing $\zeta > 0$, this moduli space is isomorphic to the following super-GIT quotient,
\begin{align}
    \widetilde{\mathfrak{M}}_{n,k} =  \{ (B_1, B_2, I, J) \in X \mid \mu_\mathbb{C} = 0, \text{stability for $K_0$, co-stability for $K_1$} \}/\!\!/\mathrm{GL}(K)
    \, .
\end{align}
For $\zeta < 0$, the stability and co-stability conditions are exchanged.

\subsection{Equivariant localization}

Based on the description of the instanton moduli space, we apply the equivariant localization formalism to compute the partition function of supergroup gauge theory.
See, e.g., \cite{Pestun:2016zxk} for details of the localization calculus.

In the Euclidean path integral formalism, the partition function of $G$-gauge theory is given as follows,
\begin{align}
    Z = \int \mathfrak{D}\!A \, \ee^{-S[A]}
    \, ,
\end{align}
where we consider the case $M = \mathbb{R}^4$, and the action consists of the YM action and the $\theta$-term.
Writing the one-form connection in the form of $A = A_k + \delta A$, where we denote the $k$-instanton configuration by $A_k$ and the deviation by $\delta A$ (we assume no $k$-dependence), we have the following decomposition of the path integral measure, $\mathfrak{D} A = \sum_{k=0}^\infty \mathfrak{D} A_k \mathfrak{D}(\delta A)$.
In this decomposition, the weight with the action is written by $\ee^{-S[A]} = \mathfrak{q}^k \ee^{-S_\text{dev}[\delta A]}$, where $\mathfrak{q} = \exp (\ii \theta - \frac{8 \pi^2}{g^2}) =: \exp (2 \pi \ii \tau)$, and the deviation part $S_\text{dev}[\delta A]$ starts with $O(\delta A^2)$ since $A_k$ gives a solution to the equation of motion. 
Noticing that the integral over $A_k$ can be identified with the integral over the moduli space $\mathfrak{M}_{G,k}$, we have the following form of the partition function,%
\footnote{%
The anti-instanton partition function shall be obtained by the complex conjugate of the instanton partition function: 
The weight in this case is given by $\bar{\mathfrak{q}}^k = \exp k\left(-\ii \theta - \frac{8 \pi^2}{g^2} \right)$, and the moduli space integral will be $\overline{Z_k}$ since it has an opposite orientation compared with the original one.
}
\begin{align}
    Z = Z_\text{inst} Z_\text{pert}
    \, , \qquad 
    Z_\text{inst} = \sum_{k = 0}^\infty \mathfrak{q}^k Z_k 
    \, ,
\end{align}
where $Z_\text{pert}$ is the deviation (perturbative) part, while $Z_k$ is called the \emph{instanton partition function},
\begin{align}
    Z_k %= \int \mathfrak{D} A_k 
    = \int_{\mathfrak{M}_{G,k}} 1 = \operatorname{vol} (\mathfrak{M}_{G,k})
    \, .
    \label{eq:Zk_as_volume}
\end{align}
As mentioned in Sec.~\ref{sec:ADHM_const}, we should replace the moduli space $\mathfrak{M}_{G,k}$ with the regularized version.
Moreover, noticing that there exist several group actions on the moduli space, the integral is performed as the equivariant integral, and thus $\operatorname{vol} (\mathfrak{M}_{G,k})$ is understood as the equivariant volume, which can be computed based on the equivariant localization formula,
\begin{align}
    \int_{\mathfrak{M}_{G,k}} \alpha = \sum_{\lambda \in \mathfrak{M}_{G,k}^\textsf{T}} \frac{\iota^* \alpha}{e(T_\lambda \mathfrak{M}_{G,k})}
    \, ,
    \label{eq:localization_formula}
\end{align}
where $\alpha$ is an equivariant cohomology class, $\iota^*$ is a pull-back of the inclusion map, $\iota$ : $o \times \mathfrak{M}_{G,k} \hookrightarrow \mathbb{C}^2 \times \mathfrak{M}_{G,k}$, $e$ is the equivariant Euler class, $\mathfrak{M}_{G,k}^\mathsf{T}$ is a set of the equivariant $\mathsf{T}$-fixed points, and $T_\lambda \mathfrak{M}_{G,k}$ is the tangent bundle to the moduli space at the fixed point $\lambda$.
The instanton partition function corresponds to the trivial insertion $\alpha = 1$.
In the presence of the fundamental matter, we add the equivariant Euler class of the matter bundle $\mathsf{M}$, whose fiber is the space of virtual zero modes of the Dirac operator in the instanton background.
For $\mathcal{N}=2^*$ theory involving the adjoint matter, we insert the shifted tangent bundle, $\mathsf{M}_\text{adj} \otimes T\mathfrak{M}_{G,k}$, where $\mathsf{M}_\text{adj}$ is a line bundle whose Chern root is identified with the adjoint mass parameter.%
\footnote{%
In the five-dimensional (equivariant K-theory) convention discussed below, this case corresponds to Hirzebruch's $\chi_y$-genus of the instanton moduli space with identifying $y = \ee^m$.
}

\subsubsection{Index functor}

The formalism discussed above is generalized to five and six dimensional theories by replacing the equivariant Euler class with the corresponding multiplicative index functor:
For the vector bundle $\mathbf{X}$ with the character, $\operatorname{ch} \mathbf{X} = \sum_{i=1}^{\operatorname{rk} \mathbf{X}} n_i \ee^{x_i}$ with multiplicity $n_i \in \mathbb{Z}$, we define
\begin{align}
    \mathbb{I}[\mathbf{X}] = \prod_{i \in [\operatorname{rk} \mathbf{X}]} [x_i]^{n_i}
    \, ,
\end{align}
where we apply the notation,%
\footnote{%
Not to be confused with the notation~\eqref{eq:set_N}.
}
\begin{align}
    [x] = 
    \begin{cases}
        x & (4d) \\
        (1 - \ee^{-x}) & (5d) \\
        \theta(e^{-x};p) & (6d)
    \end{cases}
    \label{eq:index_notation}
\end{align}
The instanton partition function is then given by summation over the topological sectors,
\begin{align}
    Z_k = \sum_{\lambda \in \mathfrak{M}_{G,k}^\mathsf{T}} Z_\lambda
    \, , \qquad
    Z_\lambda = \mathbb{I}[- T_\lambda \mathfrak{M}_{G,k}]
    \, .
\end{align}
The five and six dimensional conventions correspond to the equivariant K-theory and the equivariant elliptic cohomology, respectively.

The five-dimensional index is given by the character of alternating sum of anti-symmetrizations of the dual bundle,%
\footnote{%
For the vector bundle $\mathbf{X}$ with the character, $\operatorname{ch} \mathbf{X} = \sum_{i=1}^{\operatorname{rk} \mathbf{X}} n_i \ee^{x_i}$, the dual bundle is defined to have the character, $\operatorname{ch} \mathbf{X}^\vee = \sum_{i=1}^{\operatorname{rk} \mathbf{X}} n_i \ee^{-x_i}$.
}
\begin{align}
    \mathbb{I}[\mathbf{X}] \ \stackrel{5d}{=} \  \operatorname{ch} \wedge \mathbf{X}^\vee = \sum_{i=0}^{\infty} (-1)^i \operatorname{ch} \wedge^i \mathbf{X}^\vee
    \, .
\end{align}
In this case, we have the K-theoretic instanton partition function as an equivariant Euler characteristic over the instanton moduli space~\cite{Nakajima:2003pg,Nakajima:2005fg},
\begin{align}
    Z_k = \sum_{i=0}^\infty (-1)^i \operatorname{ch} H^i(\mathfrak{M}_{G,k},\mathcal{O})
    \, ,
    \label{eq:K-theoretic_partition_function}
\end{align}
where we denote the structure sheaf of the moduli space by $\mathcal{O}$.
From the point of view of the localization formula~\eqref{eq:localization_formula}, we may rewrite the partition function as follows,%
\footnote{%
The $\widehat{A}$ genus is typically used for the five-dimensional convention instead of the Todd genus.
See, e.g.,~\cite{Nekrasov:2002qd}.
}
\begin{align}
    Z_k = \int_{\mathfrak{M}_{G,k}} \operatorname{td}(T\mathfrak{M}_{G,k})
    \, ,
\end{align}
where we define the Todd class,
\begin{align}
    \operatorname{td}(\operatorname{\mathbf{X}}) = \prod_{i=1}^{\operatorname{rk} \mathbf{X} }\frac{x_i}{1 - \ee^{- x_i}} = e(\mathbf{X}) \mathbb{I}[\mathbf{X}]^{-1} = e(\mathbf{X}) \mathbb{I}[-\mathbf{X}]
    \, .
\end{align}
Recalling the Hirzebruch--Riemann--Roch formula for the holomorphic Euler characteristic of a holomorphic vector bundle $E$ on $\mathfrak{M}$,%
\footnote{%
We may write the K-theoretic partition function~\eqref{eq:K-theoretic_partition_function} as an alternating sum of higher direct images and the pushforward of the projection map from the moduli space to the point (derived pushforward), which can be understood as the Grothendieck--Hirzebruch--Riemann--Roch formula.
}
\begin{align}
    \chi(\mathfrak{M},E) = \int_\mathfrak{M} \operatorname{ch} E \, \operatorname{td}(T\mathfrak{M})
    \, ,
\end{align}
the K-theoretic partition function is understood as an integral of ``1'' analogously to the cohomological version~\eqref{eq:Zk_as_volume}.
We remark that the equivariant Euler class contribution in the denominator of the localization formula~\eqref{eq:localization_formula} is canceled with that in the numerator of the Todd class.
See also~\cite{Pestun:2016qko} for details.

For the six-dimensional case, the top form insertion is necessary to be compatible with the modular property.
Hence, for $\mathrm{U}(n)$ gauge theory, we need to incorporate $2n$ fundamental matters, or a single adjoint matter, which correspond to the superconformal matter content in the four-dimensional convention.
We remark that these index formulations are understood as the Witten index of the supersymmetric ADHM quiver quantum mechanics on the circle $\mathbb{S}^1$ and the elliptic genus of the ADHM quiver gauge theory on the elliptic curve $\mathcal{E}$ with the nome $p \in \mathbb{C}^\times$.

\subsubsection{Equivariant fixed points}

We turn to more details on the instanton moduli space.
We denote the fiber of the cotangent bundle at the marked point $o \in \mathbb{C}^2$ by 
\begin{align}
    Q = T_o^\vee \mathbb{C}^2 = Q_1 \oplus Q_2
\end{align}
with the character $\operatorname{ch} Q_i = q_i = \ee^{\epsilon_i}$ ($i=1,2$).
We use the notation
\begin{subequations}
    \begin{align}
        Q_{12} = Q_1 \otimes Q_2 = \det Q
        \, , \qquad &
        q_{12} = q_1 q_2 
        \quad (\epsilon_{12} = \epsilon_1 + \epsilon_2)
        \, , \\
        P_i = \wedge Q_i = 1 - Q_i \quad (i = 1,2)
        \, , \qquad &
        P_{12} = P_1 P_2
        \, .
    \end{align}
\end{subequations}
Then, the equivariant group actions on the moduli space are given by
\begin{subequations}
    \begin{align}
    g \cdot (B_1, B_2, I, J) & = (g B_1 g^{-1}, g B_2 g^{-1}, g I, J g^{-1})
    \, , && 
    g \in \mathrm{GL}(K)
    \, , \\
    h \cdot (B_1, B_2, I, J) & = (B_1, B_2, I h^{-1}, h J )
    \, , &&
    h \in \mathrm{GL}(N)
    \, , \\
    (q_1,q_2) \cdot (B_1, B_2, I, J) & = (q_1^{-1} B_1, q_2^{-1} B_2, I, q_{12}^{-1} J ) 
    \, , &&
    (q_1,q_2) \in \mathsf{T}_Q \subset \mathrm{GL}(Q)
    \, .
    \end{align}
\end{subequations}
Parametrizing these group elements as $g = \ee^\phi$, $h = \ee^a$, $q_i = \ee^{\epsilon_i}$ ($i=1,2$), namely $\phi \in \operatorname{Lie} \mathrm{GL}(K)$, $h \in \operatorname{Lie} \mathrm{GL}(N)$, $(\epsilon_1,\epsilon_2) \in \operatorname{Lie} \mathsf{T}_Q$, the fixed point equations are given as follows,
\begin{subequations}\label{eq:fixed_pt_eq}
    \begin{align}
        [\phi, B_i] - \epsilon_i B_i & = 0 
        \, ,
        \qquad (i = 1, 2)
        \label{eq:fixed_point_B}
        \\
        \phi I - I a & = 0
        \, , \\
        - J \phi + (a - \epsilon_{12}) J & = 0
        \, .
    \end{align}
\end{subequations}
We apply the basis diagonalizing $a = \bigoplus_{\alpha \in [n]} a_\alpha$, which corresponds to $N = \bigoplus_{\alpha \in [n]} N_\alpha$, and $(a_\alpha)_{\alpha \in [n]} \in \operatorname{Lie} \mathsf{T}_N$.
We have the (left/right) eigenvalue equations, $\phi I_\alpha = a_\alpha I_\alpha$ and $J_\alpha \phi = J_\alpha (a_\alpha - \epsilon_{12})$.
Moreover, using~\eqref{eq:fixed_point_B}, we obtain 
\begin{subequations}
    \begin{align}
    \phi B_1^{i-1} B_2^{j-1} I_\alpha
    & = (a_\alpha + (i-1)\epsilon_1 + (j-1) \epsilon_2 ) B_1^{i-1} B_2^{j-1} I_\alpha
    \, , \\
    J_\alpha B_1^{i-1} B_2^{j-1} \phi & = J_\alpha B_1^{i-1} B_2^{j-1} (a_\alpha - i \epsilon_1 - j \epsilon_2 )
    \, .
\end{align}
\end{subequations}
Applying the stability condition, $K = \mathbb{C}[B_1, B_2] I(N)$, and recalling $\dim K = k$, there exist only $k$ eigenvectors in the form of $B_1^{i-1} B_2^{j-1} I_\alpha$, which are parametrized by $n$-tuple partition, $\lambda = (\lambda_\alpha)_{\alpha \in [n]} \in \mathfrak{M}_{n,k}^\mathsf{T}$ with $|\lambda| = \sum_{\alpha \in [n]} |\lambda_\alpha| = k$, and hence $B_1^{i-1} B_2^{j-1} I_\alpha = 0$ for $(i, j) \not\in \lambda_\alpha$ and also $J_\alpha = 0$, $\alpha \in [n]$.
Namely, we have $K = \operatorname{Span}\{ B_1^{i-1} B_2^{j-1} I_\alpha \}_{\alpha \in [n], (i,j) \in \lambda_\alpha}$.
Starting with the co-stability condition, we have the left eigenvectors in the form of $J_\alpha B_1^{i-1} B_2^{j-1}$ with $I_\alpha = 0$.
We remark that together with the ADHM equation, $\mu_\mathbb{C} = [B_1, B_2] + IJ = 0$, we have $[B_1, B_2] = 0$ at the fixed point.
Therefore, we obtain the characters of the vector spaces $(N,K)$ as follows,
\begin{align}
    \operatorname{ch} K\Big|_\lambda = \sum_{\alpha \in [n]} \operatorname{ch} K_\alpha\Big|_{\lambda_\alpha}
     % = \sum_{I \in [k]} \ee^{\phi_I} 
    \, , \qquad
    \operatorname{ch} N = \sum_{\alpha \in [n]} e^{a_\alpha}
    \, ,
\end{align}
where we have
\begin{align}
    \operatorname{ch} K_\alpha\Big|_{\lambda_\alpha} = \sum_{(i,j) \in \lambda_\alpha} 
    \begin{cases}
        \ee^{a_\alpha} q_1^{i-1} q_2^{j-1} & (\text{with stability}) \\
        \ee^{a_\alpha} q_1^{-i} q_2^{-j} & (\text{with co-stability}) \\
    \end{cases}
\end{align}
This structure has a connection with the ideal $\mathsf{I}_\lambda$ generated by all monomials outside the partition, $\{ z_1^{i-1} z_2^{j-1} \}_{(i,j) \not\in \lambda}$ as follows,
\begin{align}
    K_\alpha\Big|_{\lambda_\alpha} = N_\alpha \otimes
    \begin{cases}
    \mathsf{I}_\emptyset / \mathsf{I}_{\lambda_\alpha}
    & (\text{stability}) \\
    Q_{12}^\vee (\mathsf{I}_\emptyset / \mathsf{I}_{\lambda_\alpha})^\vee
    & (\text{co-stability}) 
    \end{cases}
    \, .
    \label{eq:K_sp_ideal_rep}
\end{align}
We remark that $\mathsf{I}_\emptyset = \mathbb{C}[z_1,z_2]$.

\subsubsection{Tangent bundle}\label{sec:tangent_bundle}

We next describe the tangent bundle to the moduli space.
From the ADHM construction, we have the following chain complex,
\begin{equation}
    \begin{tikzcd}
    % 0 \arrow[r] & 
    & \operatorname{Hom}(K,K) \arrow[r,"d_1"] & \substack{\displaystyle \operatorname{Hom}(K,K) \otimes Q^\vee \\[.5em] \displaystyle \oplus \\[.5em] \displaystyle  \operatorname{Hom}(N,K) \\[.5em] \displaystyle \oplus \\[.5em] \displaystyle \operatorname{Hom}(K,N) \otimes Q_{12}^\vee }  \arrow[r,"d_2"] & \operatorname{Hom}(K,K) \otimes Q_{12}^\vee & \\ %\arrow[r] & 0 \\
    ( & % 0 \arrow[r] & 
    C_0 \arrow[r] & C_1 \arrow[r]& C_2 % \arrow[r] & 0 
    & )
    \end{tikzcd}
\end{equation}
where $C_{0,1,2}$ correspond to GL$(K)$ action, ADHM variables, and the moment map $\mu_\mathbb{C}$, and we define
\begin{align}
    d_1(\xi) = 
    \begin{pmatrix}
        [\xi, B_1] \\ [\xi, B_2] \\ \xi I \\ - J \xi
    \end{pmatrix}
    \, , \qquad
    d_2 
    \begin{pmatrix}
        b_1 \\ b_2 \\ i \\ j
    \end{pmatrix}
    = [B_1, b_2] + [b_1, B_2] + Ij + iJ
    \, .
\end{align}
These maps are the infinitesimal GL$(K)$ action parametrized by $\xi \in \operatorname{Lie} \mathrm{GL}(K)$ and the differential of the moment map $\mu_\mathbb{C}$.
We can check that $d_2 \circ d_1(\xi) = [\xi, \mu_\mathbb{C}] = 0$ due to the ADHM equation, $\mu_\mathbb{C} = 0$.
Then, the tangent bundle is identified with the middle cohomology with respect to this sequence,
\begin{align}
    T\mathfrak{M}_{n,k} = \operatorname{Ker} d_2 / \operatorname{Im} d_1 = C_1 - C_0 - C_2 = N^\vee K + Q_{12}^\vee K^\vee N - P_{12}^\vee K^\vee K
    \, .
\end{align}

There exists another construction of the tangent bundle based on $\mathsf{Y}_o$ the observable sheaf on $\mathfrak{M}_{n,k} \times o$ obtained from $\mathsf{Y}_{\mathbb{C}^2}$ the universal sheaf on $\mathfrak{M}_{n,k} \times \mathbb{C}^2$ via the localization,
\begin{align}
    \mathsf{Y} \equiv \mathsf{Y}_o = N - P_{12} K
    \, , \qquad
    \mathsf{Y}_{\mathbb{C}^2} = \frac{\mathsf{Y}}{P_{12}}
    \, ,
\end{align}
where we use the notation that $(N,K)$ also means a bundle whose fiber is given by the vector space $N$ and $K$, respectively.
Several equivalent expressions are available for the character of the observable sheaf at the fixed point $\lambda \in \mathfrak{M}^\mathsf{T}_{n,k}$.
The first expression is given by
\begin{align}
    %\operatorname{ch} \mathsf{Y}_\lambda = 
    \operatorname{ch} \mathsf{Y}\Big|_\lambda = \operatorname{ch} (P_1 \mathsf{X})
\end{align}
where we define the partial reduction of the universal sheaf $\mathsf{X} = \mathsf{Y}/P_1 = \mathsf{Y}_{\mathbb{C}^2} P_2$, and the character
\begin{align}
    \operatorname{ch} \mathsf{X} % = \sum_{\alpha \in [n]} \sum_{k \in \mathbb{N}} \ee^{a_\alpha} q_1^{k-1} q_2^{\lambda_{\alpha,k}}
    = \sum_{x \in \mathcal{X}_\lambda} x
    \, , \qquad
    \mathcal{X}_\lambda = \{ x_{\alpha,i} = \ee^{a_\alpha} q_1^{i-1} q_2^{\lambda_{\alpha,i}} \}_{\alpha \in [n], i \in \mathbb{N}}
    \, .
    \label{eq:obs_sheaf_def}
\end{align}
In order to obtain this expression, we need the condition $|q_1| < 1$ to have $\sum_{i = 1}^\infty q_1^{i-1} = 1/(1 - q_1)$.
If $|q_2| < 1$, we may apply another expression based on the transposed partition of $\lambda$ denoted by $\check{\lambda}$,
\begin{align}
    \operatorname{ch} \mathsf{Y}\Big|_\lambda = \operatorname{ch} (P_2 \check{\mathsf{X}})
    \, , \qquad
    \operatorname{ch} \check{\mathsf{X}} % = \sum_{\alpha \in [n]} \sum_{k \in \mathbb{N}} \ee^{a_\alpha} q_1^{k-1} q_2^{\lambda_{\alpha,k}}
    = \sum_{x \in \check{\mathcal{X}}_\lambda} x
    \, , \qquad
    \check{\mathcal{X}}_\lambda = \{ x_{\alpha,i} = \ee^{a_\alpha} q_1^{\check{\lambda}_{\alpha,j}} q_2^{j-1} \}_{\alpha \in [n], j \in \mathbb{N}}
    \, .
\end{align}
Applying the co-stability condition, we instead obtain 
\begin{align}
    \operatorname{ch} \mathsf{Y}\Big|_\lambda =
    \begin{cases}
    \displaystyle
    \operatorname{ch} (P_1^\vee \mathsf{X})
    \, , \quad
    \operatorname{ch} \mathsf{X} % = \sum_{\alpha \in [n]} \sum_{k \in \mathbb{N}} \ee^{a_\alpha} q_1^{k-1} q_2^{\lambda_{\alpha,k}}
    = \sum_{x \in \mathcal{X}_\lambda} x
    \, , \quad
    \mathcal{X}_\lambda = \{ x_{\alpha,i} = \ee^{a_\alpha} q_1^{-i+1} q_2^{-\lambda_{\alpha,i}} \}_{\alpha \in [n], i \in \mathbb{N}} 
    & (|q_1|>1) \\
    \displaystyle
    \operatorname{ch} (P_2^\vee \check{\mathsf{X}})
    \, , \quad
    \operatorname{ch} \check{\mathsf{X}} % = \sum_{\alpha \in [n]} \sum_{k \in \mathbb{N}} \ee^{a_\alpha} q_1^{k-1} q_2^{\lambda_{\alpha,k}}
    = \sum_{x \in \check{\mathcal{X}}_\lambda} x
    \, , \quad
    \check{\mathcal{X}}_\lambda = \{ x_{\alpha,i} = \ee^{a_\alpha} q_1^{-\check{\lambda}_{\alpha,j}} q_2^{-j+1} \}_{\alpha \in [n], j \in \mathbb{N}}
    & (|q_2|>1)
    \end{cases}
    \, .
\end{align}

Another expression is given as a finite sum,
\begin{align}
    \operatorname{ch} \mathsf{Y}\Big|_\lambda =
    \begin{cases}
    \displaystyle
    \sum_{x \in \mathcal{X}_{\partial_+ \lambda}} x - \sum_{x \in \mathcal{X}_{\partial_- \lambda}} x q_{12} 
    & (\text{stability})
    \\ \displaystyle
    \sum_{x \in \mathcal{X}_{\partial_+ \lambda}} x q_{12} - \sum_{x \in \mathcal{X}_{\partial_- \lambda}} x 
    & (\text{co-stability})
    \end{cases}
    \label{eq:obs_sheaf_comb}
\end{align}
where we define the addable and removable boundary of the partition,
\begin{align}
    \partial_+ \lambda = \{ (i,\lambda_{\alpha,i}+1) \mid \lambda_{\alpha,i-1} > \lambda_{\alpha,i} \}_{\alpha \in [n], i \in \mathbb{N}}
    \, , \quad
    \partial_- \lambda = \{ (i,\lambda_{\alpha,i}) \mid \lambda_{\alpha,i+1} < \lambda_{\alpha,i} \}_{\alpha \in [n], i \in \mathbb{N}}
\end{align}
where we put $\lambda_{\alpha,0} = \infty$, and we define
\begin{subequations}
\begin{align}
    \mathcal{X}_{\partial_+ \lambda} & =
    \begin{cases}
    \displaystyle
    \{ \ee^{a_\alpha} q_1^{i-1} q_2^{j-1} \mid (i,j) \in \partial_+ \lambda \}
    & (\text{stability})
    \\ \displaystyle
    \{ \ee^{a_\alpha} q_1^{-i} q_2^{-j} \mid (i,j) \in \partial_+ \lambda \}
    & (\text{co-stability})
    \end{cases}
    \, , \\ 
    \mathcal{X}_{\partial_- \lambda} & = 
    \begin{cases}
    \displaystyle
    \{ \ee^{a_\alpha} q_1^{i-1} q_2^{j-1} \mid (i,j) \in \partial_- \lambda \}
    & (\text{stability})
    \\ \displaystyle
    \{ \ee^{a_\alpha} q_1^{-i} q_2^{-j} \mid (i,j) \in \partial_- \lambda \}
    & (\text{co-stability})
    \end{cases}
    \, .
\end{align}
\end{subequations}

The tangent bundle is then obtained from the observable sheaf as follows,
\begin{align}
    \mathsf{V} := \frac{\mathsf{Y}^\vee \mathsf{Y}}{P_{12}} = - T\mathfrak{M}_{n,k} + \frac{N^\vee N}{P_{12}}
    \, .
\end{align}
The last term $N^\vee N/P_{12}$ is understood as the perturbative contribution.
The combination of $\mathsf{Y}^\vee \mathsf{Y}$ corresponds to the adjoint representation of GL group. 
In order to construct the SO/Sp gauge theory, we consider the anti-symmetric and symmetric tensor product, which yield the adjoint representation of SO/Sp group~\cite{Marino:2004cn,Nekrasov:2004vw,Shadchin:2005mx},
\begin{align}
    \mathsf{V}_\text{SO} = \frac{1}{2} \frac{\mathsf{Y}^2 - \mathsf{Y}^{[2]}}{P_{12}}
    \, , \qquad
    \mathsf{V}_\text{Sp} = \frac{1}{2} \frac{\mathsf{Y}^2 + \mathsf{Y}^{[2]}}{P_{12}}
    \, ,
    \label{eq:O_SP_adj_rep}
\end{align}
where we denote the degree-$p$ Adams operation to $X$ by $X^{[p]}$.
Hence, the tangent bundle at the fixed point is given by
\begin{align}
    \mathsf{V}_\lambda = - T_\lambda\mathfrak{M}_{n,k} + \frac{N^\vee N}{P_{12}} = \frac{\mathsf{Y}^\vee \mathsf{Y}}{P_{12}}\Bigg|_{\lambda} = \frac{P_1^\vee}{P_2} \mathsf{X}^\vee \mathsf{X} 
    \, .
\end{align}
Denoting the total fixed point configuration space by $\mathfrak{M}^\mathsf{T} = \bigsqcup_{k=0}^\infty \mathfrak{M}_{n,k}^\mathsf{T}$, the full partition function is given as follows.
\begin{itembox}{Equivariant index formula}
The full partition function of pure SYM gauge theory is given by summation over the fixed point contributions,
\begin{align}
    Z = \sum_{\lambda \in \mathfrak{M}^\mathsf{T}} \mathfrak{q}^{|\lambda|} \, \mathbb{I}[ \mathsf{V}_\lambda ]
    \ \stackrel{(5d)}{=} \
    \sum_{\lambda \in \mathfrak{M}^\mathsf{T}} \mathfrak{q}^{|\lambda|} \prod_{(x,x') \in \mathcal{X}_\lambda \times \mathcal{X}_\lambda} 
    \frac{(x'/x;q_2^{-1})_\infty}{(q_{1} x'/x;q_2^{-1})_\infty}
    % \frac{(q_{12} x'/x;q_2)_\infty}{(q_{2} x'/x;q_2)_\infty}
    \, , \qquad
    |q_2| > 1
    \, ,
    \label{eq:full_part_fn_pureSYM_5d}
\end{align}
where $(z;q)_\infty$ is the $q$-factorial~\eqref{eq:q-factorial} interpreted as the $q$-deformation of the gamma function.
\end{itembox}
For four and six-dimensional cases, we obtain a similar infinite product formula based on the gamma and elliptic gamma functions, respectively.
See, e.g.,~\cite{Kimura:2020jxl}.

\subsubsection{Combinatorial formula}\label{sec:combinatorial_formula}

We have seen that the partition function is obtained by applying the index functor to the tangent bundle.
The full partition function is written in the closed form, whereas it involves infinite product as in~\eqref{eq:full_part_fn_pureSYM_5d}.
In fact, the infinite product contribution originates from the perturbative part, and one can subtract the finite rational contribution to be identified with the instanton partition function.
For this purpose, we evaluate the character of the tangent bundle with the fix point configuration (see, e.g.~\cite{Nakajima:1999}),
\begin{align}
    \operatorname{ch} T_\lambda \mathfrak{M}_{n,k} = 
    \sum_{\alpha, \beta \in [n]} \ee^{a_{\alpha \beta}} \Xi(\lambda_\alpha, \lambda_\beta; q_1, q_2)
    \, ,
\end{align}
where we denote $a_{\alpha \beta} = a_\alpha - a_\beta$.
The combinatorial factor is given by
\begin{align}
    \Xi(\lambda_\alpha, \lambda_\beta; q_{1,2}) & = \operatorname{ch} \left( \mathsf{I}_\emptyset/\mathsf{I}_{\lambda_\beta} + Q_{12}^\vee (\mathsf{I}_\emptyset/\mathsf{I}_{\lambda_\alpha})^\vee - P_{12}^\vee (\mathsf{I}_\emptyset/\mathsf{I}_{\lambda_\alpha})^\vee (\mathsf{I}_\emptyset/\mathsf{I}_{\lambda_\beta}) \right)
    \nonumber \\
    & = \sum_{s \in \lambda_\alpha} q_1^{\mathbf{l}_\beta(s)} q_2^{- \mathbf{a}_\alpha(s) - 1} + \sum_{s \in \lambda_\beta} q_1^{-\mathbf{l}_\alpha(s) - 1} q_2^{\mathbf{a}_\beta(s)}
    \, ,
    \label{eq:combinatorial_Xi}
\end{align}
where we define the arm and leg lengths,
\begin{align}
    \mathbf{a}_\alpha(i,j) = \lambda_{\alpha,i} - j
    \, , \qquad
    \mathbf{l}_\alpha(i,j) = \check{\lambda}_{\alpha,j} - i
    \, .
\end{align}
We remark that the same combinatorial factor is obtained starting from the co-stability condition.
We observe the following relation,
\begin{align}
    \Xi(\lambda_\alpha, \lambda_\beta; q_{1,2}) 
    = \Xi(\check{\lambda}_\alpha, \check{\lambda}_\beta; q_{2,1})
    = q_{12}^{-1} \Xi(\lambda_\beta, \lambda_\alpha; q_{1,2}^{-1})
    \, .
\end{align}
Hence, applying the index functor, we obtain the $k$-instanton partition function, a.k.a., \emph{Nekrasov partition function}.
\begin{itembox}{Nekrasov partition function (combinatorial formula)~\cite{Nekrasov:2002qd,Nekrasov:2003rj}}
The instanton partition function has the following combinatorial formula (Nekrasov partition function),
\begin{align}
    Z_k = \sum_{\lambda \in \mathfrak{M}_{n,k}^\mathsf{T}} \mathbb{I}[-T_\lambda\mathfrak{M}_{n,k}] = \sum_{\lambda \in \mathfrak{M}_{n,k}^\mathsf{T}} \prod_{\alpha,\beta \in [n]} \frac{1}{\mathsf{N}_{\lambda_\alpha\lambda_\beta}(a_{\alpha\beta};\epsilon_{1,2})}
    \, ,
\end{align}
where we define the combinatorial factor,
\begin{align}
    \mathsf{N}_{\lambda_\alpha\lambda_\beta}(z;\epsilon_{1,2}) & = \prod_{s \in \lambda_\alpha }[z + \epsilon_1 \mathbf{l}_\beta(s) - \epsilon_2 (\mathbf{a}_\alpha(s)+1)]
    \nonumber \\
    & \qquad \times \prod_{s \in \lambda_\beta }[z - \epsilon_1 (\mathbf{l}_\beta(s)+1) + \epsilon_2 \mathbf{a}_\alpha(s)]
    \, .
    \label{eq:Nekrasov_factor}
\end{align}
\end{itembox}

\subsubsection{Contour integral formula}

It has been known that the instanton partition function can be described in the form of contour integral, a.k.a., LMNS (Losev--Moore--Nekrasov--Shatashvili) formula.
For this purpose, we use the vector bundles on the moduli space without substituting the fixed point value, i.e., $\operatorname{ch} K = \sum_{I \in [k]} \ee^{\phi_I}$.
Then, we compute the index functor of the tangent bundle to the $k$-instanton moduli space, which gives rise to the contour integral over the maximal Cartan torus $\mathsf{T}_K \subset \mathrm{GL}(K)$.
\begin{itembox}{LMNS formula (contour integral formula)~\cite{Losev:1997tp,Moore:1997dj,Lossev:1997bz}}
The instanton partition function has the following contour integral formula (LMNS formula),
\begin{align}
    Z_k = \mathbb{I}[-T\mathfrak{M}_{n,k}] = \frac{1}{k!} \frac{[-\epsilon_{12}]^k}{[-\epsilon_{1,2}]^k} \oint_{\mathsf{T}_K} \dd{\underline{\phi}} \frac{1}{P(\underline{\phi}) \widetilde{P}(\underline{\phi}+\epsilon_{12})} \prod_{I \neq J}^k \mathscr{S}(\phi_{IJ})^{-1}
    \, ,
\end{align}
where we define the gauge polynomials
\begin{align}
    P(\underline{\phi}) = \mathbb{I}[N^\vee K] = \prod_{I \in [k], \alpha \in [n]} [\phi_I - a_\alpha] 
    \, , \quad
    \widetilde{P}(\underline{\phi}) = \mathbb{I}[K^\vee N] = \prod_{I \in [k], \alpha \in [n]} [- \phi_I + a_\alpha]
\end{align}
and the $\mathscr{S}$-function,
\begin{align}
    \mathscr{S}(z) = \frac{[z - \epsilon_{1,2}]}{[z][z - \epsilon_{12}]}
    \, .
\end{align}
\end{itembox}
We use the notation, $[z - \epsilon_{1,2}] = [z - \epsilon_{1}][z - \epsilon_{2}]$.
The integral measure is defined as $\dd{\underline{\phi}} = \prod_{I \in [k]} \dd{\phi}_I / 2 \pi \ii$, which is understood as the index of the zero mode appearing from $K^\vee K$.
We have the Weyl group volume, $k! = |\mathfrak{S}_k|$.
In fact, the poles of the integrand are consistent with the fixed points, such that we take the residues at $\phi_{I} = \phi_{I-1} + \epsilon_{1,2}$ or $\phi_I = a_\alpha$, which yields the map, $\{\phi_I\}_{I \in [k]} \to \{a_\alpha + (i - 1) \epsilon_1 + (j-1) \epsilon_2\}_{\alpha \in [n], (i,j) \in \lambda_{alpha}}$.
This multi-variable contour integral is also understood as the Jeffrey--Kirwan residue prescription~\cite{Jeffrey:1995}.
See, e.g.,~\cite{Benini:2013xpa,Hori:2014tda,Hwang:2014uwa,Nakamura:2015zsa} for details.

\subsubsection{Quiver gauge theory}\label{sec:quiver_gauge_theory}

Using the observable sheaf, we can similarly obtain the hypermultiplet contribution to the partition function.
For the matter bundle $\mathsf{M}$ and $\widetilde{\mathsf{M}}$, we have the (anti)fundamental hypermultiplet contribution as follows,
\begin{align}
    \mathsf{H} = - \frac{\mathsf{M}^\vee \mathsf{Y}}{P_{12}}
    \, , \qquad
    \widetilde{\mathsf{H}} = - \frac{\mathsf{Y}^\vee \widetilde{\mathsf{M}}}{P_{12}}
    \, ,
    \label{eq:fund_hyp_contribution}
\end{align}
where the corresponding Chern roots are identified with the fundamental mass parameters,
\begin{align}
    \operatorname{ch} \mathsf{M} = \sum_{f \in [n^F]} \ee^{m_f}
    \, , \qquad
    \operatorname{ch} \widetilde{\mathsf{M}} = \sum_{f \in [\tilde{n}^F]} \ee^{\tilde{m}_f}
    \, .
\end{align}

For quiver gauge theory of type $\Gamma = (\Gamma_0,\Gamma_1)$ with $\Gamma_0 = \{ \text{nodes} \}$ and $\Gamma_1 = \{ \text{edges} \}$,%
\footnote{%
Since we are considering eight supercharge gauge theories, we also have the dual edge for each edge $e : i \to j$ given by $e^\vee : j \to i$, and the set of the dual edges denoted by $\Gamma_1^\vee = \{ e^\vee \}$.
In order to realize the oriented multiple arrow to construct non-simply-laced quiver, we need to properly assign the $\Omega$-background parameters~\cite{Kimura:2017hez}.
}
we have the instanton moduli space for each node, and the total moduli space is given by $\mathfrak{M}_{\underline{G},\underline{k}} = \bigsqcup_{i \in \Gamma_0} \mathfrak{M}_{G_i,k_i}$~\cite{Nekrasov:2012xe,Nekrasov:2013xda}.
In this case, we have the observable sheaf for each node, $(\mathsf{Y}_i)_{i \in \Gamma_0}$, and the vector multiplet and the bifundamental hypermultiplet contributions are given by
\begin{align}
    \mathsf{V}_i = \frac{\mathsf{Y}_i^\vee \mathsf{Y}_i}{P_{12}}
    \, , \qquad
    \mathsf{H}_{e:i\to j} = - \mathsf{M}_e \frac{\mathsf{Y}_i^\vee \mathsf{Y}_j}{P_{12}}
    \, ,
\end{align}
where $\mathsf{M}_e$ is a line bundle assigned to the edge $e \in \Gamma_1$ with the Chern root identified with the bifundamental mass parameter $m_e$.
For a generic quiver (except for the loop/cyclic cases), we can put $m_e$ to be zero by shift of Coulomb moduli parameters.
The total tangent bundle is thus given by~\cite{Kimura:2015rgi}
\begin{align}
    - T\mathfrak{M}_{\underline{G},\underline{k}} = \sum_{i \in \Gamma_0} \mathsf{V}_i + \sum_{e: i \to j} \mathsf{H}_{e:i\to j}
    = \sum_{i,j \in \Gamma_0} \frac{\mathsf{Y}_i^\vee c_{ij}^+ \mathsf{Y}_j}{P_{12}}
    \, ,
    \label{eq:total_tangent_bundle}
\end{align}
where we define the half $q$-Cartan matrix,
\begin{align}
    c_{ij}^+ = \delta_{ij} - \sum_{e:i \to j} \mathsf{M}_e
    \, .
    \label{eq:half-q-Cartan}
\end{align}
We remark that this total tangent bundle involves also the perturbative contributions in addition to the instanton part.

\subsubsection*{Contour integral}

Given the tangent bundle, we obtain the contour integral form of the instanton partition function.
Let us discuss the examples.

\paragraph{$A_p$ quiver theory}

We consider the linear quiver that consists of $p$ gauge nodes:
\dynkin[label,labels={1,2,p-1,p}]{A}{}\\
In this case, we can put all the bifundamental mass parameters to be zero without loss of generality.
The bifundamental hypermultiplet contribution is given by
\begin{align}
    \mathsf{H}_{i \to i+1} = - \frac{\mathsf{Y}_i^\vee \mathsf{Y}_{i+1}}{P_{12}} = - \frac{\mathsf{N}_i^\vee \mathsf{N}_{i+1}}{P_{12}} + \mathsf{N}_i^\vee \mathsf{K}_{i+1} + Q_{12}^\vee \mathsf{K}_i^\vee \mathsf{N}_{i+1} - P_{12}^\vee \mathsf{K}_i^\vee \mathsf{K}_{i+1}
    \, ,
\end{align}
where the first term is interpreted as the perturbative contribution.
Hence, applying the index functor to the instanton part, we obtain the contour integral formula for the instanton partition function,
\begin{align}
    Z_{\underline{k}} = \frac{1}{\underline{k}!} \frac{[-\epsilon_{12}]^{k}}{[-\epsilon_{1,2}]^{k}} \oint_{\mathsf{T}_K} \dd{\underline{\phi}} \frac{\prod_{i \in [p-1]} P_i(\underline{\phi}_{i+1})\widetilde{P}_{i+1}(\underline{\phi}_i + \epsilon_{12})}{\prod_{i \in [p]} P_i(\underline{\phi}_i) \widetilde{P}_i(\underline{\phi}_i + \epsilon_{12})}
    \frac{\prod_{i \in [p-1]} \prod^{I \in [k_{i+1}]}_{J \in [k_{i}]} \mathscr{S}(\phi_{i+1,I} - \phi_{i,J})}{\prod_{i \in [p]} \prod_{I \neq J}^{k_i} \mathscr{S}(\phi_{i,I} - \phi_{i,J})}
    \, ,
\end{align}
where we denote $\underline{k}! = \prod_{i \in \Gamma_0} k_i!$ and $k = \sum_{i \in \Gamma_0} k_i$.
The gauge polynomials are given by $P_i(z) = \prod_{\alpha \in [n_i]} [z - a_{i,\alpha}]$ and $\widetilde{P}_i(z) = \prod_{\alpha \in [n_i]} [ a_{i,\alpha} - z]$.

\paragraph{$\widehat{A}_{p}$ quiver theory}

The next example is a cyclic quiver $\widehat{A}_{p}$ with $p+1$ nodes:
\dynkin[label,labels={0,1,2,p-1,p},extended]{A}{}\\
In this case, we can put all the bifundamental mass parameters to be the same, $m_{i \to i+1} = m$. 
Then, the contour integral form of the instanton partition function is given by
\begin{align}
    Z_{\underline{k}} = \frac{1}{\underline{k}!} \frac{[-\epsilon_{12}]^{k}}{[-\epsilon_{1,2}]^{k}} \oint_{\mathsf{T}_K} \dd{\underline{\phi}} \prod_{i \in \mathbb{Z}_{p+1}} \left[ \frac{P_i(\underline{\phi}_{i+1} + m)\widetilde{P}_{i+1}(\underline{\phi}_i - m + \epsilon_{12})}{P_i(\underline{\phi}_i) \widetilde{P}_i(\underline{\phi}_i + \epsilon_{12})} 
    \frac{\prod^{I \in [k_{i+1}]}_{J \in [k_{i}]} \mathscr{S}(\phi_{i+1,I} - \phi_{i,J} + m)}{\prod_{I \neq J}^{k_i} \mathscr{S}(\phi_{i,I} - \phi_{i,J})} \right]
    \, .
\end{align}
The case $p = 0$ involves a single node with a loop edge describing the hypermultiplet in the adjoint representation, which corresponds to four-dimensional $\mathcal{N} = 2^*$ theory.
The instanton partition function is given by
\begin{align}
    Z_k = \frac{1}{k!} \left( \frac{[-\epsilon_{12}]}{[-\epsilon_{1,2}]} \mathscr{S}(m) \right)^k \oint_{\mathsf{T}_K} \dd{\underline{\phi}} \frac{P(\underline{\phi} + m) \widetilde{P}(\underline{\phi} - m + \epsilon_{12})}{P(\underline{\phi}) \widetilde{P}(\underline{\phi} + \epsilon_{12})}
    \prod_{I \neq J}^k \frac{\mathscr{S}(\phi_{IJ}+m)}{\mathscr{S}(\phi_{IJ})}
    \, ,
\end{align}
where $m$ is the adjoint mass parameter.

\subsubsection*{Fixed points}

The fixed points are similarly described for quiver gauge theory.
For $\underline{G} = \prod_{i \in \Gamma_0} \mathrm{U}(n_i)$, we have a set of partitions characterizing the fixed points,
\begin{align}
    \lambda = (\lambda_{i,\alpha})_{i \in \Gamma_0, \alpha \in [n_i]} \in \mathfrak{M}^\mathsf{T}
    \, ,
\end{align}
and we define
\begin{align}
    \mathcal{X}_\lambda = \bigsqcup_{i \in \Gamma_0} \mathcal{X}_{\lambda_i}
    \, , \qquad
    \mathcal{X}_{\lambda_i} = \{ x_{i,\alpha,k} = \ee^{a_{i,\alpha}} q_1^{k-1} q_2^{\lambda_{i,\alpha,k}} \}_{\alpha \in [n], k \in \mathbb{N}}
    \, ,
\end{align}
under the stability condition and $|q_1| < 1$.
Then, the total tangent space at the fixed point $\lambda \in \mathfrak{M}^\mathsf{T}$ is given by
\begin{align}
    - T_\lambda \mathfrak{M}_{\underline{G},\underline{k}} = \sum_{i,j \in \Gamma_0} \frac{P_1^\vee}{P_2} \mathsf{X}_i^\vee c_{ij}^+ \mathsf{X}_j
    \, , \qquad 
    \operatorname{ch} \mathsf{X}_i = \sum_{x \in \mathcal{X}_{\lambda_i}} x
    \, ,
\end{align}
and thus the full partition function is given as follows,
\begin{align}
    Z = \sum_{\lambda \in \mathfrak{M}^\mathsf{T}} \underline{\mathfrak{q}}^{|\lambda|} \, \mathbb{I}[- T_\lambda \mathfrak{M}_{\underline{G},\underline{k}}]
\end{align}
where we denote the instanton counting parameter,
\begin{align}
    \underline{\mathfrak{q}}^{|\lambda|} = \prod_{i \in \Gamma_0} \mathfrak{q}_i^{|\lambda_i|}
    \, .
\end{align}

\subsubsection{Supergroup gauge theory}

We apply the formalism discussed above to supergroup gauge theory.
The discussion of this part is mainly based on~\cite{Kimura:2019msw}.
First of all, we consider the supercharacter of the supervector spaces instead of the ordinary characters,
\begin{subequations}
    \begin{align}
        \operatorname{sch} K & = \operatorname{ch} K_0 - \operatorname{ch} K_1 = \sum_{I \in [k_0]} \ee^{\phi_I^0} - \sum_{I \in [k_1]} \ee^{\phi_I^1}
        \, , \\
        \operatorname{sch} N & = \operatorname{ch} N_0 - \operatorname{ch} N_1 = \sum_{\alpha \in [n_0]} \ee^{a_\alpha^0} - \sum_{\alpha \in [n_1]} \ee^{a_\alpha^1}
        \, .
    \end{align}
\end{subequations}
In order to apply the localization formula, we then analyze the fixed points in the moduli space $\mathfrak{M}_{n,k}$ under the equivariant actions.
In the case of supergroup gauge theory, we still have the same form of the fixed point equations~\eqref{eq:fixed_pt_eq}.
Recalling that we should assign the both stability and co-stability conditions, 
\begin{align}
    K_0 = \mathbb{C}[B_1, B_2] I(N_0)
    \, , \qquad
    K_1 = \mathbb{C}[B_1, B_2] J(N_1)^\vee
    \, ,
\end{align}
the fixed points under the equivariant actions are parametrized by $\lambda = (\lambda^\sigma_\alpha)_{\sigma = 0, 1, \alpha \in [n_\sigma]}$ with $|\lambda^\sigma| = \sum_{\alpha \in [n_\sigma]} |\lambda_\alpha^\sigma| = k_\sigma$.
Hence, each subspace is given by $K_0 = \operatorname{Span}\{ B_1^{i-1} B_2^{j-1} I_{\alpha^0} \}_{\alpha \in [n_0],(i,j) \in \lambda^0_\alpha}$ and $K_1 = \operatorname{Span}\{ J_{\alpha^1} B_1^{i-1} B_2^{j-1} \}_{\alpha \in [n_1],(i,j) \in \lambda^1_\alpha}$ at the fixed point $\lambda \in \mathfrak{M}_{n,k}^\mathsf{T}$.
Therefore, the supercharacter is given by
\begin{align}
    \operatorname{sch} K\Big|_{\lambda} = \sum_{\alpha \in [n_0]} \sum_{(i,j) \in \lambda_\alpha^0} \ee^{a_\alpha^0} q_1^{i-1} q_2^{j-1} - \sum_{\alpha \in [n_1]} \sum_{(i,j) \in \lambda_\alpha^1} \ee^{a_\alpha^1} q_1^{-i} q_2^{-j}
    \, .
\end{align}
Namely, we assign the GL($Q$)-equivariant parameters $(\epsilon_{1,2})$ for the positive node and $(-\epsilon_{1,2})$ for the negative node.
This assignment is similarly understood as the analytic continuation to obtain supergroup theory as discussed in Sec.~\ref{sec:quiver_realization}, which is also consistent with the $\Omega$-background in the context of supermatrix model~\eqref{eq:Omega_background_matrix_parameter}.

\subsubsection*{Contour integral formula}

We construct the observable sheaf as before, $\mathsf{Y} = N - P_{12} K$.
Obtaining the vector multiplet contribution, associated with the tangent bundle, 
\begin{align}
    \mathsf{V} & = \frac{\mathsf{Y}^\vee \mathsf{Y}}{P_{12}} = - T\mathfrak{M}_{n,k} + \frac{\mathsf{N}^\vee \mathsf{N}}{P_{12}}
    \nonumber \\
    & \xrightarrow{\text{sch}}
    \sum_{\sigma,\sigma' = 0, 1} (-1)^{\sigma + \sigma'} 
    \left[ \frac{N_\sigma^\vee N_{\sigma'}}{P_{12}} - N_\sigma^\vee K_{\sigma'} - Q_{12}^\vee K_{\sigma}^\vee N_{\sigma'} + P_{12} K_{\sigma}^\vee K_{\sigma'} \right]
    \, ,
\end{align}
we have the contour integral form of the instanton partition function.
\begin{itembox}{LMNS formula for supergroup gauge theory}
The instanton partition function of $\mathrm{U}(n_0|n_1)$ gauge theory is given by the following contour integral over the Cartan torus $\mathsf{T}_K$ with $k_{01} = k_0 + k_1$,
\begin{align}
    Z_k & = \mathbb{I}[-T\mathfrak{M}_{n,k}]
    \nonumber \\
    & = \frac{1}{k_{0,1}!} \frac{[-\epsilon_{12}]^{k_{01}}}{[-\epsilon_{1,2}]^{k_{01}}} \oint_{\mathsf{T}_K} \dd{\underline{\phi}} \, \frac{P_{0,1}(\underline{\phi}^{1,0}) \widetilde{P}_{0,1}(\underline{\phi}^{1,0}+\epsilon_{12})}{P_{0,1}(\underline{\phi}^{0,1}) \widetilde{P}_{0,1}(\underline{\phi}^{0,1}+\epsilon_{12})}
    \frac{\prod^{I \in [k_0]}_{J \in [k_1]} \mathscr{S}(\phi^0_I - \phi^1_J) \mathscr{S}(\phi^1_J - \phi^0_I) }{\prod_{I \neq J}^{k_{0}} \mathscr{S}(\phi^0_{IJ}) \prod_{I \neq J}^{k_{1}} \mathscr{S}(\phi^1_{IJ})}
    \, .
    \label{eq:LMNS_formula_supergroup}
\end{align}
\end{itembox}
This formula agrees with the contour integral formula of $\widehat{A}_1$ quiver gauge theory obtained in Sec.~\ref{sec:quiver_gauge_theory}, which is consistent with the quiver gauge theory realization of supergroup gauge theory presented in Sec.~\ref{sec:quiver_realization}.
In this case, we should take the residues of the following poles, $\phi_{I}^0 = \phi_{I-1}^0 + \epsilon_{1,2}$ or $\phi_I^0 = a_\alpha^0$ and $\phi_{I}^1 = \phi_{I-1}^1 - \epsilon_{1,2}$ or $\phi_I^1 = a_\alpha^1 - \epsilon_{12}$, to be compatible with the stability and the co-stability conditions.
From the point of view of the JK residue prescription, we may use the reference vectors $\pm \eta$ for the even and the odd sectors.
The formula~\eqref{eq:LMNS_formula_supergroup} is originally given as an integral over $\mathrm{GL}(K)$, which is evaluated by the maximal torus $\mathsf{T}_K$.
Hence, it should be understood as a normalized integral by the supergroup volume as discussed for the supermatrix models in Secs.~\ref{sec:Hermitian_supermatrix_model} and \ref{sec:real-quaternion_supermatrix_model}.

\subsubsection*{Equivariant index formula}

Recalling that we assign the stability and the co-stability conditions for the even and the odd nodes, we obtain for $|q_1| < 1$ and $|q_2| > 1$ as
\begin{align}
    \mathsf{Y}\Big|_\lambda & = (P_1 X_0)_0 \oplus (P_2^\vee \check{X}_1)_1
%    \nonumber \\ &
    \xrightarrow{\text{sch}} (1 - q_1) \sum_{x \in \mathcal{X}_{\lambda}^0} x - (1 - q_2^{-1}) \sum_{x \in \check{\mathcal{X}}_{\lambda}^1} x
\end{align}
where we define the sets,
\begin{align}
    \mathcal{X}_\lambda^0 = \{ \ee^{a^0_\alpha} q_1^{i-1} q_2^{\lambda_{\alpha,i}^0} \}_{\alpha \in [n_0], i \in \mathbb{N}}
    \, , \qquad
    \check{\mathcal{X}}_\lambda^1 = \{ \ee^{a^1_\alpha} q_1^{- \check{\lambda}_{\alpha,j}^1} q_2^{-j+1} \}_{\alpha \in [n_0], j \in \mathbb{N}}
    \, .
\end{align}
Imposing the fixed point configuration, the vector multiplet contribution is given as follows,
\begin{align}
    \mathsf{V}_{\lambda} =
    \left( \frac{P_1^\vee}{P_2} X_0^\vee X_0 + \frac{P_2^\vee}{P_1} \check{X}_1^\vee \check{X}_1  \right)_0 \oplus \left( Q_{12}^\vee X_0^\vee \check{X}_1 + \check{X}_1^\vee X_0 \right)_1
    \, .
\end{align}
Applying the index functor with taking care of the conditions $|q_1| < 1$ and $|q_2| > 1$, we obtain the partition function.
\begin{itembox}{Equivariant index formula for supergroup partition function}
The partition function of $\mathrm{U}(n_0|n_1)$ gauge theory is given by summation over the fixed point contributions,
\begin{align}
    Z = \sum_{\lambda \in \mathfrak{M}^\mathsf{T}} \mathfrak{q}^{|\lambda^0| - |\lambda^1|} \, Z_\lambda
    \, ,
\end{align}
where the contribution at each fixed point $\lambda \in \mathfrak{M}^\mathsf{T}$ is given in the five-dimensional convention for $|q_1|<1$ and $|q_2|>1$ by
\begin{align}
    Z_\lambda & = 
    \prod_{(x,x') \in \mathcal{X}^0_\lambda \times \mathcal{X}^0_\lambda} \frac{(x'/x;q_2^{-1})_\infty}{(q_1 x'/x;q_2^{-1})_\infty}
    \prod_{(x,x') \in \check{\mathcal{X}}^1_\lambda \times \check{\mathcal{X}}^1_\lambda} \frac{(q_{12} x'/x;q_1)_\infty}{(q_1 x'/x;q_1)_\infty}
    \nonumber \\
    & \qquad \times
    \prod_{(x,x') \in {\mathcal{X}}^0_\lambda \times \check{\mathcal{X}}_\lambda^1}
    \left( 1 - \frac{x'}{x} \right)^{-1} \left( 1 - q_{12} \frac{x}{x'} \right)^{-1}
    \, .
\end{align}
\end{itembox}
We remark a similarity of this partition function to the supermatrix model discussed in Sec.~\ref{sec:super_matrix}.
We will see further similarities from the underlying algebraic point of view.
See Sec.~\ref{sec:quiver_W-algebra}.

\subsubsection*{Combinatorial formula}

We consider the combinatorial formula, which gives rise to the finite rational expression,
\begin{align}
    \operatorname{sch} T_\lambda\mathfrak{M}_{n,k} =  \sum_{\sigma,\sigma'=0,1} (-1)^{\sigma + \sigma'} \sum_{\alpha \in [n_\sigma], \beta \in [n_{\sigma'}]} \ee^{a^\sigma_\alpha - a^{\sigma'}_\beta} \Xi_{\sigma\sigma'}(\lambda_\alpha,\lambda_\beta;q_{1,2})
    \, ,
\end{align}
where the diagonal combinatorial factors are given by
\begin{align}
    \Xi_{00}(\lambda_\alpha,\lambda_\beta;q_{1,2}) = \Xi(\lambda_\alpha,\lambda_\beta;q_{1,2})
    \, , \qquad
    \Xi_{11}(\lambda_\alpha,\lambda_\beta;q_{1,2}) = \Xi(\lambda_\beta,\lambda_\alpha;q_{1,2})
    \, .
\end{align}
This is clear from the original expression of $\Xi$ in terms of the polynomial ideals~\eqref{eq:K_sp_ideal_rep} and \eqref{eq:combinatorial_Xi}.
See Sec.~\ref{sec:combinatorial_formula}.
The off-diagonal factors are given as follows (see~\cite{Kimura:2019msw,Kimura:2020jxl}),
\begin{subequations}
\begin{align}
    \Xi_{01}(\lambda_\alpha,\lambda_\beta;q_{1,2}) & = \operatorname{ch} \left( Q_{12}^\vee (\mathsf{I}_\emptyset/\mathsf{I}_{\lambda_\beta})^\vee + Q_{12}^\vee (\mathsf{I}_\emptyset/\mathsf{I}_{\lambda_\alpha})^\vee - P_{12}^\vee Q_{12}^\vee (\mathsf{I}_\emptyset/\mathsf{I}_{\lambda_\alpha})^\vee (\mathsf{I}_\emptyset/\mathsf{I}_{\lambda_\beta})^\vee \right)
    \nonumber \\
    & =
    - \sum_{i=1}^{\check\lambda_{\alpha,1}} \sum_{j'=1}^{\lambda_{\beta,1}}
 \left[
 q_1^{-\check\lambda_{\beta,j'}-i} q_2^{-\lambda_{\alpha,i}-j'}
 - q_1^{-i} q_2^{-j'}
 \right]
  \nonumber \\
 & \qquad 
 + \sum_{(i,j) \in \lambda_\alpha} q_1^{-i} q_2^{-\lambda_{\beta,1}-j}
 + \sum_{(i',j') \in \lambda_\beta} q_1^{-\check\lambda_{\alpha,1}-i'} q_2^{-j'}
 \, , \\
    \Xi_{10}(\lambda_\alpha,\lambda_\beta;q_{1,2}) & = \operatorname{ch} \left( \mathsf{I}_\emptyset/\mathsf{I}_{\lambda_\beta} + \mathsf{I}_\emptyset/\mathsf{I}_{\lambda_\alpha} - P_{12}^\vee (\mathsf{I}_\emptyset/\mathsf{I}_{\lambda_\alpha}) (\mathsf{I}_\emptyset/\mathsf{I}_{\lambda_\beta}) \right)
    \nonumber \\
& =  - \sum_{i=1}^{\check\lambda_{\alpha,1}} \sum_{j'=1}^{\lambda_{\beta,1}}
 \left[
 q_1^{\check\lambda_{\beta,j'}+i-1} q_2^{\lambda_{\alpha,i}+j'-1}
 - q_1^{i-1} q_2^{j'-1}
 \right]
 \nonumber \\
 & \qquad 
 + \sum_{(i,j) \in \lambda_\alpha} q_1^{i-1} q_2^{\lambda_{\beta,1}+j-1}
 + \sum_{(i',j') \in \lambda_\beta} q_1^{\check\lambda_{\alpha,1}+i'-1} q_2^{j'-1}    
 \, .
\end{align}
\end{subequations}
In contrast to the diagonal part $\Xi_{00(11)}$, further simplification does not occur for these off-diagonal ones.
We remark that these off-diagonal factors are symmetric under $\lambda_\alpha \leftrightarrow \lambda_\beta$,
\begin{align}
 \Xi_{01}(\lambda_\alpha,\lambda_\beta;q_{1,2}) = \Xi_{01}(\lambda_\beta,\lambda_\alpha;q_{1,2})
 \, , \qquad
 \Xi_{10}(\lambda_\alpha,\lambda_\beta;q_{1,2}) = \Xi_{10}(\lambda_\beta,\lambda_\alpha;q_{1,2})
 \, ,
\end{align}
and we have the relation,
\begin{align}
 q_{12} \, \Xi_{01}(\lambda_\alpha,\lambda_\beta;q_{1,2})
 = \Xi_{10}(\lambda_\alpha,\lambda_\beta;q_{1,2}^{-1})
 \, .
\end{align}
Applying the index functor to this expression, we obtain the instanton partition function.
In addition to the Nekrasov factor for the diagonal part~\eqref{eq:Nekrasov_factor}, we also define the off-diagonal combinatorial factor,
\begin{align}
    \mathsf{N}^{01}_{\lambda_\alpha \lambda_\beta}(z;\epsilon_{1,2}) & =
    \prod_{i=1}^{\check\lambda_{\alpha,1}} \prod_{j'=1}^{\lambda_{\beta,1}} \frac{[z - i \epsilon_1 - j' \epsilon_2]}{[z - (\check{\lambda}_{\beta,j'} + i) \epsilon_1 - (\lambda_{\alpha,i} + j')]}
    \nonumber \\
    & \qquad \times
    \prod_{(i,j) \in \lambda_\alpha} [z - i \epsilon_1 - (\lambda_{\beta,1} + j)\epsilon_2]
    \prod_{(i',j') \in \lambda_\alpha} [z - (\check{\lambda}_{\alpha,1} + i') \epsilon_1 - j' \epsilon_2]
    \, , \\
    \mathsf{N}^{10}_{\lambda_\alpha \lambda_\beta}(z;\epsilon_{1,2}) & =
    \prod_{i=1}^{\check\lambda_{\alpha,1}} \prod_{j'=1}^{\lambda_{\beta,1}} \frac{[z + (i-1) \epsilon_1 + (j'-1) \epsilon_2]}{[z + (\check{\lambda}_{\beta,j'} + i - 1) \epsilon_1 + (\lambda_{\alpha,i} + j' - 1)]}
    \nonumber \\
    & \qquad \times
    \prod_{(i,j) \in \lambda_\alpha} [z + (i-1) \epsilon_1 +  (\lambda_{\beta,1} + j - 1)\epsilon_2]
    \prod_{(i',j') \in \lambda_\alpha} [z + (\check{\lambda}_{\alpha,1} + i' - 1) \epsilon_1 + (j' - 1) \epsilon_2]
    \, .
\end{align}
The combinatorial form of the $k$-instanton partition function for $\mathrm{U}(n_0|n_1)$ supergroup gauge theory is given as follows.
\begin{itembox}{Nekrasov partition function for $\mathrm{U}(n_0|n_1)$ gauge theory}
The instanton partition function of $\mathrm{U}(n_0|n_1)$ gauge theory has the following combinatorial expression,
\begin{align}
    Z_k = \sum_{\lambda \in \mathfrak{M}_{n,k}^\mathsf{T}} \frac{\prod_{\alpha \in [n_0], \beta \in [n_1]} \mathsf{N}^{01}_{\lambda_\alpha^0 \lambda_\beta^1}(a_\alpha^0 - a_\beta^1;\epsilon_{1,2}) \mathsf{N}^{10}_{\lambda_\beta^1 \lambda_\alpha^0}(a_\beta^1-a_\alpha^0;\epsilon_{1,2})}{\prod_{\sigma = 0, 1} \prod_{\alpha,\beta \in [n_\sigma]} \mathsf{N}_{\lambda_\alpha^\sigma \lambda_\beta^\sigma}(a_\alpha^\sigma - a_\beta^\sigma;\epsilon_{1,2})}
    \, .
\end{align}
\end{itembox}
We see that the denominator factors are the standard gauge node contributions, whereas the numerator factors are analogous, but not identical to the bifundamental hypermultiplet contributions.

\section{Non-perturbative aspects of supergroup gauge theory}\label{sec:non-perturbative_study}

In this section, we explore non-perturbative aspects of supergroup gauge theory based on the instanton partition function obtained in the previous section. 

\subsection{Topological string theory approach}

We have seen in Sec.~\ref{sec:string_theory_perspective} that string/M-theory provides a geometric description of $\mathcal{N}=2$ gauge theory. 
Choosing the Calabi--Yau (CY) three-fold constructed from resolution of specific singularities, one can obtain the corresponding Seiberg--Witten geometry.
This process is known as geometric engineering~\cite{Katz:1996fh,Katz:1997eq}.
This description can be further pursued at the level of the microscopic instanton counting: 
The A-model topological string amplitude of the corresponding non-compact toric CY three-fold computes the five-dimensional instanton partition function~\cite{Iqbal:2003ix,Iqbal:2003zz,Eguchi:2003sj}.
The primary example is the local $\mathbb{P}^1 \times \mathbb{P}^1$ geometry, which corresponds to pure SU(2) SYM theory.
In this case, we have the following web diagram, which is dual to the toric diagram,
\begin{align}
%    \rem{To be added: web diagram and toric diagram of P1 * P1}
    \begin{tikzpicture}[baseline=(current bounding  box.center),scale = 1]
    \draw (-.5,-.5) rectangle (.5,.5);
    \draw (.5,.5) -- ++(45:.5);
    \draw (-.5,.5) -- ++(135:.5);
    \draw (.5,-.5) -- ++(-45:.5);
    \draw (-.5,-.5) -- ++(225:.5);
    \draw[very thick,blue,latex-latex] (1.5,0) -- ++(2,0);
    \node at (0,-1.75) {web diagram};
    \begin{scope}[shift={(5.5,0)}]
    \draw[dotted] (0,1) -- (1,0) -- (0,-1) --(-1,0) -- cycle;
    \draw[dotted] (-1,0) -- ++(2,0);
    \draw[dotted] (0,-1) -- ++(0,2);
    \filldraw (0,0) circle (.05);
    \filldraw (1,0) circle (.05);
    \filldraw (0,1) circle (.05);
    \filldraw (0,-1) circle (.05);
    \filldraw (-1,0) circle (.05);
    \node at (0,-1.75) {toric diagram};
    \end{scope}
    \end{tikzpicture}
\end{align}
This web diagram can be identified with the configuration of $(p,q)$-brane in type IIB theory obtained from the T-dual of the Hanany--Witten configuration discussed in Sec.~\ref{sec:Hany-Witten_const}.
In order to compute the topological string amplitude, we may apply a systematic approach, called the topological vertex~\cite{Aganagic:2003db}.
The idea is to describe the toric CY three-fold by gluing the local $\mathbb{C}^3$ patches, corresponding to each trivalent vertex in the web diagram.
We denote the topological vertex function with generic boundary conditions parametrized by the partitions by
\begin{align}
    C_{\lambda \mu \nu}(q) = q^{\frac{\kappa(\mu)}{2}} s_{\check{\nu}}(q^{-\rho}) \sum_{\eta} s_{\check{\lambda}/\eta}(q^{-\nu - \rho})
    s_{\mu/\eta}(q^{-\check{\nu} - \rho}) = \quad
    \begin{tikzpicture}[baseline=(current  bounding  box.center),scale=.7]
    \draw[-latex] (0,0) -- ++(1,0) node [right] {$\mu$};
    \draw[-latex] (0,0) -- ++(0,1) node [above] {$\lambda$};
    \draw[-latex] (0,0) -- ++(225:1) node [left] {$\nu$};
    \end{tikzpicture}
%    \, ,
\end{align}
where we denote the skew Schur function by $s_{\lambda/\eta}(X) = \sum_{\mu} c_{\eta\mu}^\lambda s_\mu(X)$ with the Littlewood--Richardson coefficient $c_{\eta\mu}^\lambda$.
The combinatorial factor is defined by $\kappa(\lambda)/2 = \sum_{(i,j) \in \lambda} (j-i)$.
We denote the Weyl vector by $\rho$, such that $q^{-\lambda - \rho} = \{ q^{-\lambda_i + i - \frac{1}{2}} \}_{i \in \mathbb{N}}$.
The string coupling $g_s$ with $q = \ee^{g_s}$ would be identified with the $\Omega$-background parameters at the self-dual point (unrefined situation), $g_s = \epsilon_2 = - \epsilon_1$.
In order to reproduce generic $\epsilon_{1,2}$, we need the refinement of the topological vertex~\cite{Awata:2005fa,Iqbal:2007ii}.
From this point of view, recalling that the web diagram describes $(p,q)$-branes, we introduce the anti-topological vertex (anti-vertex for short) involving the negative coupling to realize supergroup gauge theory~\cite{Vafa:2001qf,Kimura:2020lmc},
\begin{align}
    \overline{C}_{\lambda \mu \nu} (q) = {C}_{\lambda \mu \nu} (q^{-1}) 
    = \quad
    \begin{tikzpicture}[baseline=(current  bounding  box.center),scale=.7]
    \draw[-latex,dashed] (0,0) -- ++(1,0) node [right] {$\mu$};
    \draw[-latex,dashed] (0,0) -- ++(0,1) node [above] {$\lambda$};
    \draw[-latex,dashed] (0,0) -- ++(225:1) node [left] {$\nu$};
    \end{tikzpicture}
%    \, .
\end{align}
We can check that this vertex--anti-vertex formalism reproduces the instanton partition function of supergroup gauge theory~\cite{Kimura:2020lmc}.

Moreover, this vertex formalism exhibits further interesting features.
As discussed in Sec.~\ref{sec:Hany-Witten_const}, the brane realization is not unique for supergroup gauge theory.
In fact, we can see that the same instanton partition function can be obtained from different configurations corresponding to the possible Dynkin diagrams~\eqref{eq:HW_U(2|1)}.
Another remark is the relation to non-supergroup gauge theory.
From the combinatorial property of the Schur function, flipping the coupling $q \to q^{-1}$ yields the transposition of the partition $\lambda \to \check{\lambda}$, etc, which leads to the relation
\begin{align}
    \begin{tikzpicture}[baseline=(current  bounding  box.center),scale=.7]
    \draw[-latex] (0,0) -- ++(1,0) node [right] {$\mu$};
    \draw[-latex] (0,0) -- ++(0,1) node [above] {$\lambda$};
    \draw[-latex] (0,0) -- ++(225:1) node [left] {$\nu$};
    \end{tikzpicture}
    \quad \simeq \quad
    \begin{tikzpicture}[baseline=(current  bounding  box.center),xscale=-1,scale=.7]
    \draw[-latex,dashed] (0,0) -- ++(1,0) node [left] {$\mu$};
    \draw[-latex,dashed] (0,0) -- ++(0,1) node [above] {$\lambda$};
    \draw[-latex,dashed] (0,0) -- ++(225:1) node [right] {$\nu$};
    \end{tikzpicture}
\end{align}
up to the framing factor.
We remark that such a reflection relation holds only for the unrefined case.
Hence, we have realization of $\mathrm{U}(1|1)$ theory in terms of the ordinary vertices,
\begin{align}
    \begin{tikzpicture}[baseline=(current  bounding  box.center)]
    \draw (0,0) -- (1,0);
    \draw (0,0) -- ++(225:.5);
    \draw (1,0) -- ++(-45:.5);
    \draw[dashed] (225:.5) -- ++(225:.5) -- ++(2.414,0) -- ++(135:.5);
    \draw (0,0) -- ++(90:.5);
    \draw (1,0) -- ++(90:.5);
    \draw[dashed] (1,0)++(-45:1) -- ++(-30:.5);
    \draw[dashed] (225:1) -- ++(210:.5);
    \end{tikzpicture}
    \qquad \simeq \qquad
    \begin{tikzpicture}[baseline=(current  bounding  box.center)]
    \draw (0,0) -- (1,0);
    \draw (0,0) -- ++(225:1);
    \draw (1,0) -- ++(-45:1);
   \draw (0,0) -- ++(90:.5);
    \draw (1,0) -- ++(90:.5);
    \draw (1,0)++(-45:1)++(.5,0) -- ++(-.5,0) -- ++(0,-.5);
    \draw (225:1)++(-.5,0) -- ++(.5,0) -- ++(0,-.5);
    \end{tikzpicture}    
\end{align}
This configuration gives rise to $\mathrm{U}(1)$ theory with two fundamental matters, which is consistent with the gauging trick discussed in Sec.~\ref{sec:gauging_trick}.

Let us comment on a algebraic interpretation of the anti-vertex.
It has been known that the refined topological vertex is understood as the intertwiner of quantum troidal algebra of $\mathfrak{gl}_1$~\cite{Awata:2011ce}, known as Ding--Iohara--Miki algebra~\cite{Ding:1996mq,Miki:2007JMP}.
In this context, the anti-vertex is realized as the intertwiner involving the negative level~\cite{Bourgine:2018fjy,Noshita:2022dxv}.
See also~\cite{Rapcak:2019wzw} for a related realization of superalgebra from the CY geometry.

\subsection{Non-perturbative Schwinger--Dyson equation}\label{sec:SD_eq}

The instanton partition function is given by summation over the fixed point configurations, which can be related to each other by the process to add/remove instantons.
We discuss the behavior of instanton partition function under such a non-perturbative process of adding/removing instantons, which gives rise to functional relations that we call the \emph{non-perturbative Schwinger--Dyson equations}~\cite{Kanno:2012hk,Kanno:2013aha,Bourgine:2015szm,Nekrasov:2015wsu,Kimura:2015rgi}.

\subsubsection{Adding/removing instantons}

We consider instanton-adding process by shifting the vector space, $K \to K + V$ with $V = \mathbb{C}^v$, where $v$ is the number of instantons that we add to the configuration.
Under this shift, the tangent bundle at the fixed point $\lambda \in \mathfrak{M}_{n,k}^\mathsf{T}$ behaves as follows,
\begin{align}
    \delta_V T_\lambda \mathfrak{M}_{n,k} 
    & = \mathsf{Y}^\vee V + Q_{12}^\vee V^\vee \mathsf{Y} - P_{12}^{-1} V^\vee V
    \, .
    \label{eq:tangent_bundle_shift+}
\end{align}
Hence, this shift defines a map, $\delta_V$ : $\mathfrak{M}_{n,k} \to \mathfrak{M}_{n,k+v}$.
Applying the index functor, we obtain
\begin{align}
    \delta_V Z_\lambda = \mathbb{I}[-\delta_V T_\lambda \mathfrak{M}_{n,k}] = \frac{Z_{\lambda+v}}{Z_\lambda} = \frac{1}{v!} \frac{[-\epsilon_{12}]^v}{[-\epsilon_{1,2}]^v} \oint \dd{\underline{\phi}} \frac{1}{\mathscr{Y}_\lambda(\underline{\phi}) \widetilde{\mathscr{Y}}_\lambda(\underline{\phi}+\epsilon_{12})} \prod_{I \neq J}^v \mathscr{S}(\phi_{IJ})^{-1}
    \, ,
\end{align}
where we denote by $Z_{\lambda+v}$ the instanton partition function contribution with the charge $k+v$ that can be obtained from the configuration $\lambda$ by adding $v$ instantons, and we define the $\mathscr{Y}$-function with the line bundle $\mathsf{x}$ whose Chern root is given by $x$,
\begin{subequations}\label{eq:Y-func_def}
\begin{align}
    \mathscr{Y}_\lambda(x) & = \mathbb{I}[\mathsf{Y}^\vee \mathbf{x}] \stackrel{\eqref{eq:obs_sheaf_def}}{=} \prod_{x' \in \mathcal{X}_\lambda^{[\log]}} \frac{[x - x']}{[x - x' - \epsilon_1]}
    \stackrel{\eqref{eq:obs_sheaf_comb}}{=} \frac{\prod_{x' \in \mathcal{X}_{\partial_+\lambda}^{[\log]}} [x - x']}{\prod_{x' \in \mathcal{X}_{\partial_-\lambda}^{[\log]}} [x - x' - \epsilon_{12}]}
    \, , \\
    \widetilde{\mathscr{Y}}_\lambda(x) & = \mathbb{I}[ \mathbf{x}^\vee \mathsf{Y}] \stackrel{\eqref{eq:obs_sheaf_def}}{=} \prod_{x' \in \mathcal{X}_\lambda^{[\log]}} \frac{[x' - x]}{[x' - x + \epsilon_1]} 
    \stackrel{\eqref{eq:obs_sheaf_comb}}{=} \frac{\prod_{x' \in \mathcal{X}_{\partial_+\lambda}^{[\log]}} [x' - x]}{\prod_{x' \in \mathcal{X}_{\partial_-\lambda}^{[\log]}} [x' - x + \epsilon_{12}]}
    \, ,
\end{align}
\end{subequations}
with the set
\begin{align}
    \mathcal{X}_\bullet^{[\log]} = \{ \log x \mid x \in \mathcal{X}_\bullet \}
    \, .
\end{align}
We have the relation for these two $\mathscr{Y}$-functions,
\begin{align}
    \mathscr{Y}(z) =
    \begin{cases}
    (-1)^n \widetilde{\mathscr{Y}}(z) & (4d) \\
    (-1)^n \ee^{- n z} \ee^{\sum_{\alpha \in [n]} a_\alpha} \widetilde{\mathscr{Y}}(z) & (5d \ \& \ 6d) \\
    \end{cases}
    .
    \label{eq:Y-fn_convert}
\end{align}
We denote $\mathscr{Y}_\lambda(\underline{\phi}) = \prod_{I \in [v]} \mathscr{Y}_\lambda(\phi_I)$ and $\dd{\underline{\phi}} = \prod_{I \in [v]} \dd{\phi}_I/2 \pi \ii$.
We focus on the simplest case, $v = 1$.
In this case, we evaluate the contour integral surrounding the poles of $1/\mathscr{Y}_\lambda(\phi)$ based on the finite product form of the $\mathscr{Y}$-function,
\begin{align}
    \delta_V Z_\lambda & = \frac{[-\epsilon_{12}]}{[-\epsilon_{1,2}]} \oint \frac{\dd{\phi}}{2 \pi \ii} \frac{1}{\mathscr{Y}_\lambda({\phi}) \widetilde{\mathscr{Y}}_\lambda({\phi}+\epsilon_{12})}
    \nonumber \\
    & = \frac{1}{[-\epsilon_{1,2}]} \sum_{x \in \mathcal{X}_{\partial_+ \lambda}^{[\log]}} \frac{\prod_{x' \in \mathcal{X}_{\partial_- \lambda}^{[\log]}} [x - x' - \epsilon_{12}][x' - x]}{\prod_{x' \in \mathcal{X}_{\partial_+ \lambda}^{[\log]}\backslash\{x\}} [x - x'][x' - x - \epsilon_{12}]}
%    \nonumber \\ &
    = \sum_{x \in \mathcal{X}_{\partial_+ \lambda}^{[\log]}} \frac{-1}{\mathscr{Y}_\lambda(x) \widetilde{\mathscr{Y}}_{\lambda'}(x+\epsilon_{12})}
    \, ,
\end{align}
where the configuration $\lambda'$ is obtained by shift (adding an instanton). 
%$\mathcal{X}_{\lambda'}^{[\log]} = (\mathcal{X}_\lambda^{[\log]} \backslash \{x\}) \bigsqcup \{x + \epsilon_2\}$.
The last equality is shown as follows,
\begin{align}
    \frac{1}{[-\epsilon_{1,2}]}
    \frac{\prod_{x' \in \mathcal{X}_{\partial_- \lambda}^{[\log]}} [x - x' - \epsilon_{12}][x' - x]}{\prod_{x' \in \mathcal{X}_{\partial_+ \lambda}^{[\log]}\backslash\{x\}} [x - x'][x' - x - \epsilon_{12}]}
    & = 
    \frac{1}{\mathscr{Y}_\lambda(x) \widetilde{\mathscr{Y}}_{\lambda}(x+\epsilon_{12})} \times \lim_{x'' \to x} \frac{[x - x''][x'' - x - \epsilon_{12}]}{[-\epsilon_{1,2}]}
    \nonumber \\
    & = 
    \frac{1}{\mathscr{Y}_\lambda(x) \widetilde{\mathscr{Y}}_{\lambda}(x+\epsilon_{12})}
    \nonumber \\
    & \qquad \times \lim_{x'' \to x} \frac{[x - x''][x'' - x - \epsilon_{12}][x'' - x - \epsilon_{1,2}]}{[-\epsilon_{1,2}][x'' - x][x'' - x - \epsilon_{12}]}
    \nonumber \\
    & = \frac{-1}{\mathscr{Y}_\lambda(x) \widetilde{\mathscr{Y}}_{\lambda'}(x+\epsilon_{12})}
    \, .
\end{align}
\if0
\begin{align}
    \prod_{x' \in \mathcal{X}_\lambda^{[\log]}\backslash\{x\}} \frac{[x - x' - \epsilon_1][x' - x - \epsilon_2]}{[x - x'][x' - x - \epsilon_{12}]} 
    & = \frac{1}{\mathscr{Y}_\lambda(x) \widetilde{\mathscr{Y}}_{\lambda}(x+\epsilon_{12})} \times \lim_{x'' \to x} \frac{[x - x''][x'' - x - \epsilon_{12}]}{[x -x'' - \epsilon_1][x'' - x - \epsilon_{2}]}
    \nonumber \\
    & = \frac{1}{\mathscr{Y}_\lambda(x) \widetilde{\mathscr{Y}}_{\lambda'}(x+\epsilon_{12})} \times \lim_{x'' \to x} \frac{[x - x''][x'' - x - \epsilon_{1}]}{[x -x'' - \epsilon_1][x'' - x]}
    \nonumber \\
    & = \frac{-1}{\mathscr{Y}_\lambda(x) \widetilde{\mathscr{Y}}_{\lambda'}(x+\epsilon_{12})}
    \, .
\end{align}
Although there exist formally infinitely many poles in the integrand, we have poles at the place where we can add a box to the partition $\lambda$.
\fi
From this expression, we identify $Z_{\lambda'}/Z_\lambda = -1/\mathscr{Y}_\lambda(x) \widetilde{\mathscr{Y}}_{\lambda'}(x+\epsilon_{12})$, from which we obtain
\begin{align}
    \res_{x' \to x} \left[ Z_{\lambda'} \widetilde{\mathscr{Y}}_{\lambda'}(x'+\epsilon_{12}) + Z_\lambda \frac{1}{{\mathscr{Y}}_{\lambda}(x')} \right] = 0
    \, .
\end{align}
Summing up all the configurations, we see that 
\begin{align}
    \mathscr{T}(x) := \left< \widetilde{\mathscr{Y}}(x+\epsilon_{12}) + \frac{\mathfrak{q}}{\mathscr{Y}(x)} \right>
\end{align}
is a pole-free regular function in $x$, where we define the gauge theory average of the observable by $\left< \mathcal{O} \right> = \frac{1}{Z} \sum_{\lambda \in \mathfrak{M}^\mathsf{T}} Z_\lambda \mathcal{O}_\lambda$.
The combination of the $\mathscr{Y}$-functions yielding a pole-free function is called the (average of) $qq$-character (of $A_1$ quiver in this case)~\cite{Nekrasov:2015wsu}.%
\footnote{%
The character of the fundamental representation of SL(2) is given by $y + y^{-1}$.
In general, the Seiberg--Witten geometry of quiver gauge theory is described using the fundamental representation characters of the Lie group associated with the quiver structure~\cite{Nekrasov:2012xe}.
Considering the situation $(\epsilon_1, \epsilon_2) = (\hbar,0)$, this character is promoted to the $q$-character~\cite{Nekrasov:2013xda} of Yangian/quantum affine algebra~\cite{Knight:1995JA,Frenkel:1998}.
The $qq$-character is a double quantum deformation of the character appearing in the presence of generic $\Omega$-background parameters.
Physically, the $qq$-character is realized as a codimension-four defect operator~\cite{Kim:2016qqs,Agarwal:2018tso,Haouzi:2019jzk}.
}
Indeed, the Seiberg--Witten curve is obtained in the classical limit $\epsilon_{1,2} \to 0$ of this relation, $y + \mathfrak{q}/y = \det(x - \phi)$ by identifying the characteristic polynomial by $\mathscr{T}(x)$.%
\footnote{%
Precisely speaking, we should convert $\widetilde{\mathscr{Y}}$ to $\mathscr{Y}$ using the relation~\eqref{eq:Y-fn_convert} providing an additional polynomial factor, which can be interpreted as the Chern--Simons term contribution in five dimensions.
}
We remark that in this limit, we can apply the saddle point analysis, so that the observable average can be replaced by the on-shell value with the saddle point configuration $\lambda_*$, $\left< \mathscr{Y}(x) \right> \to \mathscr{Y}_{\lambda_*}(x) = y(x)$ and $\left< 1/ \mathscr{Y}(x) \right> \to 1/\mathscr{Y}_{\lambda_*}(x) = 1/y(x)$. 

We may consider the removing process in a similar way.
In this case, we consider the shift of the form, $K \to K - V$.
The tangent bundle behaves as
\begin{align}
    \delta_{-V} T_\lambda \mathfrak{M}_{n,k} 
    & = - \mathsf{Y}^\vee V - Q_{12}^\vee V^\vee \mathsf{Y} - P_{12}^{-1} V^\vee V
    \, ,
\end{align}
where $\delta_{-V}$ : $\mathfrak{M}_{n,k} \to \mathfrak{M}_{n,k-v}$, and the index functor yields
\begin{align}
    \delta_{-V} Z_\lambda = \mathbb{I}[-\delta_{-V} T_\lambda \mathfrak{M}_{n,k}] = \frac{Z_{\lambda-v}}{Z_\lambda} = \frac{1}{v!} \frac{[-\epsilon_{12}]^v}{[-\epsilon_{1,2}]^v} \oint \dd{\underline{\phi}} \mathscr{Y}_\lambda(\underline{\phi}) \widetilde{\mathscr{Y}}_\lambda(\underline{\phi}+\epsilon_{12}) \prod_{I \neq J}^v \mathscr{S}(\phi_{IJ})^{-1}
    \, .
\end{align}
In the case of $v = 1$, we obtain
\begin{align}
    \delta_{-V} Z_\lambda & = \frac{[-\epsilon_{12}]}{[-\epsilon_{1,2}]} \oint \frac{\dd{\phi}}{2 \pi \ii} \mathscr{Y}_\lambda({\phi}) \widetilde{\mathscr{Y}}_\lambda({\phi}+\epsilon_{12})
    = \sum_{x \in \mathcal{X}_{\partial_- \lambda}^{[\log]}} - \mathscr{Y}_{\lambda'}(x) \widetilde{\mathscr{Y}}_{\lambda}(x+\epsilon_{12})
    \, ,
\end{align}
where we denote the configuration obtained by removing an instanton by $\lambda'$.
We can similarly obtain the $qq$-character from this expression as well.

\subsubsection{Supergroup analysis}

We consider the adding/removing instanton process for supergroup gauge theory.
Denoting the observable sheaf $\mathsf{Y} = \mathsf{Y}_0 \oplus \mathsf{Y}_1$, the character of even and odd part is given as follows,
\begin{align}
    \operatorname{ch} \mathsf{Y}_0\Big|_{\lambda} = 
    \sum_{x \in \mathcal{X}_{\partial_+ \lambda^0}} x - \sum_{x \in \mathcal{X}_{\partial_- \lambda^0}} x q_{12} 
    \, , \qquad
    \operatorname{ch} \mathsf{Y}_1\Big|_{\lambda} = 
    \sum_{x \in \mathcal{X}_{\partial_+ \lambda^1}} x q_{12} - \sum_{x \in \mathcal{X}_{\partial_- \lambda^1}} x 
\end{align}
where we define
\begin{align}
    \mathcal{X}_{\partial_\pm \lambda^0} = \{ \ee^{a_\alpha^0} q_1^{i-1} q_2^{j-1} \mid (i,j) \in \partial_\pm \lambda^0 \}
    \, , \qquad
    \mathcal{X}_{\partial_\pm \lambda^1} = \{ \ee^{a_\alpha^1} q_1^{-i} q_2^{-j} \mid (i,j) \in \partial_\pm \lambda^1 \}
    \, .
\end{align}
The $\mathscr{Y}$-function is then defined as follows,
\begin{align}
    \mathscr{Y}_\lambda(z) = \frac{\mathscr{Y}_{0,\lambda^0}(z)}{\mathscr{Y}_{1,\lambda^1}(z)}
    \, , \qquad
    \widetilde{\mathscr{Y}}_\lambda(z) = \frac{\widetilde{\mathscr{Y}}_{0,\lambda^0}(z)}{\widetilde{\mathscr{Y}}_{1,\lambda^1}(z)}
    \, ,
\end{align}
where each factor is given by
\begin{subequations}
\begin{align}
    \mathscr{Y}_{0,\lambda^0}(z) = \frac{\prod_{x \in \mathcal{X}_{\partial_+\lambda^0}^{[\log]}} [z - x]}{\prod_{x \in \mathcal{X}_{\partial_-\lambda^0}^{[\log]}} [z - x - \epsilon_{12}]}
    \, , \qquad &
    \mathscr{Y}_{1,\lambda^1}(z) = \frac{\prod_{x \in \mathcal{X}_{\partial_+\lambda^1}^{[\log]}} [z - x - \epsilon_{12}]}{\prod_{x \in \mathcal{X}_{\partial_-\lambda^1}^{[\log]}} [z - x]}
    \, , \\
    \widetilde{\mathscr{Y}}_{0,\lambda^0}(z) = \frac{\prod_{x \in \mathcal{X}_{\partial_+\lambda^0}^{[\log]}} [x - z]}{\prod_{x \in \mathcal{X}_{\partial_-\lambda^0}^{[\log]}} [x - z + \epsilon_{12}]}
    \, , \qquad &
    \widetilde{\mathscr{Y}}_{1,\lambda^1}(z) = \frac{\prod_{x \in \mathcal{X}_{\partial_+\lambda^1}^{[\log]}} [x - z + \epsilon_{12}]}{\prod_{x \in \mathcal{X}_{\partial_-\lambda^1}^{[\log]}} [x - z]}
    \, .
\end{align}
\end{subequations}
Under the shift $K \to K + V$ by $V = \mathbb{C}^{v_0|v_1}$, we have the same expression for the tangent bundle as before~\eqref{eq:tangent_bundle_shift+}.
The resulting contour integral is given as follows,
\begin{align}
    \delta_V Z_\lambda = \frac{1}{v_{0,1}!} \frac{[-\epsilon_{12}]^{v_{01}}}{[-\epsilon_{1,2}]^{v_{01}}} \oint \dd{\underline{\phi}} \frac{1}{\mathscr{Y}_\lambda(\underline{\phi}) \widetilde{\mathscr{Y}}_\lambda(\underline{\phi}+\epsilon_{12})} \frac{\prod^{I \in [v_0]}_{J \in [v_1]} \mathscr{S}(\phi^0_I - \phi^1_J) \mathscr{S}(\phi^1_J - \phi^0_I) }{\prod_{I \neq J}^{v_{0}} \mathscr{S}(\phi^0_{IJ}) \prod_{I \neq J}^{v_{1}} \mathscr{S}(\phi^1_{IJ})}
    \, .
\end{align}
The one-dimensional cases $v = (1|0)$ and $(0|1)$ are simultaneously formulated by
\begin{align}
    \delta_V Z_\lambda & = \frac{[-\epsilon_{12}]}{[-\epsilon_{1,2}]} \oint \frac{\dd{\phi}}{2 \pi \ii} \frac{1}{\mathscr{Y}_\lambda({\phi}) \widetilde{\mathscr{Y}}_\lambda({\phi}+\epsilon_{12})}
    = \frac{[-\epsilon_{12}]}{[-\epsilon_{1,2}]} \oint \frac{\dd{\phi}}{2 \pi \ii} \frac{\mathscr{Y}_{1,\lambda^1}({\phi}) \widetilde{\mathscr{Y}}_{1,\lambda^1}({\phi}+\epsilon_{12})}{\mathscr{Y}_{0,\lambda^0}({\phi}) \widetilde{\mathscr{Y}}_{0,\lambda^0}({\phi}+\epsilon_{12})}
    \nonumber \\ &
    = \frac{[-\epsilon_{12}]}{[-\epsilon_{1,2}]} \oint \frac{\dd{\phi}}{2 \pi \ii} \frac{\prod_{x \in \mathcal{X}_{\partial_-\lambda^0}^{[\log]}} [\phi - x - \epsilon_{12}][x - \phi]}{\prod_{x \in \mathcal{X}_{\partial_+\lambda^0}^{[\log]}} [\phi - x][x - \phi - \epsilon_{12}]} 
    \frac{\prod_{x \in \mathcal{X}_{\partial_+\lambda^1}^{[\log]}} [\phi - x - \epsilon_{12}][x - \phi]}{\prod_{x \in \mathcal{X}_{\partial_-\lambda^1}^{[\log]}} [\phi - x][x - \phi - \epsilon_{12}]}
    \, .
\end{align}
From this expression, we observe that adding an instanton to the positive node (positive instanton) is equivalent to removing an instanton from the negative node (negative instanton).
Similarly, removing a positive instanton is equivalent to adding a negative instanton from this point of view.
Therefore, we obtain the same $qq$-character expression in terms of the total $\mathscr{Y}$-functions,
\begin{align}
    \mathscr{T}(x) 
    = \left< \widetilde{\mathscr{Y}}(x+\epsilon_{12}) + \frac{\mathfrak{q}}{\mathscr{Y}(x)} \right>
    = \left< \frac{\widetilde{\mathscr{Y}}_0(x+\epsilon_{12})}{\widetilde{\mathscr{Y}}_1(x+\epsilon_{12})} + \mathfrak{q} \frac{\mathscr{Y}_1(x)}{\mathscr{Y}_0(x)} \right>
    \, ,
\end{align}
which is again consistent with $\widehat{A}_1$ quiver realization in Sec.~\ref{sec:quiver_realization} by identifying $\mathscr{Y}_{0,1}(x)$ with the $\mathscr{Y}$-functions of $\widehat{A}_1$ quiver.
Identifying the $\mathscr{T}$-function with the supercharacteristic function of the adjoint scalar, $\mathscr{T}(x) = \sdet(x - \Phi)$, we reproduce the Seiberg--Witten curve for $\mathrm{U}(n_0|n_1)$ theory discussed in Sec.~\ref{sec:SW_theory} in the classical limit $\epsilon_{1,2} \to 0$.

\subsubsection{Geometry of $qq$-character}

Let us comment on geometric representation theoretical perspectives of the $qq$-character.
We reconsider the shift $K \to K + V$ in the presence of the $\mathscr{Y}$-functions,
\begin{align}
    \delta_V \left( \mathsf{Y}^\vee W - T\mathfrak{M}_{n,k} \right)
    & = 
    - \mathsf{Y}^\vee V - Q_{12}^\vee V^\vee \mathsf{Y} - P_{12}^\vee V^\vee W + P_{12}^\vee V^\vee V
    \, ,
\end{align}
where we define $W = \mathbb{C}^w$ with the character $\operatorname{ch} W = \sum_{\alpha \in [w]} \ee^{\xi_\alpha}$, such that $\mathbb{I}[\mathsf{Y}^\vee W] = \prod_{\alpha \in [w]} \mathscr{Y}(\xi_\alpha)$.
The generic $qq$-character is then obtained by the following index formula,
\begin{align}
    \mathsf{T}_w = \sum_{v = 0}^\infty \mathfrak{q}^v \, \mathsf{T}_{w,v} 
    \, ,
\end{align}
where each contribution is given by
\begin{align}
    \mathsf{T}_{w,v} 
    & = \mathbb{I}[ \mathsf{Y}^\vee (W - (1 + Q_{12}^\vee) V) - P_{12}^\vee V^\vee W + P_{12}^\vee V^\vee V ]
    \nonumber \\
    & = \frac{1}{v!} \frac{[-\epsilon_{12}]^v}{[-\epsilon_{1,2}]^v} \oint \dd{\underline{\phi}} \frac{\mathscr{Y}(\underline{\xi})}{\mathscr{Y}(\underline{\phi}) \mathscr{Y}(\underline{\phi}-\epsilon_{12})} %\frac{\prod_{\alpha \in [w]} \mathscr{Y}(\xi_\alpha)}{\prod_{I \in [v]} \mathscr{Y}(\phi_I) \mathscr{Y}(\phi_I+\epsilon_{12})} 
    \mathscr{S}(\underline{\xi} - \underline{\phi}) \prod_{I \neq J}^v \mathscr{S}(\phi_{IJ})^{-1}
    \, .
    \label{eq:T_wv_A1}
\end{align}
We denote $\mathscr{Y}(\underline{\xi}) = \prod_{\alpha \in [w]} \mathscr{Y}(\xi_\alpha)$, $\mathscr{S}(\underline{\xi} - \underline{\phi}) = \prod_{\alpha \in [w], I \in [v]} \mathscr{S}(\xi_\alpha - \phi_I)$, etc.
We remark that the dual function $\widetilde{\mathscr{Y}}$ is converted to $\mathscr{Y}$ for convenience.
Taking the pole at $\phi_I = \xi_\alpha$ in the contour integral, we obtain the result~\cite{Nekrasov:2015wsu,Kimura:2015rgi},
\begin{align}
    \mathsf{T}_{w,v} & = \sum_{\substack{\mathsf{I} \bigsqcup \mathsf{J} = [w] \\ |\mathsf{J}| = v}} \prod_{\alpha \in \mathsf{I}, \beta \in \mathsf{J}} \mathscr{S}(\xi_\alpha - \xi_\beta) \frac{\prod_{\alpha \in \mathsf{I}} \mathscr{Y}(\xi_\alpha)}{\prod_{\beta \in \mathsf{J}} \mathscr{Y}(\xi_\beta - \epsilon_{12})}
    \, .
\end{align}
There are ${w \choose v}$ contributions in $\mathsf{T}_{w,v}$ for given $(w,v)$, which correspond to the fixed points in the cotangent bundle of the Grassmannian $\mathrm{Gr}(v,w)$ given as a quiver variety of type $A_1$.
From the representation theoretical point of view, this corresponds to the degree-$w$ tensor product of the fundamental representation of quantum affine algebra $U_q(\widehat{\mathfrak{sl}_2})$ associated with $A_1$ quiver~\cite{Nakajima:1994nid,Nakajima:1998DM,Nakajima:1999JAMS}.
In order to obtain the $qq$-character of the irreducible representation, we need to specialize the parameters $(\ee^{\xi_\alpha})_{\alpha \in [w]} \to (\ee^\xi, \ee^\xi q_1, \ldots, \ee^\xi q_1^{w-1})$, known as the $q$-segment condition~\cite{Chari:1991CMP,Chari:1994pf}.
See also~\cite{Kimura:2022spi}.

The contribution to the $qq$-character $\mathsf{T}_{w,v}$ shown in~\eqref{eq:T_wv_A1} has more direct interpretation in terms of the quiver variety of type $A_1$, which we denote by $\mathfrak{M}_{w,v} = \{ (I,J) \in \operatorname{Hom}(W,V) \oplus \operatorname{Hom}(V,W) \mid IJ = 0 \} /\!\!/ \operatorname{GL}(V)$ with the dimension, $\operatorname{dim} \mathfrak{M}_{w,v} = 2 v(w - v)$.
Hence, it becomes empty when $v > w$. 
Applying the same argument to Sec.~\ref{sec:tangent_bundle}, the tangent bundle is given by $T\mathfrak{M}_{v,w} = W^\vee V + Q_{12}^\vee V^\vee W - (1 + Q_{12}^\vee) V^\vee V$.
Denoting $c = 1 + Q_{12}^\vee$ ($q$-Cartan matrix of type $A_1$, see the definition~\eqref{eq:full-q-Cartan}), we obtain a geometric formula in the five-dimensional convention~\cite{Nekrasov:2015wsu,KPfractional},
\begin{align}
    \mathsf{T}_{w,v} = q_2^{-\frac{1}{2} \operatorname{dim} \mathfrak{M}_{w,v}}\int_{\mathfrak{M}_{w,v}} \frac{\operatorname{ch} \wedge \mathsf{Y} W^\vee}{\operatorname{ch} \wedge \mathsf{Y} c^\vee V^\vee} \operatorname{ch} \wedge_{q_2} T^\vee\mathfrak{M}_{w,v} \operatorname{td} (T\mathfrak{M}_{w,v})
    \, .
    \label{eq:qq-ch_geom}
\end{align}
Moreover, we may rewrite this formula as follows,
\begin{align}
    \mathsf{T}_{w,v} = \int_{\mathfrak{M}_{w,v}} \frac{\operatorname{ch} \wedge \mathsf{Y} W^\vee}{\operatorname{ch} \wedge \mathsf{Y} c^\vee V^\vee} \widehat{X}_y (T\mathfrak{M}_{w,v})
    \, ,
    \label{eq:qq-ch_geom_nom}
\end{align}
where we define another genus, an analog of $\widehat{A}$ genus (see, e.g.,~\cite{Gottsche:2015AG}), by
\begin{align}
    \widehat{X}_y(\mathbf{X}) = \prod_{i \in [\operatorname{rk} \mathbf{X}]} \frac{x_i ( \ee^{\frac{x_i}{2}} y^{
    -\frac{1}{2}} - \ee^{-\frac{x_i}{2}} y^{\frac{1}{2}} )}{\ee^{\frac{x_i}{2}} - \ee^{-\frac{x_i}{2}}}
    \, .
\end{align}
We remark that this genus is reduced to the $L$-genus when we take $y^{\frac{1}{2}} = \ii$, up to the normalization $(-1)^{\frac{1}{2} \operatorname{rk} \mathbf{X}}$.
From this point of view, it is clear that, if there is no $\mathscr{Y}$-function insertion, the $qq$-character is reduced to the (normalized) $\chi_{q_2}$-genus of the quiver variety $\mathfrak{M}_{w,v}$, and thus the $q$-character (the limit $q_2 \to 1$ of $qq$-character) is reduced to the Euler characteristics. 
On the other hand, another deformation of the $q$-character, a.k.a., the $t$-analog of $q$-character, is the generating function of the Betti numbers (analogous to the Poincaré polynomial) of the fixed point set of the quiver variety under the $S^1$ action~\cite{Nakajima:2001PC,Nakajima:2004AM}, which is also interpreted as another kind of deformation of Euler characteristics.

\paragraph{Generic quiver}

We can apply this formalism to generic quiver gauge theory.
In this case, we start with the total tangent bundle~\eqref{eq:total_tangent_bundle}, and consider the shift $K_i \to K_i + V_i$ ($i \in \Gamma_0$) with generic $\mathscr{Y}$-function insertions,
\begin{align}
    \delta_V\left( \sum_{i \in \Gamma_0} \mathsf{Y}_i^\vee W_i - T\mathfrak{M}_{\underline{n},\underline{k}} \right)
    & = \sum_{i,j \in \Gamma_0} \left( - \mathsf{Y}_i^\vee c_{ij}^+ V_j - Q_{12}^\vee V_i^\vee c_{ij}^+ \mathsf{Y}_j + P_{12}^\vee V_i^\vee c_{ij}^+ V_j - P_{12}^\vee V_i^\vee \delta_{ij} W_j \right)
    \, ,
\end{align}
where $\operatorname{ch} W_i = \sum_{\alpha \in [w_i]} \ee^{\xi_{i,\alpha}}$, such that $\mathbb{I}\left[\sum_{i \in \Gamma_0} \mathsf{Y}_i^\vee W_i\right] = \prod_{i \in \Gamma_0} \prod_{\alpha \in [w_i]} \mathscr{Y}_i(\xi_{i,\alpha})$.
Defining the full $q$-Cartan matrix from the half one~\eqref{eq:half-q-Cartan} as
\begin{align}
    c_{ij} = c_{ij}^+ + c_{ij}^- = (1 + Q_{12}^\vee) \delta_{ij} - \sum_{e:i \to j} \mathsf{M}_e - \sum_{e:j \to i} Q_{12}^\vee \mathsf{M}_e^\vee
    \, , \qquad 
    c_{ij}^- = Q_{12}^\vee c_{ji}^{+\vee}
    \, ,
    \label{eq:full-q-Cartan}
\end{align}
and converting $\widetilde{\mathscr{Y}}$ to $\mathscr{Y}$ as before, we have the contour integral form of the $qq$-character,
\begin{align}
    \mathsf{T}_{\underline{w}} = \sum_{\underline{v}} \underline{\mathfrak{q}}^{\underline{v}} \, \mathsf{T}_{\underline{w},\underline{v}}
    \, ,
\end{align}
where each contribution is given by
\begin{align}
    \mathsf{T}_{\underline{w},\underline{v}} & =
    \mathbb{I} \left[ \mathsf{Y}_i^\vee (W_i - c_{ij} V_j) - P_{12}^\vee V_i^\vee W_i + P_{12}^\vee V_i^\vee c_{ij}^+ V_j \right]
    \nonumber \\
    & = 
    \frac{1}{\underline{v}!} \frac{[-\epsilon_{12}]^v}{[-\epsilon_{1,2}]^v} \oint \dd{\underline{\phi}} \prod_{i \in \Gamma_0} \frac{\mathscr{Y}_i(\underline{\xi}_i)}{\mathscr{A}_i(\underline{\phi}_i)} \mathscr{S}(\underline{\xi}_i - \underline{\phi}_i) \prod_{I \neq J}^{v_i} \mathscr{S}(\phi_{i,IJ})^{-1} \prod_{e : i \to j} \mathscr{S}(\underline{\phi}_j - \underline{\phi}_i + m_e)
    % \frac{\underline{\mathscr{Y}}(\underline{\xi})}{\underline{\mathscr{A}}(\underline{\phi})}
\end{align}
and we define the $\mathscr{A}$-function for $\operatorname{ch} \mathsf{x} = \ee^x$,
\begin{align}
    \mathscr{A}_i (x) = \mathbb{I}\left[ \sum_{j \in \Gamma_0} \mathsf{Y}_j^\vee c_{ji} \mathsf{x} \right] = 
    \frac{\mathscr{Y}_i(x) \mathscr{Y}_i(x - \epsilon_{12})}{\prod_{e:j \to i} \mathscr{Y}_j(x + m_e) \prod_{e:i \to j} \mathscr{Y}_j(x - m_e - \epsilon_{12})}
    \, .
\end{align}
Therefore, the $\mathscr{Y}$-function and the $\mathscr{A}$-function are interpreted as (exponentiated) weight and root vectors, and their operator analogs are discussed in the construction of $q$-deformed W-algebras~\cite{Frenkel:1997,Kimura:2015rgi}.
We can obtain the geometric formulas, \eqref{eq:qq-ch_geom} and \eqref{eq:qq-ch_geom_nom}, for generic quiver by replacing the $q$-Cartan matrix for generic one~\eqref{eq:full-q-Cartan}~\cite{Nekrasov:2015wsu,KPfractional}.

\paragraph{Supergroup case}

The geometric formalism presented above is also applicable to supergroup case.
The representation of quiver $\Gamma$ consists of the vector spaces assigned to each node and the linear maps between them, and we denote the category of representations of quiver $\Gamma$ by $\operatorname{Rep}(\Gamma)$ (see, e.g.,~\cite{Kirillov:2016}).
Then, the \emph{super-representation of quiver} is similarly constructed by replacing the ordinary vector spaces with the supervector spaces~\cite{Thind:2010,Bovdi:2020IJAC}.
In this case, we obtain a contour integral formula for the $qq$-character similarly to the supergroup LMNS formula~\eqref{eq:LMNS_formula_supergroup}.

\subsection{Free field realization}\label{sec:quiver_W-algebra}

Similarly to the matrix model as discussed in Sec.~\ref{sec:free_field_realization}, the gauge theory partition function has a similar free field realization.
Let us discuss an operator formalism of gauge theory in this part.

\subsubsection{Holomorphic deformation}

In the context of four-dimensional $\mathcal{N}=2$ gauge theory, the holomorphic function, called the \emph{prepotential}, plays a central role to characterize the supersymmetric Lagrangian.
The ordinary SYM theory corresponds to the quadratic prepotential $\mathscr{F} = \tr \Phi^2$ at UV.
We now consider generic holomorphic deformation of the prepotential, $\mathscr{F} \to \mathscr{F} + \sum_{n=1}^\infty t_n \tr \Phi^n$.
Even after the deformation, we can still apply the localization computation, and the partition function is given as follows~\cite{Nakajima:2003uh,Marshakov:2006ii},
\begin{align}
    Z(t) = \sum_{\lambda \in \mathfrak{M}^\mathsf{T}} Z_\lambda(t)
    \, , \qquad
    Z_\lambda(t) = Z_\lambda Z^\text{pot}_\lambda(t)
\end{align}
where we define the potential term 
\begin{align}
    Z^\text{pot}_\lambda(t) = \exp \left( \sum_{n=1}^\infty t_n \mathcal{O}_{n,\lambda} \right)
    \, .
\end{align}
We denote the $\lambda$-fixed point contribution of the chiral ring operator by $\mathcal{O}_{n,\lambda} = \tr \Phi^n|_\lambda$ for 4d, and 5d and 6d analogues are obtained by the Wilson loop and the Wilson surface extending on the circle and the torus.

This $t$-deformed partition function is a generating function of the chiral ring operators,
\begin{align}
    \left< \mathcal{O}_n \right> = \pdv{}{t_n} \log Z(t)\Big|_{t \to 0}
    \, .
\end{align}
From this point of view, the $t$-dependent part plays a similar role to the potential function in the matrix model as discussed in Sec.~\ref{sec:free_field_realization}, and we see the Heisenberg algebra,
\begin{align}
    [\partial_n, t_m] = \delta_{n,m}
\end{align}
on the Fock space, $\mathsf{F} = \mathbb{C} \llbracket \partial_n, t_n \rrbracket \ket{0}$, such that the vacuum state is defined as $\partial_n \ket{0} = 0$ ($t$-constant).
The dual vacuum is then defined by $\bra{0} t_n = 0$.

\subsubsection{$Z$-state}

In the operator formalism, the deformation parameter behaves as an operator acting on the Fock space.
From this point of view, the $t$-extended partition function is also seen as an operator, and through the operator-state correspondence, we define the $Z$-state as follows,
\begin{align}
    \ket{Z} = Z(t) \ket{0}
    \, .
\end{align}
We also define $\ket{Z_\lambda} = Z_\lambda(t) \ket{0}$, such that the total $Z$-state is given by $\ket{Z} = \sum_{\lambda \in \mathfrak{M}^\mathsf{T}} \ket{Z_\lambda}$.
From this $Z$-state, the undeformed partition function can be obtained as follows,
\begin{align}
    Z = \braket{0}{Z} = \bra{0} Z(t) \ket{0}
    \, .
    \label{eq:Z-correlator}
\end{align}
Namely, the partition function is given as a correlator (of vertex operators as we will see soon) in this formalism.
Such a connection between the BPS observables and the formalism of CFT-like theory is called the BPS/CFT correspondence~\cite{Nekrasov:2004UA}, and the realization of the partition function as a correlator is interpreted as a consequence of such a correspondence.

We remark that the expression~\eqref{eq:Z-correlator} implies that the partition function is realized as a correlator on a sphere.
One can similarly consider a torus correlator (a character on the Fock module), and it turns out that such a torus correlator computes the six-dimensional partition function in this context~\cite{Kimura:2016dys}.
In the formalism of class $\mathcal{S}$ theory~\cite{Gaiotto:2009we}, on the other hand, the torus correlator is associated with the loop structure in quiver gauge theory, typically found in affine quiver gauge theory.
These two realizations are related to each other through the duality exchanging base and fiber of Seiberg--Witten geometry, and S-duality in type IIB string theory~\cite{Katz:1997eq,Aharony:1997bh}.

\subsubsection{Vertex operators}

Let us consider the vertex operator realization of the $Z$-state.
We focus on the five-dimensional convention.
See \cite{Nieri:2019mdl} and \cite{Kimura:2016dys} for four and six dimensional cases.
For this purpose, we define the screening currents,
\begin{align}
    S_\sigma(x) = {: \exp \left( s_0^\sigma \log x + \tilde{s}_0^\sigma + \sum_{n \in \mathbb{Z}_{\neq 0}} s_n^\sigma x^{-n} \right) :}
    \, , \qquad 
    \sigma = 0, 1
    \, ,
\end{align}
with the commutation relations,
\begin{subequations}
\begin{align}
    [s_n^0, s_m^0] = - \frac{1}{n} \frac{1 - q_1^n}{1 - q_2^{-n}} (1 + q_{12}^{-n}) \delta_{n+m,0}
    \, , \qquad &
    [s_n^1, s_m^1] = - \frac{1}{n} \frac{1 - q_2^n}{1 - q_1^{-n}} (1 + q_{12}^{-n}) \delta_{n+m,0}
    \, , \\
    [s_n^0, s_m^1] = \frac{1}{n} (q_1^n + q_2^{-n}) \delta_{n+m,0}
    \, , \qquad &
    [s_n^1, s_m^0] = \frac{1}{n} (q_1^{-n} + q_2^{n}) \delta_{n+m,0}
    \, , \\
    [\tilde{s}_0^0 , s_n^0] = - 2 b^2 \delta_{n,0}
    \, , \qquad
    [\tilde{s}_0^1 , s_n^1] = - 2 b^{-2} \delta_{n,0}
    \, , \qquad &
    [\tilde{s}_0^0 , s_n^1] = [\tilde{s}_0^1 , s_n^0] = 2 \delta_{n,0}
    \, ,
\end{align}
\end{subequations}
with $b^2 = - \epsilon_1/\epsilon_2$.
This parametrization is taken to be compatible with the matrix model notation~\eqref{eq:Omega_background_matrix_parameter}.
For $|q_1| < 1$, $|q_2| > 1$, we have the following OPEs,
\begin{subequations}
\begin{align}
    \frac{S_0(x) S_0(x')}{:S_0(x) S_0(x'):} & = 
    \frac{(x'/x;q_2^{-1})_\infty (q_{12}^{-1} x'/x;q_2^{-1})_\infty}{(q_1 x'/x;q_2^{-1})_\infty (q_{2}^{-1} x'/x;q_2^{-1})_\infty}
    x^{\prime -2 b^2} 
    \, , \\
    \frac{S_1(x) S_1(x')}{:S_1(x) S_1(x'):} & = 
    \frac{(x'/x;q_1)_\infty (q_{12} x'/x;q_1)_\infty}{(q_1 x'/x;q_1)_\infty (q_{2}^{-1} x'/x;q_1)_\infty}
    x^{\prime -2 b^{-2}}     
    \, , \\
    \frac{S_0(x) S_1(x')}{:S_0(x) S_1(x'):} & = 
    \frac{(-q_2 xx')}{(1 - q_1 x'/x)(1 - q_2 x/x')}
    \, .
\end{align}
\end{subequations}
Based on the screening current, we have the following realization for the $\lambda$-fixed point contribution for U($n$) gauge theory~\cite{Kimura:2015rgi},%
\footnote{%
In this case, we need infinitely many screening currents to construct five-dimensional gauge theory partition function.
A similar construction is known for three-dimensional gauge theory, which involves a finite number of the screening currents~\cite{Aganagic:2013tta,Aganagic:2014oia,Aganagic:2015cta,Nedelin:2016gwu}.
These theories are related through the Higgsing process.
See Sec.~\ref{sec:intersecting_defects}.
}
\begin{align}
    Z_\lambda(t) = \prod^\succ_{x \in \mathcal{X}_\lambda} S_0(x)
\end{align}
The product $\prod^\succ$ is the radial ordering product with the ordering in the set $\mathcal{X}_\lambda$, $\prod^\succ S_0(x) = S_0(x_1) S_0(x_2) \cdots$ for $|x_1| > |x_2| > \cdots$ with $|q_1| \ll |q_2^{-1}| < 1$ and $\ee^{a_\alpha} \sim 1$.
In order to discuss the total partition function obtained by summation over the fixed point contributions, we define the screening charge from the screening current,
\begin{align}
    Q_0(x) = \sum_{k \in \mathbb{Z}} S_0(x q_2^k)
    \, , \qquad
    Q_1(x) = \sum_{k \in \mathbb{Z}} S_1(x q_1^k)
    \, .
\end{align}
Then, the $Z$-state is constructed from the screening charge,%
\footnote{%
We also have the free field realization of the contour integral formula based on similar vertex operators.
See \cite{Kimura:2019hnw,Kimura:2022zsm} for details.
}
\begin{align}
    \ket{Z} = Z(t) \ket{0}
    \, , \qquad
    Z(t) = \prod^\succ_{x \in \mathcal{X}_\emptyset} Q_0(x)
    \, .
\end{align}
We remark that in the expansion of the screening charge product, there are contributions of pseudo-fixed points, $\lambda \in \mathfrak{M}^\mathbb{Z}\backslash \mathfrak{M}^\mathsf{T}$, where $\mathfrak{M}^\mathbb{Z} = \{ \mathbb{Z} \ni \lambda_{\alpha,i} \xrightarrow{i \gg 1} 0, \alpha \in [n], i \in \mathbb{N} \}$.
However, we have $Z_\lambda(t) = 0$ for $\lambda \in \mathfrak{M}^\mathbb{Z}\backslash \mathfrak{M}^\mathsf{T}$, and hence it does not contribute to the partition function.

We have a similar realization of supergroup gauge theory partition function.
In this case, we may use both the screening currents to obtain the $Z$-state~\cite{Kimura:2021ngu,Noshita:2022dxv}.
\begin{itembox}{$Z$-state for supergroup gauge theory}
The $Z$-state for supergroup gauge theory is constructed by two types of screening charges,
\begin{align}
    \ket{Z} = \prod^\succ_{x \in \mathcal{X}_\emptyset^0} Q_0(x) \prod^\succ_{x \in \check{\mathcal{X}}_\emptyset^1} Q_1(x) \ket{0}
    \, .
\end{align}
\end{itembox}
In fact, this expression is analogous to the operator formalism of the supermatrix model discussed in Sec.~\ref{sec:free_field_realization}.

\subsubsection{$q$-Virasoro constraint}

Similarly to the Virasoro algebra discussed in Sec.~\ref{sec:free_field_realization}, one can construct an operator that commutes with the screening charges in this case as well.
We define the vertex operator, called the $\mathsf{Y}$-operator, which yields the $\mathscr{Y}$-function in gauge theory as follows,
\begin{align}
    \mathscr{Y}(x) = \bra{0} \mathsf{Y}(x) \ket{Z}
    \, , \qquad
    \widetilde{\mathscr{Y}}(x) = \bra{Z} \mathsf{Y}(x) \ket{0}
    \, .
\end{align}
Then, we define the $\mathsf{T}$-operator using the $\mathsf{Y}$-operator, 
\begin{align}
    \mathsf{T}(x) = \mathsf{Y}(x) + \mathsf{Y}^{-1}(x q_{12}^{-1})
    \, .
\end{align}
Namely, this is an operator analog of the $qq$-character introduced in Sec.~\ref{sec:SD_eq}.
We see that this $\mathsf{T}$-operator commutes with the screening charges, $[\mathsf{T}(x), Q_\sigma(x')] = 0$.
Having the mode expansion, $\mathsf{T}(x) = \sum_{n \in \mathbb{Z}} T_n x^{-n}$, they obey the following algebraic relation,
\begin{align}
    [T_n, T_{m}] = - \sum_{k=1}^\infty f_k ( T_{n-k} T_{m+k} - T_{m-k} T_{n+k} ) - \frac{(1 - q_1^n)(1 - q_2^n)}{1 - q_{12}} (q_{12}^n - q_{12}^{-n}) \delta_{n+m,0}
    \, ,
    \label{eq:qVirasoro_rel}
\end{align}
where the coefficients $\{ f_k \}_{k \in \mathbb{N}}$ are defined from the structure function,
\begin{align}
    f(x) = \exp\left( \sum_{n = 1}^\infty \frac{(1 - q_1)(1 - q_2)}{n (1 + q_{12}^n)} x^n \right) = \sum_{k = 0}^\infty f_k x^k
    \, .
\end{align}
This structure function comes from the OPE between the $\mathsf{Y}$-operators, $\mathsf{Y}(x) \mathsf{Y}(x')/{:\mathsf{Y}(x) \mathsf{Y}(x'):} = f^{-1}(x'/x)$.
The algebraic relation~\eqref{eq:qVirasoro_rel} is equivalent to the following relation for the $\mathsf{T}$-operator,
\begin{align}
    f\left(\frac{x'}{x}\right) \mathsf{T}(x) \mathsf{T}(x') - f\left(\frac{x}{x'}\right) \mathsf{T}(x') \mathsf{T}(x) = - \frac{(1 - q_1)(1 - q_2)}{1 - q_{12}} \left( \delta\left(\frac{x'}{x} q_{12}\right) - \delta\left(\frac{x}{x'} q_{12}\right) \right)
    \, ,
    \label{eq:fTT_rel}
\end{align}
where we define the multiplicative $\delta$-function, $\delta(z) = \sum_{n \in \mathbb{Z}} z^n$.
This is called the \emph{quadratic relation} (or the $fTT$ relation) for the generating current of the $q$-deformed Virasoro algebra constructed in~\cite{Shiraishi:1995rp}.
We have remarks on this relation.
Taking the limit $q_1 \to 1$ or $q_2 \to 1$, the quadratic relation is simply reduced to $[\mathsf{T}(x), \mathsf{T}(x')] = 0$.
Hence, the $\mathsf{T}$-operator becomes a commuting operator, which is then identified with the $q$-character, and also with the transfer matrix of the corresponding quantum integrable system.
From this point of view, we have a connection with quantum integrable system in this limit.
See Sec.~\ref{sec:Bethe/gauge} for a related discussion.
Another remark is that, interpreting the $\mathsf{T}$-operator as the $qq$-character, the quadratic relation corresponds to the anti-symmetric tensor product of the fundamental representation of $A_1$ algebra, namely $\mathfrak{sl}(2)$. 
Therefore, the right hand side of~\eqref{eq:fTT_rel} is interpreted as a contribution of the trivial representation.
We can see more structures in generic tensor product.
See \cite{Kimura:2022spi} for more details.

From this point of view, the gauge theory average of $qq$-character is identified with the $\mathsf{T}$-operator correlator, $\mathscr{T}(x) = \bra{0} \mathsf{T}(x) \ket{Z}$.
In order to obtain a degree-$n$ polynomial average, we need a modification of the vacuum state.
We introduce the degree-$n$ vacuum, such that $\mathsf{T}_m \ket{n} = 0$ for $m>n$, and define the modified $Z$-state, $\ket{Z} = Z(t) \ket{n}$.
Recalling that the $\mathsf{T}$-operator commutes with the screening charges (hence with $Z(t)$), the $qq$-character average, $\mathscr{T}(x) = \bra{0} \mathsf{T}(x) \ket{Z}$, becomes a degree-$n$ polynomial in $x$.
This is an analogue of the Virasoro constraint that we call the $q$-Virasoro constraint.
See also \cite{Kanno:2012hk,Kanno:2013aha,Nedelin:2015mio,Nedelin:2016gwu,Lodin:2018lbz} for related discussions.
From this point of view, we obtain the same $q$-Virasoro constraint for supergroup gauge theory since the $\mathsf{T}$-operator commutes with the both screening charges, $Q_\sigma(x)$.
Considering generic quiver gauge theory, we will obtain the constraint with quiver W-algebra.
See~\cite{Kimura:2015rgi} for details.

\subsection{Bethe/gauge correspondence}\label{sec:Bethe/gauge}

Although we have focused on gauge theory on $\mathbb{C}^2$, we may discuss its implication to further low dimensional theory.
Taking the limit $\epsilon_2 \to 0$, while keeping $\epsilon_1$ finite, a.k.a., \emph{Nekrasov--Shatashvili (NS) limit}, the partition function asymptotically behaves as $Z \approx \exp \left( \frac{1}{\epsilon_2} \widetilde{\mathscr{W}} + \cdots \right)$, where $\widetilde{\mathscr{W}}$ is identified with the twisted superpotential of the corresponding two-dimensional $\mathcal{N}=(2,2)$ theory.
Then, the saddle point equation in this limit yields the twisted $F$-term condition,
\begin{align}
    \exp \left( \pdv{\widetilde{\mathscr{W}}}{\sigma} \right) = 1
    \, ,
\end{align}
where we denote the scalar field in the twisted chiral multiplet by $\sigma$.
For example, for four-dimensional U($n$) gauge theory with $2n$ flavors ($n$ fundamental and $n$ anti-fundamental hypermultiplets), the saddle point equation is given by
\begin{align}
    - \mathfrak{q} \frac{a(x) {d}(x + \epsilon_{12})}{\mathscr{Y}(x) {\mathscr{Y}}(x + \epsilon_{12})} = 1
    \, ,
\end{align}
where $a(x)$ and ${d}(x)$ are the matter polynomials, $a(x) = \prod_{f \in [n]} [x - m_f]$ and ${d}(x) = \prod_{f \in [n]} [x - \widetilde{m}_f]$.
%We similarly define $\widetilde{a}(x) = \prod_{f \in [n]} [m_f - x]$ and ${d}(x) = \prod_{f \in [n]} [x - \widetilde{m}_f]$.
Recalling the definition of the $\mathscr{Y}$-function~\eqref{eq:Y-func_def}, we write $\mathscr{Y}(x) = \mathscr{Q}(x) / \mathscr{Q}(x - \epsilon_1)$ with redefinition, $\mathscr{Y}(x) \to d(x) \mathscr{Y}(x)$.
Then, the saddle point equation is written in the following form in the limit $\epsilon_2 \to 0$,%
\footnote{%
In this case, the $\mathscr{Q}$-function involves infinitely many roots.
We need to impose the Higgsing condition to obtain a finite polynomial $\mathscr{Q}$-function.
See, e.g.,~\cite{Dorey:2011pa,Chen:2011sj} and also Sec.~\ref{sec:intersecting_defects} for a related discussion.
}
\begin{align}
    \frac{a(x)}{d(x)} = - \mathfrak{q} \frac{\mathscr{Q}(x + \epsilon_1)}{\mathscr{Q}(x - \epsilon_1)}
    \, ,
\end{align}
which can be identified with the Bethe equation of $\mathfrak{sl}(2)$-spin chain (choice of the index convention~\eqref{eq:index_notation} corresponds to XXX/XXZ/XYZ chain) of length $L = n$ and the roots of the $\mathscr{Q}$-function identified with the Bethe roots.
Denoting the mass parameter by $m_\alpha = \nu_\alpha + \ii s_\alpha$ and put $\widetilde{m}_\alpha = \bar{m}_\alpha$, the parameters $(\nu_\alpha,s_\alpha)$ are identified with the inhomogeneity and the spin of the site $\alpha \in [L]$.
In this context, the gauge coupling $\tau$ with $\mathfrak{q} = \exp(2 \pi \ii \tau)$ is identified with the twist boundary condition parameter.

Applying this formalism to supergroup gauge theory, we obtain the Bethe equation of the form,
\begin{align}
    \frac{a_0(x)}{a_1(x)} \frac{d_1(x)}{d_0(x)} = - \mathfrak{q} \frac{\mathscr{Q}_0(x + \epsilon_1)}{\mathscr{Q}_0(x - \epsilon_1)} \frac{\mathscr{Q}_1(x + \epsilon_1)}{\mathscr{Q}_1(x - \epsilon_1)}
    \, ,
\end{align}
which implies positive and negative magnons carrying positive and negative excitations and the sites involving positive and negative spins~\cite{Kimura:2019msw,Chen:2020rxu}.
We remark that the situation that we now consider is different from the spin chain with superalgebra symmetry.
In order to realize such a superspin chain, we need to consider quiver gauge theory that corresponds to the Dynkin diagram of the Lie superalgebra~\cite{Orlando:2010uu,Nekrasov:2018gne,Zenkevich:2018fzl,Ishtiaque:2021jan}.

The spin chain model is not a unique example to be discussed in this framework.
For example, it has been known that the Seiberg--Witten curve of $G$-SYM theory is given by the spectral curve of $\widehat{^LG}$-Toda chain~\cite{Martinec:1995by}.
In fact, one can obtain quantum Toda chain from pure SYM theory by imposing the codimension-two surface defect.
In the presence of such a defect, we need a modification of the instanton moduli space to the so-called (affine) Laumon space, which is equivalent to consider instantons on the partial orbifold space, $\mathbb{C} \times \mathbb{C}/\mathbb{Z}_n$~\cite{Feigin:2011SM,Finkelberg:2010QDZ,Kanno:2011fw}.
In this context, the $\Omega$-background parameter $\epsilon_1$ plays a role of the quantum parameter, and when $\epsilon_2$ is also finite, we obtain the non-stationary quantum integrable system, which is reduced to the stationary system in the NS limit, $\epsilon_2 \to 0$.
In this case, one can construct the quantum Toda Hamiltonian from the $qq$-character ($q$-character in the NS limit) in the presence of the surface defect.
Applying this formalism to supergroup gauge theory, we then obtain a super-Toda chain, which is associated with the root system of the corresponding Lie superalgebra~\cite{vanderLende:1994JMP} (In general, one can construct the integrable Toda chain from the root system data).
Further incorporating the hypermultiplet in the adjoint representation, namely $\mathcal{N} = 2^*$ theory, the Toda chain is promoted to the Calogero--Moser--Sutherland (CMS) system, and we obtain the ``double'' CMS system associated with the Lie superalgebra~\cite{Kerov:1998IMRN,Sergeev:2001JNMP,Sergeev:2002TMP} from supergroup gauge theory with the surface defect.
See~\cite{Chen:2020rxu} for details.

\subsection{Higgsing and intersecting defects}\label{sec:intersecting_defects}

As we discussed throughout this article, supergroup gauge theory inevitably violates the unitarity, and thus its physical realization seems not to be straightforward.
In this part, we would demonstrate that supergroup theory can be engineered from physical setups by imposing the defect operators.

In the context of supersymmetric gauge theory, there exist two branches in the moduli space of supersymmetric vacua, called the Coulomb and Higgs branches (In general, one can also consider the mixed branch).
Although Seiberg--Witten theory of four-dimensional $\mathcal{N}=2$ gauge theory describes the Coulomb branch, it has been known that there exists a root of Higgs branch locus, which is reached by tuning the fundamental mass parameter to the Coulomb moduli parameter as $m_f = a_\alpha$~\cite{Dorey:1998yh,Dorey:1999zk}.
In the presence of the $\Omega$-background, this condition is ``quantized'' to be $m_f = a_\alpha + n_{\alpha,1} \epsilon_1 + n_{\alpha,2} \epsilon_2$, and under this condition, the site $(n_{\alpha,1} + 1, n_{\alpha,2} + 1)$ cannot be included by the $\alpha$-th partition, $(n_{\alpha,1} + 1, n_{\alpha,2} + 1) \not\in \lambda_\alpha$ (the pit condition~\cite{Bershtein:2018SM}).%
\footnote{%
This can be seen as follows:
The fundamental hypermultiplet contribution is given by \eqref{eq:fund_hyp_contribution}, whose instanton part reads $\mathsf{H}_\text{inst} = \mathsf{M}^\vee \mathsf{K}$.
At the fixed point $\lambda \in \mathfrak{M}^\mathsf{T}$, we have the character $\operatorname{ch} \mathsf{H}_\text{inst} = \sum_{\alpha \in [n]} \sum_{(i,j) \in \lambda_{\alpha}} q_1^{i-1-n_{\alpha,1}} q_2^{j-1-n_{\alpha,2}}$.
Hence, if $(n_{\alpha,1} + 1, n_{\alpha,2} + 1) \in \lambda_\alpha$, we have a zero mode, and the corresponding contribution to the partition function becomes zero.
}
Namely, we have a restriction on the instanton configuration.
In fact, such a restriction on a partition is also discussed in the context of representation theory of supergroup:
Partitions with the pit condition parameterize irreducible representations of U$(n_{\alpha,1}|n_{\alpha,2})$.
In particular, for $n_{\alpha,2} = 0$, partitions have at most $n_{\alpha,1}$ rows, which yield irreducible representations of U$(n_{\alpha,1})$ group.
From physical point of view, $n_{\alpha,1,2}$ are interpreted as the numbers of fluxes or vortex defects in $\mathbb{C}_1$ and $\mathbb{C}_2$ direction, respectively.
Hence, the situation with both $n_{\alpha,1,2}$ realizes intersecting defects crossing at the origin~\cite{Dimofte:2010tz,Bonelli:2011fq,Gomis:2016ljm,Pan:2016fbl,Gorsky:2017hro,Nieri:2017ntx}.
See also~\cite{Jeong:2021rll}.
Computing the partition function of this configuration, it is given in the form of (deformation of) supermatrix model, which implies that we have an emerging supergroup structure originated from physical (non-supergroup) theory by considering the intersecting defects.

A similar realization of supergroup theory can be discussed in a higher-dimensional setup.
Starting with an eight-dimensional setup on $\mathbb{C}^4$, called the gauge origami, we incorporate D-branes extended in $\mathbb{C}_1 \times \mathbb{C}_2$ and $\mathbb{C}_2 \times \mathbb{C}_3$~\cite{Nekrasov:2016ydq,Nekrasov:2017rqy}.
In this case, we obtain the non-stationary double CMS system at finite $\epsilon_2$, and the stationary one in the limit $\epsilon_2 \to 0$~\cite{Nekrasov:2017gzb,Chen:2019vvt}.
This implies that supergroup structure emerges from intersecting defects at the $\mathbb{C}_2$-plane.

\subsection*{Acknowledgement}

I would like to thank 
Nicolas Babinet,
Heng-Yu Chen,
Norton Lee,
Fabrizio Nieri,
Go Noshita,
Vasily Pestun,
Yuji Sugimoto
for collaborations and fruitful discussions on this subject.
I am in particular grateful to Go Noshita for careful reading of the manuscript and valuable comments.
This work was in part supported by ``Investissements d'Avenir'' program, Project ISITE-BFC (No.~ANR-15-IDEX-0003), EIPHI Graduate School (No. ANR-17-EURE-0002), and Bourgogne-Franche-Comté region.

%\appendix

%\section{Mathematical backgrounds}

%% Bibliography
%\bibliographystyle{ws-ijmpa}
\bibliographystyle{amsalpha_mod}
\bibliography{conf}

\newcommand{\etalchar}[1]{$^{#1}$}
\providecommand{\bysame}{\leavevmode\hbox to3em{\hrulefill}\thinspace}
\providecommand{\MR}{\relax\ifhmode\unskip\space\fi MR }
\providecommand{\MRhref}[2]{%
  \href{http://www.ams.org/mathscinet-getitem?mr=#1}{#2}
}
\providecommand{\href}[2]{#2}
\begin{thebibliography}{GLFMS17}

\bibitem[ABJ08]{Aharony:2008gk}
O.~Aharony, O.~Bergman, and D.~L. Jafferis, \emph{{Fractional M2-branes}},
  \href{https://doi.org/10.1088/1126-6708/2008/11/043}{JHEP \textbf{0811}
  (2008)}, 043, \href{https://arxiv.org/abs/0807.4924}{{\ttfamily
  arXiv:0807.4924 [hep-th]}}.

\bibitem[ABJM08]{Aharony:2008ug}
O.~Aharony, O.~Bergman, D.~L. Jafferis, and J.~Maldacena,
  \emph{{$\mathcal{N}=6$ superconformal Chern--Simons--matter theories,
  M2-branes and their gravity duals}},
  \href{https://doi.org/10.1088/1126-6708/2008/10/091}{JHEP \textbf{0810}
  (2008)}, 091, \href{https://arxiv.org/abs/0806.1218}{{\ttfamily
  arXiv:0806.1218 [hep-th]}}.

\bibitem[AFS12]{Awata:2011ce}
H.~Awata, B.~Feigin, and J.~Shiraishi, \emph{{Quantum Algebraic Approach to
  Refined Topological Vertex}},
  \href{https://doi.org/10.1007/JHEP03(2012)041}{JHEP \textbf{03} (2012)}, 041,
  \href{https://arxiv.org/abs/1112.6074}{{\ttfamily arXiv:1112.6074 [hep-th]}}.

\bibitem[AGIMZ92]{Alvarez-Gaume:1991vno}
L.~Alvarez-Gaume, H.~Itoyama, J.~L. Manes, and A.~Zadra, \emph{{Superloop
  equations and two-dimensional supergravity}},
  \href{https://doi.org/10.1142/S0217751X92002441}{Int. J. Mod. Phys. A
  \textbf{7} (1992)}, 5337--5368,
  \href{https://arxiv.org/abs/hep-th/9112018}{{\ttfamily
  arXiv:hep-th/9112018}}.

\bibitem[AGM91]{Alvarez-Gaume:1991ozb}
L.~Alvarez-Gaume and J.~L. Manes, \emph{{Supermatrix models}},
  \href{https://doi.org/10.1142/S0217732391002219}{Mod. Phys. Lett. A
  \textbf{6} (1991)}, 2039--2050.

\bibitem[AGPS18]{Aghaei:2018cbn}
N.~Aghaei, A.~M. Gainutdinov, M.~Pawelkiewicz, and V.~Schomerus,
  \emph{{Combinatorial Quantisation of $GL(1|1)$ Chern-Simons Theory I: The
  Torus}}, \href{https://arxiv.org/abs/1811.09123}{{\ttfamily arXiv:1811.09123
  [hep-th]}}.

\bibitem[AH15]{Aganagic:2015cta}
M.~Aganagic and N.~Haouzi, \emph{{ADE Little String Theory on a Riemann Surface
  (and Triality)}}, \href{https://arxiv.org/abs/1506.04183}{{\ttfamily
  arXiv:1506.04183 [hep-th]}}.

\bibitem[AHDM78]{Atiyah:1978ri}
M.~F. Atiyah, N.~J. Hitchin, V.~G. Drinfeld, and Y.~I. Manin,
  \emph{{Construction of instantons}},
  \href{https://doi.org/10.1016/0375-9601(78)90141-X}{Phys. Lett. \textbf{A65}
  (1978)}, 185--187.

\bibitem[AHK98]{Aharony:1997bh}
O.~Aharony, A.~Hanany, and B.~Kol, \emph{{Webs of $(p,q)$ five-branes,
  five-dimensional field theories and grid diagrams}},
  \href{https://doi.org/10.1088/1126-6708/1998/01/002}{JHEP \textbf{01}
  (1998)}, 002, \href{https://arxiv.org/abs/hep-th/9710116}{{\ttfamily
  arXiv:hep-th/9710116 [hep-th]}}.

\bibitem[AHKS13]{Aganagic:2013tta}
M.~Aganagic, N.~Haouzi, C.~Kozcaz, and S.~Shakirov, \emph{{Gauge/Liouville
  Triality}}, \href{https://arxiv.org/abs/1309.1687}{{\ttfamily arXiv:1309.1687
  [hep-th]}}.

\bibitem[AHS14]{Aganagic:2014oia}
M.~Aganagic, N.~Haouzi, and S.~Shakirov, \emph{{$A_n$-Triality}},
  \href{https://arxiv.org/abs/1403.3657}{{\ttfamily arXiv:1403.3657 [hep-th]}}.

\bibitem[AIO01]{Alishahiha:2000du}
M.~Alishahiha, H.~Ita, and Y.~Oz, \emph{{On superconnections and the tachyon
  effective action}},
  \href{https://doi.org/10.1016/S0370-2693(01)00175-7}{Phys. Lett. B
  \textbf{503} (2001)}, 181--188,
  \href{https://arxiv.org/abs/hep-th/0012222}{{\ttfamily
  arXiv:hep-th/0012222}}.

\bibitem[AK05]{Awata:2005fa}
H.~Awata and H.~Kanno, \emph{{Instanton counting, Macdonald functions and the
  moduli space of D-branes}},
  \href{https://doi.org/10.1088/1126-6708/2005/05/039}{JHEP \textbf{0505}
  (2005)}, 039, \href{https://arxiv.org/abs/hep-th/0502061}{{\ttfamily
  arXiv:hep-th/0502061 [hep-th]}}.

\bibitem[AKKS18]{Agarwal:2018tso}
P.~Agarwal, J.~Kim, S.~Kim, and A.~Sciarappa, \emph{{Wilson surfaces in
  M5-branes}}, \href{https://doi.org/10.1007/JHEP08(2018)119}{JHEP \textbf{08}
  (2018)}, 119, \href{https://arxiv.org/abs/1804.09932}{{\ttfamily
  arXiv:1804.09932 [hep-th]}}.

\bibitem[AKMT01]{Arnone:2000qd}
S.~Arnone, Y.~A. Kubyshin, T.~R. Morris, and J.~F. Tighe, \emph{{A Gauge
  invariant regulator for the ERG}},
  \href{https://doi.org/10.1142/S0217751X0100461X}{Int. J. Mod. Phys. A
  \textbf{16} (2001)}, 1989,
  \href{https://arxiv.org/abs/hep-th/0102054}{{\ttfamily
  arXiv:hep-th/0102054}}.

\bibitem[AKMV05]{Aganagic:2003db}
M.~Aganagic, A.~Klemm, M.~Mari{\~n}o, and C.~Vafa, \emph{{The Topological
  vertex}}, \href{https://doi.org/10.1007/s00220-004-1162-z}{Commun. Math.
  Phys. \textbf{254} (2005)}, 425--478,
  \href{https://arxiv.org/abs/hep-th/0305132}{{\ttfamily arXiv:hep-th/0305132
  [hep-th]}}.

\bibitem[Bab22]{Babinet:2022}
N.~Babinet, \emph{{Modèles de chaînes de matrices aléatoires Réduction et
  factorisation des modèles à deux matrices et supermatrices (Models of
  chains of random matrices Reduction and factorisation of two-matrix and
  supermatrix models)}}, Ph.D. thesis, Université Bourgogne Franche-Comté,
  2022.

\bibitem[BEHT15]{Benini:2013xpa}
F.~Benini, R.~Eager, K.~Hori, and Y.~Tachikawa, \emph{{Elliptic Genera of 2d
  ${\mathcal{N}}$ = 2 Gauge Theories}},
  \href{https://doi.org/10.1007/s00220-014-2210-y}{Commun. Math. Phys.
  \textbf{333} (2015)}, no.~3, 1241--1286,
  \href{https://arxiv.org/abs/1308.4896}{{\ttfamily arXiv:1308.4896 [hep-th]}}.

\bibitem[Ber87]{Berezin:1987}
F.~A. Berezin,
  \href{https://doi.org/10.1007/978-94-017-1963-6}{\emph{Introduction to
  superanalysis}}, Springer Netherlands, 1987.

\bibitem[BFM18]{Bershtein:2018SM}
M.~Bershtein, B.~Feigin, and G.~Merzon, \emph{Plane partitions with a
  {\textquotedblleft}pit{\textquotedblright}: generating functions and
  representation theory}, \href{https://doi.org/10.1007/s00029-018-0389-z}{Sel.
  Math. \textbf{24} (2018)}, 21--62,
  \href{https://arxiv.org/abs/1512.08779}{{\ttfamily arXiv:1512.08779
  [math.CO]}}.

\bibitem[BMRS92]{Bourdeau:1991hn}
M.~Bourdeau, E.~J. Mlawer, H.~Riggs, and H.~J. Schnitzer, \emph{{The
  Quasirational fusion structure of SU(m/n) Chern-Simons and W-Z-W theories}},
  \href{https://doi.org/10.1016/0550-3213(92)90322-3}{Nucl. Phys. B
  \textbf{372} (1992)}, 303--358.

\bibitem[BMZ16]{Bourgine:2015szm}
J.-E. Bourgine, Y.~Matsuo, and H.~Zhang, \emph{{Holomorphic field realization
  of SH$^{c}$ and quantum geometry of quiver gauge theories}},
  \href{https://doi.org/10.1007/JHEP04(2016)167}{JHEP \textbf{04} (2016)}, 167,
  \href{https://arxiv.org/abs/1512.02492}{{\ttfamily arXiv:1512.02492
  [hep-th]}}.

\bibitem[Bos24]{Bose:1924}
Bose, \emph{Plancks gesetz und lichtquantenhypothese},
  \href{https://doi.org/10.1007/bf01327326}{Z. Phys. \textbf{26} (1924)},
  no.~1, 178--181.

\bibitem[Bou19]{Bourgine:2018fjy}
J.-E. Bourgine, \emph{{Fiber-base duality from the algebraic perspective}},
  \href{https://doi.org/10.1007/JHEP03(2019)003}{JHEP \textbf{03} (2019)}, 003,
  \href{https://arxiv.org/abs/1810.00301}{{\ttfamily arXiv:1810.00301
  [hep-th]}}.

\bibitem[BS93]{Bouwknegt:1992wg}
P.~Bouwknegt and K.~Schoutens, \emph{{$\mathcal{W}$ symmetry in conformal field
  theory}}, \href{https://doi.org/10.1016/0370-1573(93)90111-P}{Phys. Rept.
  \textbf{223} (1993)}, 183--276,
  \href{https://arxiv.org/abs/hep-th/9210010}{{\ttfamily
  arXiv:hep-th/9210010}}.

\bibitem[BTZ12]{Bonelli:2011fq}
G.~Bonelli, A.~Tanzini, and J.~Zhao, \emph{{Vertices, Vortices \& Interacting
  Surface Operators}}, \href{https://doi.org/10.1007/JHEP06(2012)178}{JHEP
  \textbf{06} (2012)}, 178, \href{https://arxiv.org/abs/1102.0184}{{\ttfamily
  arXiv:1102.0184 [hep-th]}}.

\bibitem[BVW99]{Berkovits:1999im}
N.~Berkovits, C.~Vafa, and E.~Witten, \emph{{Conformal field theory of AdS
  background with Ramond-Ramond flux}},
  \href{https://doi.org/10.1088/1126-6708/1999/03/018}{JHEP \textbf{03}
  (1999)}, 018, \href{https://arxiv.org/abs/hep-th/9902098}{{\ttfamily
  arXiv:hep-th/9902098}}.

\bibitem[BZ20]{Bovdi:2020IJAC}
V.~A. Bovdi and A.~N. Zubkov, \emph{Super-representations of quivers and
  related polynomial semi-invariants},
  \href{https://doi.org/10.1142/S0218196720500241}{Int. J. Algebra Comput.
  \textbf{30} (2020)}, no.~04, 883--902,
  \href{https://arxiv.org/abs/1912.00627}{{\ttfamily arXiv:1912.00627
  [math.RT]}}.

\bibitem[CDHL11]{Chen:2011sj}
H.-Y. Chen, N.~Dorey, T.~J. Hollowood, and S.~Lee, \emph{{A New 2d/4d Duality
  via Integrability}}, \href{https://doi.org/10.1007/JHEP09(2011)040}{JHEP
  \textbf{09} (2011)}, 040, \href{https://arxiv.org/abs/1104.3021}{{\ttfamily
  arXiv:1104.3021 [hep-th]}}.

\bibitem[CDRS12]{Cristoforetti:2012su}
M.~Cristoforetti, F.~Di~Renzo, and L.~Scorzato, \emph{{New approach to the sign
  problem in quantum field theories: High density QCD on a Lefschetz thimble}},
  \href{https://doi.org/10.1103/PhysRevD.86.074506}{Phys. Rev. D \textbf{86}
  (2012)}, 074506, \href{https://arxiv.org/abs/1205.3996}{{\ttfamily
  arXiv:1205.3996 [hep-lat]}}.

\bibitem[CKL20a]{Chen:2019vvt}
H.-Y. Chen, T.~Kimura, and N.~Lee, \emph{{Quantum Elliptic Calogero-Moser
  Systems from Gauge Origami}},
  \href{https://doi.org/10.1007/JHEP02(2020)108}{JHEP \textbf{02} (2020)}, 108,
  \href{https://arxiv.org/abs/1908.04928}{{\ttfamily arXiv:1908.04928
  [hep-th]}}.

\bibitem[CKL20b]{Chen:2020rxu}
\bysame, \emph{{Quantum Integrable Systems from Supergroup Gauge Theories}},
  \href{https://doi.org/10.1007/JHEP09(2020)104}{JHEP \textbf{2009} (2020)},
  104, \href{https://arxiv.org/abs/2003.13514}{{\ttfamily arXiv:2003.13514
  [hep-th]}}.

\bibitem[CP91]{Chari:1991CMP}
V.~Chari and A.~Pressley, \emph{{Quantum affine algebras}},
  \href{https://doi.org/10.1007/bf02102063}{Commun. Math. Phys. \textbf{142}
  (1991)}, no.~2, 261--283.

\bibitem[CP95]{Chari:1994pf}
\bysame, \emph{{Quantum Affine Algebras and their Representations}},
  {Representations of Groups} (B.~N. Allison and G.~H. Cliff, eds.), CMS Conf.
  Proc., vol.~16, American Mathematical Society, 1995, pp.~59--78,
  \href{https://arxiv.org/abs/hep-th/9411145}{{\ttfamily hep-th/9411145
  [hep-th]}}.

\bibitem[CZ21]{Cassia:2021uly}
L.~Cassia and M.~Zabzine, \emph{{On refined Chern-Simons and refined ABJ matrix
  models}}, \href{https://doi.org/10.1007/s11005-022-01518-1}{Lett. Math. Phys.
  \textbf{112} (2021)}, 21, \href{https://arxiv.org/abs/2107.07525}{{\ttfamily
  arXiv:2107.07525 [hep-th]}}.

\bibitem[DE10]{Desrosiers:2009pz}
P.~Desrosiers and B.~Eynard, \emph{{Supermatrix models, loop equations, and
  duality}}, \href{https://doi.org/10.1063/1.3430564}{J. Math. Phys.
  \textbf{51} (2010)}, 123304,
  \href{https://arxiv.org/abs/0911.1762}{{\ttfamily arXiv:0911.1762
  [math-ph]}}.

\bibitem[Des12]{Desrosiers:2012S}
P.~Desrosiers, \emph{Hermite and laguerre symmetric functions associated with
  operators of calogero-moser-sutherland type},
  \href{https://doi.org/10.3842/sigma.2012.049}{SIGMA (2012)},
  \href{https://arxiv.org/abs/1103.4593}{{\ttfamily arXiv:1103.4593
  [math.QA]}}.

\bibitem[DeW92]{DeWitt:1992}
B.~DeWitt,
  \href{https://doi.org/10.1017/cbo9780511564000}{\emph{Supermanifolds}},
  Cambridge University Press, 1992.

\bibitem[DGH10]{Dimofte:2010tz}
T.~Dimofte, S.~Gukov, and L.~Hollands, \emph{{Vortex Counting and Lagrangian
  3-manifolds}}, \href{https://doi.org/10.1007/s11005-011-0531-8}{Lett. Math.
  Phys. \textbf{98} (2010)}, 225--287,
  \href{https://arxiv.org/abs/1006.0977}{{\ttfamily arXiv:1006.0977 [hep-th]}}.

\bibitem[DHJV18]{Dijkgraaf:2016lym}
R.~Dijkgraaf, B.~Heidenreich, P.~Jefferson, and C.~Vafa, \emph{{Negative
  Branes, Supergroups and the Signature of Spacetime}},
  \href{https://doi.org/10.1007/JHEP02(2018)050}{JHEP \textbf{02} (2018)}, 050,
  \href{https://arxiv.org/abs/1603.05665}{{\ttfamily arXiv:1603.05665
  [hep-th]}}.

\bibitem[DHKM02]{Dorey:2002ik}
N.~Dorey, T.~J. Hollowood, V.~V. Khoze, and M.~P. Mattis, \emph{{The Calculus
  of Many Instantons}},
  \href{https://doi.org/10.1016/S0370-1573(02)00301-0}{Phys. Rept. \textbf{371}
  (2002)}, 231--459, \href{https://arxiv.org/abs/hep-th/0206063}{{\ttfamily
  arXiv:hep-th/0206063 [hep-th]}}.

\bibitem[DHT99]{Dorey:1999zk}
N.~Dorey, T.~J. Hollowood, and D.~Tong, \emph{{The BPS spectra of gauge
  theories in two-dimensions and four-dimensions}},
  \href{https://doi.org/10.1088/1126-6708/1999/05/006}{JHEP \textbf{9905}
  (1999)}, 006, \href{https://arxiv.org/abs/hep-th/9902134}{{\ttfamily
  arXiv:hep-th/9902134 [hep-th]}}.

\bibitem[DI97]{Ding:1996mq}
J.~Ding and K.~Iohara, \emph{{Generalization and deformation of Drinfeld
  quantum affine algebras}},
  \href{https://doi.org/10.1023/A:1007341410987}{Lett. Math. Phys. \textbf{41}
  (1997)}, 181--193, \href{https://arxiv.org/abs/q-alg/9608002}{{\ttfamily
  arXiv:q-alg/9608002 [math.QA]}}.

\bibitem[Dir26]{Dirac:1926jz}
P.~A.~M. Dirac, \emph{{On the Theory of quantum mechanics}},
  \href{https://doi.org/10.1098/rspa.1926.0133}{Proc. Roy. Soc. Lond. A
  \textbf{112} (1926)}, 661--677.

\bibitem[DK97]{Donaldson:1997}
S.~K. Donaldson and P.~B. Kronheimer, \emph{{The Geometry of Four-Manifolds}},
  Oxford Univ. Press, 1997.

\bibitem[DLH11]{Dorey:2011pa}
N.~Dorey, S.~Lee, and T.~J. Hollowood, \emph{{Quantization of Integrable
  Systems and a 2d/4d Duality}},
  \href{https://doi.org/10.1007/JHEP10(2011)077}{JHEP \textbf{10} (2011)}, 077,
  \href{https://arxiv.org/abs/1103.5726}{{\ttfamily arXiv:1103.5726 [hep-th]}}.

\bibitem[Dor98]{Dorey:1998yh}
N.~Dorey, \emph{{The BPS spectra of two-dimensional supersymmetric gauge
  theories with twisted mass terms}},
  \href{https://doi.org/10.1088/1126-6708/1998/11/005}{JHEP \textbf{9811}
  (1998)}, 005, \href{https://arxiv.org/abs/hep-th/9806056}{{\ttfamily
  hep-th/9806056}}.

\bibitem[DT10]{Drukker:2009hy}
N.~Drukker and D.~Trancanelli, \emph{{A Supermatrix model for $\mathcal{N}=6$
  super Chern--Simons--matter theory}},
  \href{https://doi.org/10.1007/JHEP02(2010)058}{JHEP \textbf{1002} (2010)},
  058, \href{https://arxiv.org/abs/0912.3006}{{\ttfamily arXiv:0912.3006
  [hep-th]}}.

\bibitem[Efe96]{Efetov:1996}
K.~Efetov, \href{https://doi.org/10.1017/CBO9780511573057}{\emph{{Supersymmetry
  in Disorder and Chaos}}}, Cambridge Univ. Press, 1996.

\bibitem[EK03]{Eguchi:2003sj}
T.~Eguchi and H.~Kanno, \emph{{Topological strings and Nekrasov's formulas}},
  \href{https://doi.org/10.1088/1126-6708/2003/12/006}{JHEP \textbf{0312}
  (2003)}, 006, \href{https://arxiv.org/abs/hep-th/0310235}{{\ttfamily
  arXiv:hep-th/0310235 [hep-th]}}.

\bibitem[EKR15]{Eynard:2015aea}
B.~Eynard, T.~Kimura, and S.~Ribault, \emph{{Random matrices}},
  \href{https://arxiv.org/abs/1510.04430}{{\ttfamily arXiv:1510.04430
  [math-ph]}}.

\bibitem[Fer26]{Fermi:1999ncp}
E.~Fermi, \emph{{Sulla quantizzazione del gas perfetto monoatomico (On the
  Quantization of the Monoatomic Ideal Gas, translation by A. Zannoni)}}, Rend.
  Lincei \textbf{3} (1926), 145--149,
  \href{https://arxiv.org/abs/cond-mat/9912229}{{\ttfamily
  arXiv:cond-mat/9912229}}.

\bibitem[FFNR11]{Feigin:2011SM}
B.~Feigin, M.~Finkelberg, A.~Negut, and L.~Rybnikov, \emph{{Yangians and
  cohomology rings of Laumon spaces}},
  \href{https://doi.org/10.1007/s00029-011-0059-x}{Selecta Math. \textbf{17}
  (2011)}, 573--607, \href{https://arxiv.org/abs/0812.4656}{{\ttfamily
  arXiv:0812.4656 [math.AG]}}.

\bibitem[Fie39]{Fierz:1939}
M.~Fierz, \emph{{\"{U}ber die relativistische Theorie kr\"{a}ftefreier Teilchen
  mit beliebigem Spin}}, \href{https://doi.org/10.5169/SEALS-110930}{Helvetica
  Physica Acta \textbf{12} (1939)}, 3--37.

\bibitem[For10]{Forrester:2010}
P.~J. Forrester, \emph{Log-gases and random matrices}, Princeton University
  Press, Princeton, 2010.

\bibitem[FQS85]{Friedan:1984rv}
D.~Friedan, Z.-a. Qiu, and S.~H. Shenker, \emph{{Superconformal Invariance in
  Two-Dimensions and the Tricritical Ising Model}},
  \href{https://doi.org/10.1016/0370-2693(85)90819-6}{Phys. Lett. B
  \textbf{151} (1985)}, 37--43.

\bibitem[FR98]{Frenkel:1997}
E.~Frenkel and N.~Reshetikhin, \emph{{Deformations of {$\mathcal{W}$}-algebras
  associated to simple {L}ie algebras}}, Comm. Math. Phys. \textbf{197} (1998),
  1--32, \href{https://arxiv.org/abs/q-alg/9708006}{{\ttfamily q-alg/9708006
  [math.QA]}}.

\bibitem[FR99]{Frenkel:1998}
\bysame, \emph{{The {$q$}-characters of representations of quantum affine
  algebras and deformations of {$\mathcal{W}$}-algebras}},
  \href{https://doi.org/10.1090/conm/248/03823}{{Recent Developments in Quantum
  Affine Algebras and Related Topics}}, Contemp. Math., vol. 248, Amer. Math.
  Soc., 1999, pp.~163--205,
  \href{https://arxiv.org/abs/math/9810055}{{\ttfamily math/9810055
  [math.QA]}}.

\bibitem[FR14]{Finkelberg:2010QDZ}
M.~Finkelberg and L.~Rybnikov, \emph{{Quantization of Drinfeld Zastava in type
  $A$}}, \href{https://doi.org/10.4171/JEMS/432}{J. Eur. Math. Soc. \textbf{16}
  (2014)}, no.~2, 235--271, \href{https://arxiv.org/abs/1009.0676}{{\ttfamily
  arXiv:1009.0676 [math.AG]}}.

\bibitem[Fre12]{Freund:1986ws}
P.~G.~O. Freund,
  \href{https://doi.org/10.1017/CBO9780511564017}{\emph{{Introduction to
  Supersymmetry}}}, Cambridge Monographs on Mathematical Physics, Cambridge
  Univ. Press, Cambridge, UK, 5 2012.

\bibitem[FSS96]{Frappat:1996pb}
L.~Frappat, P.~Sorba, and A.~Sciarrino, \emph{{Dictionary on Lie
  superalgebras}}, \href{https://arxiv.org/abs/hep-th/9607161}{{\ttfamily
  arXiv:hep-th/9607161}}.

\bibitem[Gai12]{Gaiotto:2009we}
D.~Gaiotto, \emph{{${\mathcal N}=2$ dualities}},
  \href{https://doi.org/10.1007/JHEP08(2012)034}{JHEP \textbf{08} (2012)}, 034,
  \href{https://arxiv.org/abs/0904.2715}{{\ttfamily arXiv:0904.2715 [hep-th]}}.

\bibitem[GLFMS17]{Gorsky:2017hro}
A.~Gorsky, B.~Le~Floch, A.~Milekhin, and N.~Sopenko, \emph{{Surface defects and
  instanton\textendash{}vortex interaction}},
  \href{https://doi.org/10.1016/j.nuclphysb.2017.04.010}{Nucl. Phys. B
  \textbf{920} (2017)}, 122--156,
  \href{https://arxiv.org/abs/1702.03330}{{\ttfamily arXiv:1702.03330
  [hep-th]}}.

\bibitem[GLFPP17]{Gomis:2016ljm}
J.~Gomis, B.~Le~Floch, Y.~Pan, and W.~Peelaers, \emph{{Intersecting Surface
  Defects and Two-Dimensional CFT}},
  \href{https://doi.org/10.1103/PhysRevD.96.045003}{Phys. Rev. D \textbf{96}
  (2017)}, no.~4, 045003, \href{https://arxiv.org/abs/1610.03501}{{\ttfamily
  arXiv:1610.03501 [hep-th]}}.

\bibitem[GS15]{Gottsche:2015AG}
L.~G\"{o}ttsche and V.~Shende, \emph{{The $\chi_{-y}$-genera of relative
  Hilbert schemes for linear systems on Abelian and K3 surfaces}},
  \href{https://doi.org/10.14231/ag-2015-017}{Algebraic Geom. \textbf{2}
  (2015)}, no.~4, 405--421, \href{https://arxiv.org/abs/1307.4316}{{\ttfamily
  arXiv:1307.4316 [math.AG]}}.

\bibitem[Guh10]{Guhr:2010nh}
T.~Guhr, \emph{{The Oxford Handbook of Random Matrix Theory}},
  \href{https://doi.org/10.1093/oxfordhb/9780198744191.013.7}{ch.~{Supersymmetry
  in Random Matrix Theory}, pp.~135--154}, Oxford Univ. Press, 2010,
  \href{https://arxiv.org/abs/1005.0979}{{\ttfamily arXiv:1005.0979
  [math-ph]}}.

\bibitem[HK22]{Haouzi:2019jzk}
N.~Haouzi and C.~Koz\c{c}az, \emph{{Supersymmetric Wilson Loops, Instantons,
  and Deformed $\mathcal{W}$-Algebras}},
  \href{https://doi.org/10.1007/s00220-022-04375-0}{Commun. Math. Phys.
  \textbf{393} (2022)}, no.~2, 669--779,
  \href{https://arxiv.org/abs/1907.03838}{{\ttfamily arXiv:1907.03838
  [hep-th]}}.

\bibitem[HKKP15]{Hwang:2014uwa}
C.~Hwang, J.~Kim, S.~Kim, and J.~Park, \emph{{General instanton counting and 5d
  SCFT}}, \href{https://doi.org/10.1007/JHEP07(2015)063,
  10.1007/JHEP04(2016)094}{JHEP \textbf{07} (2015)}, 063,
  \href{https://arxiv.org/abs/1406.6793}{{\ttfamily arXiv:1406.6793 [hep-th]}},
  [Addendum: JHEP04,094(2016)].

\bibitem[HKY15]{Hori:2014tda}
K.~Hori, H.~Kim, and P.~Yi, \emph{{Witten Index and Wall Crossing}},
  \href{https://doi.org/10.1007/JHEP01(2015)124}{JHEP \textbf{01} (2015)}, 124,
  \href{https://arxiv.org/abs/1407.2567}{{\ttfamily arXiv:1407.2567 [hep-th]}}.

\bibitem[HL09]{Hallnas:2009CA}
M.~Halln\"{a}s and E.~Langmann, \emph{A unified construction of generalized
  classical polynomials associated with operators
  of~calogero{\textendash}sutherland type},
  \href{https://doi.org/10.1007/s00365-009-9060-4}{Constr. Approx \textbf{31}
  (2009)}, no.~3, 309--342,
  \href{https://arxiv.org/abs/math-ph/0703090}{{\ttfamily arXiv:math-ph/0703090
  [math-ph]}}.

\bibitem[Hor90]{Horne:1989ue}
J.~H. Horne, \emph{{Skein Relations and Wilson Loops in {Chern-Simons} Gauge
  Theory}}, \href{https://doi.org/10.1016/0550-3213(90)90317-7}{Nucl. Phys. B
  \textbf{334} (1990)}, 669--694.

\bibitem[HW97]{Hanany:1996ie}
A.~Hanany and E.~Witten, \emph{{Type IIB superstrings, BPS monopoles, and
  three-dimensional gauge dynamics}},
  \href{https://doi.org/10.1016/S0550-3213(97)00157-0}{Nucl. Phys. B
  \textbf{492} (1997)}, 152--190,
  \href{https://arxiv.org/abs/hep-th/9611230}{{\ttfamily
  arXiv:hep-th/9611230}}.

\bibitem[IKP04]{Iqbal:2003ix}
A.~Iqbal and A.-K. Kashani-Poor, \emph{{Instanton counting and Chern-Simons
  theory}}, \href{https://doi.org/10.4310/ATMP.2003.v7.n3.a4}{Adv. Theor. Math.
  Phys. \textbf{7} (2004)}, 457--497,
  \href{https://arxiv.org/abs/hep-th/0212279}{{\ttfamily arXiv:hep-th/0212279
  [hep-th]}}.

\bibitem[IKP06]{Iqbal:2003zz}
\bysame, \emph{{$\mathrm{SU}(N)$ geometries and topological string
  amplitudes}}, \href{https://doi.org/10.4310/ATMP.2006.v10.n1.a1}{Adv. Theor.
  Math. Phys. \textbf{10} (2006)}, 1--32,
  \href{https://arxiv.org/abs/hep-th/0306032}{{\ttfamily arXiv:hep-th/0306032
  [hep-th]}}.

\bibitem[IKV09]{Iqbal:2007ii}
A.~Iqbal, C.~Kozcaz, and C.~Vafa, \emph{{The Refined topological vertex}},
  \href{https://doi.org/10.1088/1126-6708/2009/10/069}{JHEP \textbf{0910}
  (2009)}, 069, \href{https://arxiv.org/abs/hep-th/0701156}{{\ttfamily
  arXiv:hep-th/0701156 [hep-th]}}.

\bibitem[IMRY22]{Ishtiaque:2021jan}
N.~Ishtiaque, S.~F. Moosavian, S.~Raghavendran, and J.~Yagi, \emph{{Superspin
  chains from superstring theory}},
  \href{https://doi.org/10.21468/SciPostPhys.13.4.083}{SciPost Phys.
  \textbf{13} (2022)}, no.~4, 083,
  \href{https://arxiv.org/abs/2110.15112}{{\ttfamily arXiv:2110.15112
  [hep-th]}}.

\bibitem[JK95]{Jeffrey:1995}
L.~C. Jeffrey and F.~C. Kirwan, \emph{{Localization for nonabelian group
  actions}}, \href{https://doi.org/10.1016/0040-9383(94)00028-j}{Topology
  \textbf{34} (1995)}, no.~2, 291--327,
  \href{https://arxiv.org/abs/alg-geom/9307001}{{\ttfamily
  arXiv:alg-geom/9307001 [alg-geom]}}.

\bibitem[JLN21]{Jeong:2021rll}
S.~Jeong, N.~Lee, and N.~Nekrasov, \emph{{Intersecting defects in gauge theory,
  quantum spin chains, and Knizhnik-Zamolodchikov equations}},
  \href{https://doi.org/10.1007/JHEP10(2021)120}{JHEP \textbf{10} (2021)}, 120,
  \href{https://arxiv.org/abs/2103.17186}{{\ttfamily arXiv:2103.17186
  [hep-th]}}.

\bibitem[Kim16]{Kim:2016qqs}
H.-C. Kim, \emph{{Line defects and 5d instanton partition functions}},
  \href{https://doi.org/10.1007/JHEP03(2016)199}{JHEP \textbf{03} (2016)}, 199,
  \href{https://arxiv.org/abs/1601.06841}{{\ttfamily arXiv:1601.06841
  [hep-th]}}.

\bibitem[Kim20]{Kimura:2019hnw}
T.~Kimura, \emph{{Integrating over quiver variety and BPS/CFT correspondence}},
  \href{https://doi.org/10.1007/s11005-020-01261-5}{Lett. Math. Phys.
  \textbf{110} (2020)}, no.~6, 1237--1255,
  \href{https://arxiv.org/abs/1910.03247}{{\ttfamily arXiv:1910.03247
  [hep-th]}}.

\bibitem[Kim21]{Kimura:2020jxl}
\bysame, \href{https://doi.org/10.1007/978-3-030-76190-5}{\emph{{Instanton
  Counting, Quantum Geometry and Algebra}}}, Mathematical Physics Studies,
  Springer, 2021.

\bibitem[Kim22a]{Kimura:2022zsm}
\bysame, \emph{{Double Quiver Gauge Theory and BPS/CFT Correspondence}},
  \href{https://arxiv.org/abs/2212.03870}{{\ttfamily arXiv:2212.03870
  [hep-th]}}.

\bibitem[Kim22b]{Kimura:2022spi}
\bysame, \emph{{Higgsing $qq$-character and irreducibility}},
  \href{https://arxiv.org/abs/2205.08312}{{\ttfamily arXiv:2205.08312
  [math.QA]}}.

\bibitem[Kin71]{King:1971CJM}
R.~C. King, \emph{{The Dimensions of Irreducible Tensor Representations of the
  Orthogonal and Symplectic Groups}},
  \href{https://doi.org/10.4153/cjm-1971-017-2}{Can. J. Math. \textbf{23}
  (1971)}, no.~1, 176--188.

\bibitem[Kir16]{Kirillov:2016}
A.~Kirillov, \href{https://doi.org/10.1090/gsm/174}{\emph{{Quiver
  Representations and Quiver Varieties}}}, American Mathematical Society,
  August 2016.

\bibitem[KKV97]{Katz:1996fh}
S.~H. Katz, A.~Klemm, and C.~Vafa, \emph{{Geometric engineering of quantum
  field theories}}, \href{https://doi.org/10.1016/S0550-3213(97)00282-4}{Nucl.
  Phys. \textbf{B497} (1997)}, 173--195,
  \href{https://arxiv.org/abs/hep-th/9609239}{{\ttfamily arXiv:hep-th/9609239
  [hep-th]}}.

\bibitem[KL01]{Kraus:2000nj}
P.~Kraus and F.~Larsen, \emph{{Boundary string field theory of the D anti-D
  system}}, \href{https://doi.org/10.1103/PhysRevD.63.106004}{Phys. Rev. D
  \textbf{63} (2001)}, 106004,
  \href{https://arxiv.org/abs/hep-th/0012198}{{\ttfamily
  arXiv:hep-th/0012198}}.

\bibitem[KMV98]{Katz:1997eq}
S.~Katz, P.~Mayr, and C.~Vafa, \emph{{Mirror symmetry and exact solution of 4-D
  $N=2$ gauge theories: 1.}},
  \href{https://doi.org/10.4310/ATMP.1997.v1.n1.a2}{Adv. Theor. Math. Phys.
  \textbf{1} (1998)}, 53--114,
  \href{https://arxiv.org/abs/hep-th/9706110}{{\ttfamily hep-th/9706110}}.

\bibitem[KMZ12]{Kanno:2012hk}
S.~Kanno, Y.~Matsuo, and H.~Zhang, \emph{{Virasoro constraint for Nekrasov
  instanton partition function}},
  \href{https://doi.org/10.1007/JHEP10(2012)097}{JHEP \textbf{10} (2012)}, 097,
  \href{https://arxiv.org/abs/1207.5658}{{\ttfamily arXiv:1207.5658 [hep-th]}}.

\bibitem[KMZ13]{Kanno:2013aha}
\bysame, \emph{{Extended Conformal Symmetry and Recursion Formulae for Nekrasov
  Partition Function}}, \href{https://doi.org/10.1007/JHEP08(2013)028}{JHEP
  \textbf{1308} (2013)}, 028, \href{https://arxiv.org/abs/1306.1523}{{\ttfamily
  arXiv:1306.1523 [hep-th]}}.

\bibitem[KN21]{Kimura:2021ngu}
T.~Kimura and F.~Nieri, \emph{{Intersecting defects and supergroup gauge
  theory}}, \href{https://doi.org/10.1088/1751-8121/ac2716}{J. Phys. A
  \textbf{54} (2021)}, no.~43, 435401,
  \href{https://arxiv.org/abs/2105.02776}{{\ttfamily arXiv:2105.02776
  [hep-th]}}.

\bibitem[Kni95]{Knight:1995JA}
H.~Knight, \emph{{Spectra of Tensor Products of Finite Dimensional
  Representations of Yangians}},
  \href{https://doi.org/10.1006/jabr.1995.1123}{J. Algebra \textbf{174}
  (1995)}, 187--196.

\bibitem[KOO98]{Kerov:1998IMRN}
S.~Kerov, A.~Okounkov, and G.~Olshanski,
  \href{https://doi.org/10.1155/s1073792898000154}{Int. Math. Res. Not.
  \textbf{1998} (1998)}, no.~4, 173,
  \href{https://arxiv.org/abs/q-alg/9703037}{{\ttfamily arXiv:q-alg/9703037
  [math.QA]}}.

\bibitem[KP18a]{Kimura:2017hez}
T.~Kimura and V.~Pestun, \emph{{Fractional quiver W-algebras}},
  \href{https://doi.org/10.1007/s11005-018-1087-7}{Lett. Math. Phys.
  \textbf{108} (2018)}, 2425--2451,
  \href{https://arxiv.org/abs/1705.04410}{{\ttfamily arXiv:1705.04410
  [hep-th]}}.

\bibitem[KP18b]{Kimura:2016dys}
\bysame, \emph{{Quiver elliptic W-algebras}},
  \href{https://doi.org/10.1007/s11005-018-1073-0}{Lett. Math. Phys.
  \textbf{108} (2018)}, 1383--1405,
  \href{https://arxiv.org/abs/1608.04651}{{\ttfamily arXiv:1608.04651
  [hep-th]}}.

\bibitem[KP18c]{Kimura:2015rgi}
\bysame, \emph{{Quiver W-algebras}},
  \href{https://doi.org/10.1007/s11005-018-1072-1}{Lett. Math. Phys.
  \textbf{108} (2018)}, 1351--1381,
  \href{https://arxiv.org/abs/1512.08533}{{\ttfamily arXiv:1512.08533
  [hep-th]}}.

\bibitem[KP19]{Kimura:2019msw}
\bysame, \emph{{Super instanton counting and localization}},
  \href{https://arxiv.org/abs/1905.01513}{{\ttfamily arXiv:1905.01513
  [hep-th]}}.

\bibitem[KP23]{KPfractional}
\bysame, \emph{{Fractionalization of quiver variety and $qq$-character}}, to
  appear (2023).

\bibitem[KS20]{Kimura:2020lmc}
T.~Kimura and Y.~Sugimoto, \emph{{Topological Vertex/anti-Vertex and Supergroup
  Gauge Theory}}, \href{https://doi.org/10.1007/JHEP04(2020)081}{JHEP
  \textbf{04} (2020)}, 081, \href{https://arxiv.org/abs/2001.05735}{{\ttfamily
  arXiv:2001.05735 [hep-th]}}.

\bibitem[KT11]{Kanno:2011fw}
H.~Kanno and Y.~Tachikawa, \emph{{Instanton counting with a surface operator
  and the chain-saw quiver}},
  \href{https://doi.org/10.1007/JHEP06(2011)119}{JHEP \textbf{06} (2011)}, 119,
  \href{https://arxiv.org/abs/1105.0357}{{\ttfamily arXiv:1105.0357 [hep-th]}}.

\bibitem[LNS98]{Losev:1997tp}
A.~Losev, N.~Nekrasov, and S.~L. Shatashvili, \emph{{Issues in topological
  gauge theory}}, \href{https://doi.org/10.1016/S0550-3213(98)00628-2}{Nucl.
  Phys. \textbf{B534} (1998)}, 549--611,
  \href{https://arxiv.org/abs/hep-th/9711108}{{\ttfamily arXiv:hep-th/9711108
  [hep-th]}}.

\bibitem[LNS99]{Lossev:1997bz}
\bysame, \emph{{Testing Seiberg--Witten solution}},
  \href{https://doi.org/10.1007/978-94-011-4730-9_13}{{Strings, Branes and
  Dualities. NATO ASI Series}}, vol. 520, Springer Netherlands, 1999,
  pp.~359--372, \href{https://arxiv.org/abs/hep-th/9801061}{{\ttfamily
  arXiv:hep-th/9801061 [hep-th]}}.

\bibitem[LPSZ20]{Lodin:2018lbz}
R.~Lodin, A.~Popolitov, S.~Shakirov, and M.~Zabzine, \emph{{Solving q-Virasoro
  constraints}}, \href{https://doi.org/10.1007/s11005-019-01216-5}{Lett. Math.
  Phys. \textbf{110} (2020)}, no.~1, 179--210,
  \href{https://arxiv.org/abs/1810.00761}{{\ttfamily arXiv:1810.00761
  [hep-th]}}.

\bibitem[Meh04]{Mehta:2004RMT}
M.~L. Mehta, \href{https://doi.org/10.1016/S0079-8169(04)80088-6}{\emph{{Random
  Matrices}}}, 3rd ed., Pure and Applied Mathematics, vol. 142, Academic Press,
  2004.

\bibitem[Mik07]{Miki:2007JMP}
K.~Miki, \emph{{A ($q,\gamma$) analog of the W$_{1+\infty}$ algebra}},
  \href{https://doi.org/10.1063/1.2823979}{J. Math. Phys. \textbf{48} (2007)},
  no.~12, 123520.

\bibitem[Mik15]{Mikhaylov:2015nsa}
V.~Mikhaylov, \emph{{Analytic Torsion, 3d Mirror Symmetry And Supergroup
  Chern-Simons Theories}}, \href{https://arxiv.org/abs/1505.03130}{{\ttfamily
  arXiv:1505.03130 [hep-th]}}.

\bibitem[Miy66]{Miyazawa:1966mfa}
H.~Miyazawa, \emph{{Baryon Number Changing Currents}},
  \href{https://doi.org/10.1143/PTP.36.1266}{Prog. Theor. Phys. \textbf{36}
  (1966)}, no.~6, 1266--1276.

\bibitem[Miy68]{Miyazawa:1968zz}
\bysame, \emph{{Spinor Currents and Symmetries of Baryons and Mesons}},
  \href{https://doi.org/10.1103/PhysRev.170.1586}{Phys. Rev. \textbf{170}
  (1968)}, 1586--1590.

\bibitem[MN07]{Marshakov:2006ii}
A.~Marshakov and N.~A. Nekrasov, \emph{{Extended Seiberg--Witten theory and
  integrable hierarchy}},
  \href{https://doi.org/10.1088/1126-6708/2007/01/104}{JHEP \textbf{01}
  (2007)}, 104, \href{https://arxiv.org/abs/hep-th/0612019}{{\ttfamily
  arXiv:hep-th/0612019}}.

\bibitem[MNS00]{Moore:1997dj}
G.~W. Moore, N.~Nekrasov, and S.~Shatashvili, \emph{{Integrating over Higgs
  branches}}, \href{https://doi.org/10.1007/PL00005525}{Commun. Math. Phys.
  \textbf{209} (2000)}, 97--121,
  \href{https://arxiv.org/abs/hep-th/9712241}{{\ttfamily
  arXiv:hep-th/9712241}}.

\bibitem[MP10]{Marino:2009jd}
M.~Mari{\~n}o and P.~Putrov, \emph{{Exact Results in ABJM Theory from
  Topological Strings}}, \href{https://doi.org/10.1007/JHEP06(2010)011}{JHEP
  \textbf{1006} (2010)}, 011, \href{https://arxiv.org/abs/0912.3074}{{\ttfamily
  arXiv:0912.3074 [hep-th]}}.

\bibitem[MQ86]{Mathai:1986tc}
V.~Mathai and D.~G. Quillen, \emph{{Superconnections, Thom classes and
  equivariant differential forms}},
  \href{https://doi.org/10.1016/0040-9383(86)90007-8}{Topology \textbf{25}
  (1986)}, 85--110.

\bibitem[MSS22]{Marino:2022rpz}
M.~Marino, R.~Schiappa, and M.~Schwick, \emph{{New Instantons for Matrix
  Models}}, \href{https://arxiv.org/abs/2210.13479}{{\ttfamily arXiv:2210.13479
  [hep-th]}}.

\bibitem[MW96]{Martinec:1995by}
E.~J. Martinec and N.~P. Warner, \emph{{Integrable systems and supersymmetric
  gauge theory}}, \href{https://doi.org/10.1016/0550-3213(95)00588-9}{Nucl.
  Phys. \textbf{B459} (1996)}, 97--112,
  \href{https://arxiv.org/abs/hep-th/9509161}{{\ttfamily arXiv:hep-th/9509161
  [hep-th]}}.

\bibitem[MW04]{Marino:2004cn}
M.~Mari{\~n}o and N.~Wyllard, \emph{{A Note on instanton counting for
  $\mathcal{N}=2$ gauge theories with classical gauge groups}},
  \href{https://doi.org/10.1088/1126-6708/2004/05/021}{JHEP \textbf{05}
  (2004)}, 021, \href{https://arxiv.org/abs/hep-th/0404125}{{\ttfamily
  arXiv:hep-th/0404125 [hep-th]}}.

\bibitem[MW15]{Mikhaylov:2014aoa}
V.~Mikhaylov and E.~Witten, \emph{{Branes And Supergroups}},
  \href{https://doi.org/10.1007/s00220-015-2449-y}{Commun. Math. Phys.
  \textbf{340} (2015)}, no.~2, 699--832,
  \href{https://arxiv.org/abs/1410.1175}{{\ttfamily arXiv:1410.1175 [hep-th]}}.

\bibitem[Nak94a]{Nakajima:1994nid}
H.~Nakajima, \emph{{Instantons on ALE spaces, quiver varieties, and Kac--Moody
  algebras}}, \href{https://doi.org/10.1215/S0012-7094-94-07613-8}{Duke Math.
  J. \textbf{76} (1994)}, no.~2, 365--416.

\bibitem[Nak94b]{Nakajima:1993jg}
\bysame, \emph{{Resolutions of Moduli Spaces of Ideal Instantons on R$^4$}},
  \href{https://doi.org/10.1142/2407}{{Topology, Geometry and Field Theory}},
  World Scientific, 1994, pp.~129--136.

\bibitem[Nak98]{Nakajima:1998DM}
\bysame, \emph{{Quiver varieties and Kac--Moody algebras}},
  \href{https://doi.org/10.1215/s0012-7094-98-09120-7}{Duke Math. J.
  \textbf{91} (1998)}, 515--560.

\bibitem[Nak99]{Nakajima:1999}
\bysame, \href{https://doi.org/10.1090/ulect/018}{\emph{{Lectures on Hilbert
  Schemes of Points on Surfaces}}}, American Mathematical Society, 1999.

\bibitem[Nak01a]{Nakajima:1999JAMS}
\bysame, \emph{{Quiver varieties and finite-dimensional representations of
  quantum affine algebras}},
  \href{https://doi.org/10.1090/S0894-0347-00-00353-2}{J. Amer. Math. Soc.
  \textbf{14} (2001)}, 145--238,
  \href{https://arxiv.org/abs/math/9912158}{{\ttfamily math/9912158}}.

\bibitem[Nak01b]{Nakajima:2001PC}
\bysame, \emph{{$t$-analogue of the $q$-characters of finite dimensional
  representations of quantum affine algebras}},
  \href{https://doi.org/10.1142/9789812810007_0009}{Physics and Combinatorics
  (A.~N. Kirillov and N.~Liskova, eds.)}, {World Scientific}, 2001,
  pp.~196--219, \href{https://arxiv.org/abs/math/0009231}{{\ttfamily
  arXiv:math/0009231 [math.QA]}}.

\bibitem[Nak04]{Nakajima:2004AM}
\bysame, \emph{{Quiver varieties and $t$-analogs of $q$-characters of quantum
  affine algebras}}, \href{https://doi.org/10.4007/annals.2004.160.1057}{Ann.
  Math. \textbf{160} (2004)}, no.~3, 1057--1097,
  \href{https://arxiv.org/abs/math/0105173}{{\ttfamily arXiv:math/0105173
  [math.QA]}}.

\bibitem[Nak15]{Nakamura:2015zsa}
S.~Nakamura, \emph{{On the Jeffrey--Kirwan residue of BCD-instantons}},
  \href{https://doi.org/10.1093/ptep/ptv085}{PTEP \textbf{2015} (2015)}, no.~7,
  073B02, \href{https://arxiv.org/abs/1502.04188}{{\ttfamily arXiv:1502.04188
  [hep-th]}}.

\bibitem[Nek04a]{Nekrasov:2004UA}
N.~Nekrasov, \emph{{On the BPS/CFT correspondence}}, {Lecture at the University
  of Amsterdam String Theory Group Seminar}, 2004.

\bibitem[Nek04b]{Nekrasov:2002qd}
\bysame, \emph{{Seiberg--Witten Prepotential from Instanton Counting}},
  \href{https://doi.org/10.4310/ATMP.2003.v7.n5.a4}{Adv. Theor. Math. Phys.
  \textbf{7} (2004)}, 831--864,
  \href{https://arxiv.org/abs/hep-th/0206161}{{\ttfamily
  arXiv:hep-th/0206161}}.

\bibitem[Nek16]{Nekrasov:2015wsu}
\bysame, \emph{{BPS/CFT correspondence: non-perturbative Dyson--Schwinger
  equations and qq-characters}},
  \href{https://doi.org/10.1007/JHEP03(2016)181}{JHEP \textbf{03} (2016)}, 181,
  \href{https://arxiv.org/abs/1512.05388}{{\ttfamily arXiv:1512.05388
  [hep-th]}}.

\bibitem[Nek17]{Nekrasov:2017gzb}
\bysame, \emph{{BPS/CFT correspondence V: BPZ and KZ equations from
  qq-characters}}, \href{https://arxiv.org/abs/1711.11582}{{\ttfamily
  arXiv:1711.11582 [hep-th]}}.

\bibitem[Nek18]{Nekrasov:2016ydq}
\bysame, \emph{{BPS/CFT Correspondence III: Gauge Origami partition function
  and qq-characters}}, \href{https://doi.org/10.1007/s00220-017-3057-9}{Commun.
  Math. Phys. \textbf{358} (2018)}, no.~3, 863--894,
  \href{https://arxiv.org/abs/1701.00189}{{\ttfamily arXiv:1701.00189
  [hep-th]}}.

\bibitem[Nek19a]{Nekrasov:2017rqy}
\bysame, \emph{{BPS/CFT correspondence IV: sigma models and defects in gauge
  theory}}, \href{https://doi.org/10.1007/s11005-018-1115-7}{Lett. Math. Phys.
  \textbf{109} (2019)}, no.~3, 579--622,
  \href{https://arxiv.org/abs/1711.11011}{{\ttfamily arXiv:1711.11011
  [hep-th]}}.

\bibitem[Nek19b]{Nekrasov:2018gne}
\bysame, \emph{{Superspin chains and supersymmetric gauge theories}},
  \href{https://doi.org/10.1007/JHEP03(2019)102}{JHEP \textbf{03} (2019)}, 102,
  \href{https://arxiv.org/abs/1811.04278}{{\ttfamily arXiv:1811.04278
  [hep-th]}}.

\bibitem[NNZ17]{Nedelin:2016gwu}
A.~Nedelin, F.~Nieri, and M.~Zabzine, \emph{{$q$-Virasoro modular double and 3d
  partition functions}},
  \href{https://doi.org/10.1007/s00220-017-2882-1}{Commun. Math. Phys.
  \textbf{353} (2017)}, no.~3, 1059--1102,
  \href{https://arxiv.org/abs/1605.07029}{{\ttfamily arXiv:1605.07029
  [hep-th]}}.

\bibitem[NO06]{Nekrasov:2003rj}
N.~Nekrasov and A.~Okounkov,
  \href{https://doi.org/10.1007/0-8176-4467-9_15}{\emph{{Seiberg--Witten Theory
  and Random Partitions}}}, The Unity of Mathematics (P.~Etingof, V.~Retakh,
  and I.~M. Singer, eds.), Progress in Mathematics, vol. 244, Birkh\"auser
  Boston, 2006, pp.~525--596,
  \href{https://arxiv.org/abs/hep-th/0306238}{{\ttfamily arXiv:hep-th/0306238
  [hep-th]}}.

\bibitem[Nos22]{Noshita:2022dxv}
G.~Noshita, \emph{{5d AGT correspondence of supergroup gauge theories from
  quantum toroidal $\mathfrak{gl}_{1}$}},
  \href{https://doi.org/10.1007/JHEP12(2022)157}{JHEP \textbf{12} (2022)}, 157,
  \href{https://arxiv.org/abs/2209.08313}{{\ttfamily arXiv:2209.08313
  [hep-th]}}.

\bibitem[NP12]{Nekrasov:2012xe}
N.~Nekrasov and V.~Pestun, \emph{{Seiberg--Witten geometry of four dimensional
  $\mathcal{N}=2$ quiver gauge theories}},
  \href{https://arxiv.org/abs/1211.2240}{{\ttfamily arXiv:1211.2240 [hep-th]}}.

\bibitem[NPS18]{Nekrasov:2013xda}
N.~Nekrasov, V.~Pestun, and S.~Shatashvili, \emph{{Quantum Geometry and Quiver
  Gauge Theories}}, \href{https://doi.org/10.1007/s00220-017-3071-y}{Commun.
  Math. Phys. \textbf{357} (2018)}, no.~2, 519--567,
  \href{https://arxiv.org/abs/1312.6689}{{\ttfamily arXiv:1312.6689 [hep-th]}}.

\bibitem[NPZ18]{Nieri:2017ntx}
F.~Nieri, Y.~Pan, and M.~Zabzine, \emph{{3d Expansions of 5d Instanton
  Partition Functions}}, \href{https://doi.org/10.1007/JHEP04(2018)092}{JHEP
  \textbf{04} (2018)}, 092, \href{https://arxiv.org/abs/1711.06150}{{\ttfamily
  arXiv:1711.06150 [hep-th]}}.

\bibitem[NS98]{Nekrasov:1998ss}
N.~Nekrasov and A.~S. Schwarz, \emph{{Instantons on Noncommutative
  $\mathbb{R}^4$ and $(2,0)$ Superconformal Six Dimensional Theory}},
  \href{https://doi.org/10.1007/s002200050490}{Commun. Math. Phys. \textbf{198}
  (1998)}, 689--703, \href{https://arxiv.org/abs/hep-th/9802068}{{\ttfamily
  arXiv:hep-th/9802068}}.

\bibitem[NS04]{Nekrasov:2004vw}
N.~Nekrasov and S.~Shadchin, \emph{{ABCD of instantons}},
  \href{https://doi.org/10.1007/s00220-004-1189-1}{Commun. Math. Phys.
  \textbf{252} (2004)}, 359--391,
  \href{https://arxiv.org/abs/hep-th/0404225}{{\ttfamily
  arXiv:hep-th/0404225}}.

\bibitem[NY99]{Noumi:1999NMJ}
M.~Noumi and Y.~Yamada, \emph{{Symmetries in the fourth Painlev\'e equation and
  Okamoto polynomials}},
  \href{https://doi.org/10.1017/S0027763000006899}{Nagoya Math. J. \textbf{153}
  (1999)}, 53--86, \href{https://arxiv.org/abs/q-alg/9708018}{{\ttfamily
  arXiv:q-alg/9708018 [math.QA]}}.

\bibitem[NY03]{Nakajima:2003uh}
H.~Nakajima and K.~Yoshioka, \emph{{Lectures on instanton counting}},
  \href{https://doi.org/10.1090/crmp/038/02}{CRM Proc. Lec. Notes \textbf{38}
  (2003)}, 31--102, \href{https://arxiv.org/abs/math/0311058}{{\ttfamily
  arXiv:math/0311058 [math.AG]}}.

\bibitem[NY05a]{Nakajima:2003pg}
\bysame, \emph{{Instanton counting on blowup. I. 4-dimensional pure gauge
  theory}}, \href{https://doi.org/10.1007/s00222-005-0444-1}{Invent. Math.
  \textbf{162} (2005)}, 313--355,
  \href{https://arxiv.org/abs/math/0306198}{{\ttfamily arXiv:math/0306198
  [math.AG]}}.

\bibitem[NY05b]{Nakajima:2005fg}
\bysame, \emph{{Instanton counting on blowup. II: K-theoretic partition
  function}}, \href{https://doi.org/10.1007/s00031-005-0406-0}{Transf. Groups
  \textbf{10} (2005)}, 489--519,
  \href{https://arxiv.org/abs/math/0505553}{{\ttfamily arXiv:math/0505553}}.

\bibitem[NZ17]{Nedelin:2015mio}
A.~Nedelin and M.~Zabzine, \emph{{q-Virasoro constraints in matrix models}},
  \href{https://doi.org/10.1007/JHEP03(2017)098}{JHEP \textbf{03} (2017)}, 098,
  \href{https://arxiv.org/abs/1511.03471}{{\ttfamily arXiv:1511.03471
  [hep-th]}}.

\bibitem[NZ20]{Nieri:2019mdl}
F.~Nieri and Y.~Zenkevich, \emph{{Quiver $\text{W}_{\epsilon_1,\epsilon_2}$
  algebras of 4d $\mathcal{N}=2$ gauge theories}},
  \href{https://doi.org/10.1088/1751-8121/ab9275}{J. Phys. A \textbf{53}
  (2020)}, no.~27, 275401, \href{https://arxiv.org/abs/1912.09969}{{\ttfamily
  arXiv:1912.09969 [hep-th]}}.

\bibitem[OR10]{Orlando:2010uu}
D.~Orlando and S.~Reffert, \emph{{Relating Gauge Theories via Gauge/Bethe
  Correspondence}}, \href{https://doi.org/10.1007/JHEP10(2010)071}{JHEP
  \textbf{10} (2010)}, 071, \href{https://arxiv.org/abs/1005.4445}{{\ttfamily
  arXiv:1005.4445 [hep-th]}}.

\bibitem[OS18]{Okazaki:2017sbc}
T.~Okazaki and D.~J. Smith, \emph{{Matrix supergroup Chern-Simons models for
  vortex-antivortex systems}},
  \href{https://doi.org/10.1007/JHEP02(2018)119}{JHEP \textbf{02} (2018)}, 119,
  \href{https://arxiv.org/abs/1712.01370}{{\ttfamily arXiv:1712.01370
  [hep-th]}}.

\bibitem[Osb78]{Osborn:1978rn}
H.~Osborn, \emph{{Solutions of the Dirac Equation for General Instanton
  Solutions}}, \href{https://doi.org/10.1016/0550-3213(78)90312-7}{Nucl. Phys.
  \textbf{B140} (1978)}, 45--53.

\bibitem[OT06]{Okuda:2006fb}
T.~Okuda and T.~Takayanagi, \emph{{Ghost D-branes}},
  \href{https://doi.org/10.1088/1126-6708/2006/03/062}{JHEP \textbf{03}
  (2006)}, 062, \href{https://arxiv.org/abs/hep-th/0601024}{{\ttfamily
  arXiv:hep-th/0601024 [hep-th]}}.

\bibitem[Pau40]{Pauli:1940zz}
W.~Pauli, \emph{{The Connection Between Spin and Statistics}},
  \href{https://doi.org/10.1103/PhysRev.58.716}{Phys. Rev. \textbf{58} (1940)},
  716--722.

\bibitem[Pes17]{Pestun:2016qko}
V.~Pestun, \emph{{Review of localization in geometry}},
  \href{https://doi.org/10.1088/1751-8121/aa6161}{J. Phys. \textbf{A50}
  (2017)}, no.~44, 443002, \href{https://arxiv.org/abs/1608.02954}{{\ttfamily
  arXiv:1608.02954 [hep-th]}}.

\bibitem[PP17]{Pan:2016fbl}
Y.~Pan and W.~Peelaers, \emph{{Intersecting Surface Defects and Instanton
  Partition Functions}}, \href{https://doi.org/10.1007/JHEP07(2017)073}{JHEP
  \textbf{07} (2017)}, 073, \href{https://arxiv.org/abs/1612.04839}{{\ttfamily
  arXiv:1612.04839 [hep-th]}}.

\bibitem[PS79]{Parisi:1979ka}
G.~Parisi and N.~Sourlas, \emph{{Random Magnetic Fields, Supersymmetry and
  Negative Dimensions}},
  \href{https://doi.org/10.1103/PhysRevLett.43.744}{Phys. Rev. Lett.
  \textbf{43} (1979)}, 744.

\bibitem[PS82]{Parisi:1982ud}
\bysame, \emph{{Supersymmetric Field Theories and Stochastic Differential
  Equations}}, \href{https://doi.org/10.1016/0550-3213(82)90538-7}{Nucl. Phys.
  B \textbf{206} (1982)}, 321--332.

\bibitem[PZ{\etalchar{+}}17]{Pestun:2016zxk}
V.~Pestun, M.~Zabzine, et~al., \emph{{Localization techniques in quantum field
  theories}}, \href{https://doi.org/10.1088/1751-8121/aa63c1}{J. Phys.
  \textbf{A50} (2017)}, no.~44, 440301,
  \href{https://arxiv.org/abs/1608.02952}{{\ttfamily arXiv:1608.02952
  [hep-th]}}.

\bibitem[QS13]{Quella:2013oda}
T.~Quella and V.~Schomerus, \emph{{Superspace conformal field theory}},
  \href{https://doi.org/10.1088/1751-8113/46/49/494010}{J. Phys. \textbf{A46}
  (2013)}, 494010, \href{https://arxiv.org/abs/1307.7724}{{\ttfamily
  arXiv:1307.7724 [hep-th]}}.

\bibitem[Qui85]{Quillen:1985vya}
D.~Quillen, \emph{{Superconnections and the Chern character}},
  \href{https://doi.org/10.1016/0040-9383(85)90047-3}{Topology \textbf{24}
  (1985)}, no.~1, 89--95.

\bibitem[Rap20]{Rapcak:2019wzw}
M.~Rap\v{c}\'ak, \emph{{On extensions of $
  \mathfrak{gl}\widehat{\left(\left.m\right|n\right)} $ Kac-Moody algebras and
  Calabi-Yau singularities}},
  \href{https://doi.org/10.1007/JHEP01(2020)042}{JHEP \textbf{01} (2020)}, 042,
  \href{https://arxiv.org/abs/1910.00031}{{\ttfamily arXiv:1910.00031
  [hep-th]}}.

\bibitem[RS94]{Rozansky:1992zt}
L.~Rozansky and H.~Saleur, \emph{{Reidemeister torsion, the Alexander
  polynomial and U(1,1) Chern--Simons Theory}},
  \href{https://doi.org/10.1016/0393-0440(94)90022-1}{J. Geom. Phys.
  \textbf{13} (1994)}, 105--123,
  \href{https://arxiv.org/abs/hep-th/9209073}{{\ttfamily arXiv:hep-th/9209073
  [hep-th]}}.

\bibitem[Sch87]{Schlottmann:1987zz}
P.~Schlottmann, \emph{{Integrable narrow-band model with possible relevance to
  heavy-fermion systems}},
  \href{https://doi.org/10.1103/PhysRevB.36.5177}{Phys. Rev. B \textbf{36}
  (1987)}, 5177--5185.

\bibitem[Ser01]{Sergeev:2001JNMP}
A.~Sergeev, \emph{{Superanalogs of the Calogero Operators and Jack
  Polynomials}}, \href{https://doi.org/10.2991/jnmp.2001.8.1.7}{J. Nonlin.
  Math. Phys. \textbf{8} (2001)}, 59--64,
  \href{https://arxiv.org/abs/math/0106222}{{\ttfamily arXiv:math/0106222
  [math.RT]}}.

\bibitem[Ser02]{Sergeev:2002TMP}
A.~N. Sergeev, \emph{{Calogero Operator and Lie Superalgebras }},
  \href{https://doi.org/10.1023/a:1015968505753}{Theor. Math. Phys.
  \textbf{131} (2002)}, no.~3, 747--764.

\bibitem[Sha05]{Shadchin:2005mx}
S.~Shadchin, \emph{{On certain aspects of string theory/gauge theory
  correspondence}}, Ph.D. thesis, Ecole Polytechnique, 2005,
  \href{https://arxiv.org/abs/hep-th/0502180}{{\ttfamily arXiv:hep-th/0502180
  [hep-th]}}.

\bibitem[SKAO96]{Shiraishi:1995rp}
J.~Shiraishi, H.~Kubo, H.~Awata, and S.~Odake, \emph{{A Quantum deformation of
  the Virasoro algebra and the Macdonald symmetric functions}},
  \href{https://doi.org/10.1007/BF00398297}{Lett. Math. Phys. \textbf{38}
  (1996)}, 33--51, \href{https://arxiv.org/abs/q-alg/9507034}{{\ttfamily
  arXiv:q-alg/9507034 [math.QA]}}.

\bibitem[SST23]{Schiappa:2023ned}
R.~Schiappa, M.~Schwick, and N.~Tamarin, \emph{{All the D-Branes of
  Resurgence}}, \href{https://arxiv.org/abs/2301.05214}{{\ttfamily
  arXiv:2301.05214 [hep-th]}}.

\bibitem[Sut75]{Sutherland:1975vr}
B.~Sutherland, \emph{{A General Model for Multicomponent Quantum Systems}},
  \href{https://doi.org/10.1103/PhysRevB.12.3795}{Phys. Rev. B \textbf{12}
  (1975)}, 3795--3805.

\bibitem[SV05]{Sergeev:2005AM}
A.~Sergeev and A.~Veselov, \emph{{Generalised discriminants, deformed
  Calogero{\textendash}Moser{\textendash}Sutherland operators and super-Jack
  polynomials}}, \href{https://doi.org/10.1016/j.aim.2004.04.009}{Adv. Math.
  \textbf{192} (2005)}, no.~2, 341--375,
  \href{https://arxiv.org/abs/math-ph/0307036}{{\ttfamily arXiv:math-ph/0307036
  [math-ph]}}.

\bibitem[SW99]{Seiberg:1999vs}
N.~Seiberg and E.~Witten, \emph{{String theory and noncommutative geometry}},
  \href{https://doi.org/10.1088/1126-6708/1999/09/032}{JHEP \textbf{09}
  (1999)}, 032, \href{https://arxiv.org/abs/hep-th/9908142}{{\ttfamily
  arXiv:hep-th/9908142 [hep-th]}}.

\bibitem[Tan09]{Taniguchi:2009zz}
T.~Taniguchi, \emph{{ADHM construction of super Yang-Mills instantons}},
  \href{https://doi.org/10.1016/j.geomphys.2009.06.003}{J. Geom. Phys.
  \textbf{59} (2009)}, 1199--1209.

\bibitem[Thi10]{Thind:2010}
J.~Thind, \emph{{Quiver Representations in the Super-Category and Gabriel's
  Theorem for $A(m,n)$}}, \href{https://arxiv.org/abs/1010.3056}{{\ttfamily
  arXiv:1010.3056 [math.RT]}}.

\bibitem[TTU01]{Takayanagi:2000rz}
T.~Takayanagi, S.~Terashima, and T.~Uesugi, \emph{{Brane - anti-brane action
  from boundary string field theory}},
  \href{https://doi.org/10.1088/1126-6708/2001/03/019}{JHEP \textbf{03}
  (2001)}, 019, \href{https://arxiv.org/abs/hep-th/0012210}{{\ttfamily
  arXiv:hep-th/0012210}}.

\bibitem[Vaf01]{Vafa:2001qf}
C.~Vafa, \emph{{Brane/anti-brane systems and U($N|M$) supergroup}},
  \href{https://arxiv.org/abs/hep-th/0101218}{{\ttfamily
  arXiv:hep-th/0101218}}.

\bibitem[Vaf14]{Vafa:2014iua}
\bysame, \emph{{Non-Unitary Holography}},
  \href{https://arxiv.org/abs/1409.1603}{{\ttfamily arXiv:1409.1603 [hep-th]}}.

\bibitem[{van}94]{vanderLende:1994JMP}
E.~{van der Lende}, \emph{{Super-Toda lattices}},
  \href{https://doi.org/10.1063/1.530586}{J. Math. Phys. \textbf{35} (1994)},
  no.~3, 1233--1251.

\bibitem[Var04]{Varadarajan:2004yz}
V.~S. Varadarajan, \href{https://doi.org/10.1090/cln/011}{\emph{{Supersymmetry
  for mathematicians: An introduction}}}, Courant Lecture Notes, vol.~11,
  American Mathematical Society, Providence, R.I, 2004.

\bibitem[Vor15]{Voronov:2015cya}
T.~Voronov, \emph{{On volumes of classical supermanifolds}},
  \href{https://doi.org/10.1070/SM8705}{Sb. Math. \textbf{207} (2015)},
  1512--1536, \href{https://arxiv.org/abs/1503.06542}{{\ttfamily
  arXiv:1503.06542 [math.DG]}}.

\bibitem[Weg16]{Wegner:2016ahw}
F.~Wegner,
  \href{https://doi.org/10.1007/978-3-662-49170-6}{\emph{{Supermathematics and
  its Applications in Statistical Physics}: {Grassmann Variables and the Method
  of Supersymmetry}}}, Lect. Notes Phys., vol. 920, Springer, 2016.

\bibitem[Wit97]{Witten:1997sc}
E.~Witten, \emph{{Solutions of four-dimensional field theories via M-theory}},
  \href{https://doi.org/10.1016/S0550-3213(97)00416-1}{Nucl. Phys.
  \textbf{B500} (1997)}, 3--42,
  \href{https://arxiv.org/abs/hep-th/9703166}{{\ttfamily
  arXiv:hep-th/9703166}}.

\bibitem[Wit98]{Witten:1998cd}
\bysame, \emph{{D-branes and K-theory}},
  \href{https://doi.org/10.1088/1126-6708/1998/12/019}{JHEP \textbf{12}
  (1998)}, 019, \href{https://arxiv.org/abs/hep-th/9810188}{{\ttfamily
  arXiv:hep-th/9810188}}.

\bibitem[Wit11]{Witten:2010cx}
\bysame, \emph{{Analytic Continuation Of Chern--Simons Theory}},
  \href{https://doi.org/10.1090/amsip/050/19}{AMS/IP Stud. Adv. Math.
  \textbf{50} (2011)}, 347--446,
  \href{https://arxiv.org/abs/1001.2933}{{\ttfamily arXiv:1001.2933 [hep-th]}}.

\bibitem[Yos92]{Yost:1991ht}
S.~A. Yost, \emph{{Supermatrix models}},
  \href{https://doi.org/10.1142/S0217751X92002775}{Int. J. Mod. Phys. A
  \textbf{7} (1992)}, 6105--6120,
  \href{https://arxiv.org/abs/hep-th/9111033}{{\ttfamily
  arXiv:hep-th/9111033}}.

\bibitem[Zen18]{Zenkevich:2018fzl}
Y.~Zenkevich, \emph{{Higgsed network calculus}},
  \href{https://arxiv.org/abs/1812.11961}{{\ttfamily arXiv:1812.11961
  [hep-th]}}.

\end{thebibliography}

\end{document}